\documentclass[preprint,3p,12pt]{elsarticle}
\usepackage{mathrsfs}
\usepackage{amsmath}
\usepackage{stmaryrd}
\usepackage{bbding}
\usepackage{dcolumn}
\usepackage{graphicx}
\usepackage{amsfonts}
\usepackage{amssymb}
\usepackage{psfrag}
\usepackage{wrapfig}
\usepackage{makeidx}
\usepackage{tabularx}
\usepackage{bm}
\usepackage{diagbox}
\usepackage{epsf}
\usepackage{epsfig}
\usepackage{setspace}
\usepackage{graphicx}
\usepackage{epstopdf}
\usepackage{psfrag}
\usepackage{subcaption}
\usepackage{color}
\usepackage{multirow}
\usepackage{comment}
\usepackage{hyperref}
\usepackage{algorithm}
\usepackage{algpseudocode}

\begin{document}
\title{Very High-order Compact Gas-kinetic Scheme With Discontinuity Feedback Factor}
\author[XJTU]{Junlei Mu}
\ead{mujl@stu.xjtu.edu.cn}
		
\author[XJTU]{Hong Zhang}
\ead{zhanghong2001@stu.xjtu.edu.cn}
		
\author[XJTU]{Xing Ji\corref{cor1}}
\ead{jixing@xjtu.edu.cn}

\author[XJTU]{Yang Zhang}
\ead{youngz@xjtu.edu.cn}

\author[XJTU]{Gang Chen}
\ead{aachengang@xjtu.edu.cn}
		
\author[HKUST1,HKUST2,HKUST3]{Kun Xu}
\ead{makxu@ust.hk}

\address[XJTU]{State Key Laboratory for Strength and Vibration of Mechanical Structures, Shaanxi Key Laboratory
	of Environment and Control for Flight Vehicle, School of Aerospace Engineering, Xi'an Jiaotong
	University, Xi'an, Shaanxi, China}
\address[HKUST1]{Department of Mathematics, Hong Kong University of Science and Technology, Clear Water Bay, Kowloon, Hong Kong}
\address[HKUST2]{Department of Mechanical and Aerospace Engineering, Hong Kong University of Science and Technology, Clear Water Bay, Kowloon, Hong Kong}
\address[HKUST3]{Shenzhen Research Institute, Hong Kong University of Science and Technology, Shenzhen, China}
\cortext[cor1]{Corresponding author}	

\begin{abstract}

This paper presents a robust and efficient very high-order scheme for compressible flow simulation, addressing key limitations of existing high-order methods. The proposed approach integrates the compact gas-kinetic scheme (CGKS) with an adaptive stencil extension reconstruction with a discontinuity feedback factor (ASE-DFF), resulting in significant improvements in both robustness and computational efficiency. Traditional weighted essentially non-oscillatory (WENO) schemes often exhibit reduced robustness at higher orders and require computationally expensive smoothness indicators for large stencils. On the other hand, compact methods based on Discontinuous Galerkin (DG) and Flux Reconstruction (FR) suffer from low time-marching efficiency.
In contrast, the ASE-DFF-CGKS introduces two major innovations: (1) a unified framework that enables arbitrarily high-order compact gas-kinetic schemes without sacrificing a large CFL number, and (2) a discontinuity feedback factor that eliminates the need for costly smoothness indicator calculations while maintaining first-order robustness near discontinuities. The advantages of the proposed scheme are demonstrated through benchmark simulations, where it sustains a CFL number above 0.5 for schemes up to ninth order—significantly higher than conventional compact methods. Moreover, it delivers high-resolution results for flows involving strong shocks and rarefaction waves. Overall, this work offers a practically impactful solution for high-fidelity compressible flow simulation, achieving an effective balance between computational efficiency, high-order accuracy, and robustness in challenging flow regimes.

\end{abstract}


\maketitle
\section{Introduction}

Over the past few decades, high-order computational fluid dynamics (CFD) methods for solving the Euler and Navier-Stokes equations in compressible flows have seen significant advancements. Among these methods, weighted essentially non-oscillatory (WENO) schemes~\cite{JIANG1996202,LIU1994200} on structured meshes have gained prominence for their simplicity, robustness, and ability to achieve high-order accuracy~\cite{BALSARA2016780}. However, when applied to unstructured meshes, WENO schemes face challenges associated with large stencils, which complicate their implementation in complex geometries.
In contrast, high-order compact methods such as discontinuous Galerkin (DG)\cite{10.1007/978-3-642-59721-3_1,COCKBURN198990}, flux reconstruction (FR)\cite{VINCENT20118134}, and correction procedure via reconstruction (CPR)~\cite{LUO20088875,WANG20098161} frameworks offer greater geometric flexibility and efficient parallelization. Despite these advantages, their stringent CFL conditions impose severe restrictions on the allowable time step, thus limiting computational efficiency. Additionally, these methods continue to face challenges in accurately capturing shock discontinuities.
To address these limitations, the compact gas-kinetic scheme (CGKS)~\cite{JI2018446,Zhao2019,JI2020109367,doi:10.2514/1.J060208,Ji09082021} has been developed, enabling the use of larger CFL numbers (typically around 0.5 or higher) while stably resolving strong shocks in complex compressible flows.

The compact gas-kinetic scheme (CGKS)\cite{JI2018446,Zhao2019,li2021compact} represents an innovative advancement in computational fluid dynamics. This method demonstrates exceptional capability in achieving third-order accuracy within a compact framework on three-dimensional unstructured meshes\cite{JI2020109367,doi:10.2514/1.J060208,Ji09082021}. Extensive numerical experiments involving complex three-dimensional compressible flows have consistently validated its robustness and accuracy~\cite{YANG2024112902,YANG2025113534}.
A fundamental distinction sets CGKS apart from conventional CFD methods. Building upon the classical two-stage fourth-order gas-kinetic scheme (GKS) formulation introduced by Pan and Xu~\cite{PAN2016197}, CGKS departs from the widely used fourth-order Runge-Kutta (RK4) temporal discretization~\cite{doi:10.2514/6.1981-1259}, instead incorporating concepts from the two-stage fourth-order generalized Riemann problem (GRP) approach developed by Li~\cite{doi:10.1137/15M1052512}. This implementation leverages the time-dependent gas-kinetic flux distribution function, while spatial discretization employs traditional WENO reconstruction to determine conservative variables at cell interfaces for the gas-kinetic flux function.
Recognizing the inherent complexity of gas-kinetic flux evaluation, Ji et al. have proposed enhanced variants, such as the WENO-AO GKS, to streamline computations~\cite{YANG2022110706,CiCP-28-539}. While standard high-order gas-kinetic schemes with WENO reconstruction may underutilize the method's full potential, CGKS effectively combines three key elements: the GKS flux formulation, two-stage fourth-order temporal discretization, and Hermite WENO reconstruction. The solution evolution process in CGKS shares conceptual similarities with discontinuous Galerkin (DG) methods and Hermite WENO (HWENO) schemes~\cite{QIU2004115}, requiring simultaneous updates of both conservative variables and their gradients within each computational cell.
The relative simplicity of the CGKS framework arises primarily from the unique properties of the gas-kinetic flux. By fully exploiting the spatiotemporal evolution characteristics of the gas-kinetic distribution function at interfaces, CGKS enables the instantaneous determination of conservative variables through moment integration of the distribution function. This elegant formulation also permits straightforward gradient updates via Gauss's theorem, offering notable computational efficiency.

The WENO scheme has undergone extensive development and has found widespread application in computational fluid dynamics. For high-speed turbulence simulations, direct numerical simulation (DNS) approaches~\cite{LELE199216,HATTORI201334} predominantly employ high-order finite difference methods that incorporate WENO schemes. Building on this foundation, Qiu and Zhu introduced the Hermite WENO scheme~\cite{QIU2004115,Liu2015,zhao2020hybrid}, which utilizes the gradient information of conservative variables within each cell. Similar to conventional WENO methods, Hermite WENO constructs high-order polynomials using cell-centered information and employs smoothness indicators to weight sub-stencils for effective discontinuity resolution. The key distinction lies in its ability to achieve enhanced accuracy on more compact stencils by incorporating gradient data. Physically, these compact stencils demonstrate superior performance in resolving discontinuities across computational domains.
To facilitate the integration of WENO schemes within the finite volume framework, the development of the WENO-AO scheme has effectively addressed issues related to negative weights during Gaussian point value reconstruction~\cite{BALSARA2016780,CiCP-28-539}. Recent advances involving the discontinuity feedback factor~\cite{Ji09082021} have demonstrated remarkable success when implemented in high-order numerical methods~\cite{Ji09082021,Zhang2024}, exhibiting exceptional robustness and shock-capturing capability in three-dimensional supersonic and hypersonic flow simulations. Further refinements have led to the development of adaptive stencil extension with discontinuity feedback reconstruction on structured meshes~\cite{zhang2024adaptivereconstructionmethodarbitrary}, which maintains solution robustness while efficiently achieving seventh- and ninth-order accuracy through non-compact stencils combined with the WENO scheme. This advancement is particularly valuable, as the computation of smoothness indicators for high-order stencils typically entails significant computational expense.

For high-order schemes, it is crucial to design methods capable of maintaining high accuracy in the presence of shock waves or turbulence~\cite{pirozzoli2011numerical}. Achieving seventh-, ninth-, or even higher-order accuracy is essential for resolving flows that contain shocks or turbulence. However, increasing the order of accuracy inevitably poses severe challenges to robustness. The primary reason is that higher-order reconstructions substantially reduce numerical dissipation, making it more difficult to suppress non-physical oscillations near shocks or strong discontinuities~\cite{BALSARA2000405,SURESH199783}. The typical solution involves optimizing the nonlinear stability mechanism and appropriately reducing the order at discontinuity points.
For WENO and TENO schemes, adaptive-accuracy variants such as WENO-AO~\cite{BALSARA2016780}, TENO-AA~\cite{FU2021114193}, and other hybrid WENO schemes~\cite{ADAMS199627,PIROZZOLI2006489} can deliver ninth- or tenth-order accuracy. However, beyond the significant computational cost of evaluating smoothness indicators~\cite{baeza2019efficient}, these schemes often require complex stencil-based smoothness assessment mechanisms or rely on positivity-preserving fluxes to pass stringent benchmark cases. In the case of DG formulations, the incorporation of limiters or artificial viscosity is necessary, yet designing high-order limiters remains notoriously challenging and often leads to excessive loss of accuracy~\cite{KRIVODONOVA2007879}. Moreover, the large stencils required by very high-order schemes complicate the imposition of boundary conditions and limit adaptability to unstructured grids, thereby undermining robustness.

This work employs the discontinuity feedback factor to achieve higher-order spatial accuracy within the compact gas-kinetic scheme (CGKS) framework~\cite{JI2018446,Zhao2019}. The proposed CGKS with a discontinuity feedback factor attains seventh- and ninth-order spatial accuracy using a central stencil. Compared to non-compact high-order gas-kinetic schemes, it utilizes the same stencil size as the fifth-order non-compact HGKS, while overcoming the primary limitations of previous compact GKS methods, including sixth- and eighth-order compact GKS.
Earlier CGKS formulations constructed the non-equilibrium and equilibrium states separately on both sides of the interface. However, both subsequent non-compact and compact HGKS abandoned this approach, instead obtaining the equilibrium state directly through kinetic weighting of the non-equilibrium state. The drawback of using larger stencils is that equilibrium state construction at the interface still requires a non-compact stencil. By introducing the discontinuity feedback factor strategy, it becomes possible to construct fifth-, seventh-, and ninth-order CGKS using compact stencils, obtaining the equilibrium state through kinetic weighting of higher-order non-equilibrium states. In addition, the tangential reconstruction process in the two-dimensional CGKS is simplified when employing the discontinuity feedback factor. Furthermore, the high-order CGKS based on this strategy eliminates the need for computationally expensive high-order smoothness indicators, thereby reducing overall computational cost. The discontinuity feedback factor plays a crucial role in shock capturing, enabling accurate resolution of shock discontinuities and effective compression of higher-order reconstruction polynomial slopes, thus imparting strong robustness to the CGKS.

The organization of this paper is as follows:
Section~\ref{sec2} introduces the gas-kinetic scheme and the two-stage fourth-order time advancement method. Section~\ref{sec3} presents the discontinuity feedback factor and a series of compact stencils for achieving high-order accuracy in the compact gas-kinetic scheme; the two-dimensional reconstruction framework of the CGKS is also detailed. Section~\ref{sec4} demonstrates the performance of the scheme, including its accuracy and robustness, through various numerical cases. Finally, the paper concludes with a summary of the current work.

\section{Gas-kinetic compact framework with two-stage fourth-order solution update}
\label{sec2}
This section outlines the core numerical framework, including:(i) the gas-kinetic flux solver,
(ii) the two-stage fourth-order temporal discretization method, and
(iii) the update procedures for conservative variables and their spatial gradients.
For comprehensive theoretical derivations and implementation details, readers are referred to the original
compact GKS formulation in~\cite{JI2018446}.

\subsection{Gas-kinetic flux solver}
\label{sec2.1}
The gas-kinetic evolution model in the gas-kinetic flux is based on the BGK model equation ~\cite{PhysRev.94.511}, which is given
in two-dimensional case as follows
\begin{equation}\label{equ2.1}
{{f}_{t}}+\mathbf{u}\cdot \nabla f=\frac{g-f}{\tau },
\end{equation}
where $f$ is the gas distribution function, $\mathbf{u}=\left( u,v \right)$ is
the particle velocity, and $\tau $ is the collision time. $g$ is the corresponding Maxwellian distribution
for equilibrium state and written as follows,
\begin{equation}\label{equ2.2}
g=\rho {{\left( \frac{\lambda }{\pi } \right)}^{\frac{K+3}{2}}}{{e}^{-\lambda \left( {{\left( u-U \right)}
^{2}}+{{\left( v-V \right)}^{2}}+{{\xi }^{2}} \right)}},
\end{equation}
where $\lambda$ is defined as $m/2kT$, $m$ represents the molecular mass, $T$ denotes the temperature, and $k$ is
the Boltzmann constant. $\rho$ is the density, while $U$ and $V$ correspond to the macroscopic velocity in the
x-direction and y-direction, respectively. The term $K=\left( 5-3\gamma  \right)/\left( \gamma -1 \right)+1$, which
represents the internal degree of freedom, with $\gamma$ being the specific heat ratio. In equilibrium, the internal
variable  ${{\xi }^{2}}=\xi _{1}^{2}+\xi _{2}^{2}+\cdots +\xi _{K}^{2}$.
The collision term satisfies the compatibility condition
\begin{equation}\label{equ2.3}
\int{\frac{g-f}{\tau }}\boldsymbol{\psi} d\Xi =0,
\end{equation}
where $\boldsymbol{\psi} ={{\left( 1,u,v,\frac{1}{2}\left( {{u}^{2}}+{{v}^{2}}+{{\xi }^{2}} \right) \right)}^{T}}$,
$d\Xi =dudvd{{\xi }_{1}}\ldots d{{\xi }_{K}}$. \par
With the integral solution of BGK equation~\cite{XU2001289}, the gas distribution function at the cell
interface ${{x}_{i+1/2}}$ can be constructed as follows is
\begin{equation}\label{equ2.4}
f\left( {{x}_{i+1/2}},y,t,u,v,\xi  \right)=\frac{1}{\tau }\int_{0}^{t}{g\left( {x}',{y}',{t}',u,v,\xi  \right)}{{e}^{-\left( t-{t}' \right)/\tau }}d{t}'+{{e}^{-t/\tau }}{{f}_{0}}\left( -ut,-vt,u,v,\xi  \right),
\end{equation}
where ${x}'={{x}_{i+1/2}}-u\left( t-{t}' \right)$ and ${y}'=y-v\left( t-{t}' \right)$ are the particle
trajectory, ${{x}_{i+1/2}}=0$ is the location of the cell interface. ${{f}_{0}}$ is the initial gas
distribution function $f$ at the beginning of each time step $t=0$.\par
Based on the Chapman-Enskog expansion for the Boltzmann-BGK equation~\cite{xu2014direct}, the macroscopic
governing equations can be derived. In the continuum regime, the gas distribution function can be expanded as
\begin{equation*}
f=g-\tau {{D}_\mathbf{u}}g+\tau {{D}_\mathbf{u}}\left( \tau {{D}_\mathbf{u}}
\right)g-\tau {{D}_\mathbf{u}}\left[ \tau {{D}_\mathbf{u}}\left( \tau {{D}_\mathbf{u}} \right)g
\right]+\cdots,
\end{equation*}
where ${{D}_\mathbf{u}}=\partial /\partial t+\mathbf{u}\cdot \nabla $. For the Euler equations, the zeroth
order truncation is taken, i.e. $f=g$. For the Navier-Stokes equations, the first order truncation is used and the distribution function is
\begin{equation*}
f=g-\tau \left( u{{g}_{x}}+v{{g}_{y}}+{{g}_{t}} \right).
\end{equation*}
By reconstructing the macroscopic variables, the second-order gas distribution function at the cell
interface
can be formulated as
\begin{equation}\label{equ2.5}
\begin{aligned}
f\left( {{x}_{i+1/2}},y,t,u,v,\xi  \right) &=\left( 1-{{e}^{-t/\tau }} \right){{g}^{0}}+\left( \left( t+\tau  \right){{e}^{-t/\tau }}-\tau  \right)\left( {\overline{a}_1}u+{\overline{a}_2}v \right){{g}^{0}} \\
& +\left( t-\tau +\tau {{e}^{-t/\tau }} \right)\overline{A}{{g}^{0}} \\
& +{{e}^{-t/\tau }}{{g}^{r}}\left[ 1-\left( \tau +t \right)\left( a_{1}^{r}u+a_{2}^{r}v \right)-\tau {{A}^{r}} \right]H\left( u \right) \\
& +{{e}^{-t/\tau }}{{g}^{l}}\left[ 1-\left( \tau +t \right)\left( a_{1}^{l}u+a_{2}^{l}v \right)-\tau {{A}^{l}} \right]\left( 1-H\left( u \right) \right),
\end{aligned}
\end{equation}
where $H(u)$ is the Heaviside function. Derivations related to gas kinetic scheme can be found
in ~\cite{XU2001289}.
To develop the compact gas-kinetic scheme, the third-order gas-kinetic distribution function is required.
The most widely used form is the simplified third-order gas-kinetic distribution function~\cite{JI2018446,Zhao2019,CHEN2016708},
which is given by
\begin{equation}\label{equ2.6}
\begin{aligned}
f(x_{i+1/2}, y, t, u, v, \xi) = & \ {g}^{0} + \overline{A} {g}^{0} t + \frac{1}{2} \overline{a}_{tt} {g}^{0} t^2 \\
& - \tau \left[ (\overline{a}_1 u + \overline{a}_2 v + \overline{A}) {g}^{0} + (\overline{a}_{1t} u + \overline{a}_{2t} v + \overline{a}_{tt}) {g}^{0} t \right] \\
& - e^{-t/\tau} {g}^{0} \left[ 1 - (\overline{a}_1 u + \overline{a}_2 v) t \right] \\
& + e^{-t/\tau} {g}^{l} \left[ 1 - (a_{1}^{l} u + a_{2}^{l} v) t \right] H(u) \\
& + e^{-t/\tau} {g}^{r} \left[ 1 - (a_{1}^{l} u + a_{2}^{l} v) t \right] (1 - H(u)),
\end{aligned}
\end{equation}
The equilibrium state ${g}^{0}$, along with the corresponding conservative variables ${Q}^{0}$ and the spatial
derivatives in the local coordinates at the quadrature point, can be ascertained through the compatibility condition
as defined by Eq.~\eqref{equ2.3}.
\begin{equation}\label{equ2.7}
\begin{aligned}
& \int{\psi {{g}^{0}}d\Xi ={{Q}^{0}}=}\int_{u>0}{\boldsymbol{\psi} {{g}^{l}}d\Xi +}\int_{u<0}{\boldsymbol{\psi} {{g}^{r}}d\Xi }, \\
& \frac{\partial {{Q}^{0}}}{\partial {x}}=\int_{u>0}{\boldsymbol{\psi} a_{1}^{l}{{g}^{l}}d\Xi +}\int_{u<0}{\boldsymbol{\psi} a_{1}^{r}{{g}^{r}}d\Xi }, \\
& \frac{\partial {{Q}^{0}}}{\partial {y}}=\int_{u>0}{\boldsymbol{\psi} a_{2}^{l}{{g}^{l}}d\Xi +}\int_{u<0}{\boldsymbol{\psi} a_{2}^{r}{{g}^{r}}d\Xi },\\
& \frac{\partial^2 {{Q}^{0}}}{\partial {x^2}}=\int_{u>0}{\boldsymbol{\psi} a_{11}^{l}{{g}^{l}}d\Xi +}\int_{u<0}{\boldsymbol{\psi} a_{11}^{r}{{g}^{r}}d\Xi }, \\
& \frac{\partial^2 {{Q}^{0}}}{\partial {y^2}}=\int_{u>0}{\boldsymbol{\psi} a_{22}^{l}{{g}^{l}}d\Xi +}\int_{u<0}{\boldsymbol{\psi} a_{22}^{r}{{g}^{r}}d\Xi },\\
& \frac{\partial^2 {{Q}^{0}}}{\partial {x}\partial {y}}=\int_{u>0}{\boldsymbol{\psi} a_{12}^{l}{{g}^{l}}d\Xi +}\int_{u<0}{\boldsymbol{\psi} a_{12}^{r}{{g}^{r}}d\Xi }.\\
\end{aligned}\
\end{equation}
The coefficients in Eq.~\eqref{equ2.6} are ascertainable based on the spatial derivatives of the macroscopic flow
variables and the compatibility condition, as detailed below.
\begin{equation}\label{equ2.8}
\begin{aligned}
& \left\langle a_{1}^{k} \right\rangle =\frac{\partial {{Q}^{k}}}{\partial {x}},\left\langle a_{2}^{k} \right\rangle =\frac{\partial {{Q}^{k}}}{\partial {{y}}},\left\langle {{{\overline{a}}}_{1}} \right\rangle =\frac{\partial {{Q}^{0}}}{\partial {x}},\left\langle {{{\overline{a}}}_{2}} \right\rangle =\frac{\partial {{Q}^{0}}}{\partial {y}}, \\
& \left\langle a_{11}^{k} \right\rangle = \frac{\partial^2 {{Q}^{k}}}{\partial x^2},\left\langle a_{12}^{k} \right\rangle=\left\langle a_{21}^{k} \right\rangle = \frac{\partial^2 {Q}^{k}}{\partial x\partial y},\left\langle a_{22}^{k} \right\rangle = \frac{\partial^2 {Q}^{k}}{\partial y^2},\\
& \left\langle \overline{a}_{11} \right\rangle = \frac{\partial^2 {Q}^{0}}{\partial x^2},\left\langle \overline{a}_{12} \right\rangle=\left\langle \overline{a}_{21} \right\rangle = \frac{\partial^2 {Q}^{0}}{\partial x\partial y},\left\langle \overline{a}_{22} \right\rangle = \frac{\partial^2 {Q}^{0}}{\partial y^2},\\
& \left\langle a_{1}^{k}u+a_{2}^{k}v+{{A}^{k}} \right\rangle =0,\left\langle {{{\overline{a}}}_{1}}u+{{{\overline{a}}}_{2}}v+\overline{A} \right\rangle =0, \\
& \left\langle \overline{a}_{11}u+\overline{a}_{12}v+\overline{a}_{1t} \right\rangle =0,\left\langle \overline{a}_{12}u+\overline{a}_{22}v+\overline{a}_{2t} \right\rangle =0,\\
& \left\langle \overline{a}_{1t}u+\overline{a}_{2t}v+\overline{a}_{tt} \right\rangle =0.\\
\end{aligned}\
\end{equation}
where $k=l,r$ and $\left\langle \ldots  \right\rangle $ are the moments of the equilibrium $g$ and defined by
\begin{equation*}
\left\langle \ldots  \right\rangle =\int{g\left( \ldots  \right)}\boldsymbol{\psi} d\Xi.
\end{equation*}
For conservation laws, the semi-discrete finite volume scheme is written as
\begin{equation}\label{equ2.9}
\begin{aligned}
\frac{\mathrm{d}\mathbf{W}_{ij}}{\mathrm{d}t} = -\frac{1}{\Delta x} \left( \mathbf{F}_{i+1/2,j}(t) - \mathbf{F}_{i-1/2,j}(t) \right) - \frac{1}{\Delta y} \left( \mathbf{G}_{i,j+1/2}(t) - \mathbf{G}_{i,j-1/2}(t) \right).
\end{aligned}\
\end{equation}
And in GKS, the spatial derivatives of flow variables inside each cell are calculated through the cell interface values
with the
help of Gauss's theorem
\begin{equation}\label{equ2.10}
\begin{aligned}
(\mathbf{W}_x)_{ij} = \frac{1}{\Delta x} (\mathbf{W}_{i+1/2,j} - \mathbf{W}_{i-1/2,j})+\frac{1}{\Delta y} (\mathbf{W}_{i,j+1/2} - \mathbf{W}_{i,j-1/2}).
\end{aligned}\
\end{equation}
To solve the spatial derivatives of flow variables, the relation between the conservative variables
$\left( \rho ,\rho U,\rho V,\rho E \right)$ and the distribution function $f$ is
\begin{equation}\label{equ2.11}
{\mathbf{W}_{i+1/2,j}}\left(t \right)=\begin{aligned}
{\left( \begin{aligned}& \rho  \\
& \rho U \\
& \rho V \\
& \rho E \\ \end{aligned}\right)}
=\int{\boldsymbol{\psi }}f\left( {{x}_{i+1/2,j}},y,t,u,v,\xi  \right)d\Xi.
\end{aligned}\
\end{equation}
Upon assembling all requisite terms, these are substituted into Eq.~\eqref{equ2.5},
the solution of which yields the interfacial gas-distribution function~\cite{XU2001289,CiCP-28-539}. Subsequently, the gas-kinetic numerical
flux in the x-direction across the cell interface is evaluated as
\begin{equation}\label{equ2.12}
{\mathbf{F}_{i+1/2,j}}\left( {{\mathbf{W}}^{n}},t \right)=\int{u\left( \begin{aligned}
& 1 \\
& u \\
& v \\
& \frac{1}{2}\left( {{u}^{2}}+{{v}^{2}}+{{\xi }^{2}} \right) \\
\end{aligned} \right)f\left( {{x}_{i+1/2,j}},y,t,u,v,\xi  \right)d\Xi .}
\end{equation}
More details of the compact gas-kinetic scheme can be found in ~\cite{JI2018446}.

\subsection{Two-stage fourth-order solution update}
\label{sec2.2}
To achieve fourth-order temporal accuracy for the compact gas-kinetic scheme ~\cite{JI2018446}, the key is to ensure that both
the conservative variables and their derivatives in the cell are updated by the two-stage
fourth-order method. Below, the details about the updates for the conservative variables and their
derivatives are provided.
The conservation laws can be written as
\begin{equation*}
{{\mathbf{W}}_{t}}=-\nabla \cdot \mathbf{F}\left( \mathbf{W} \right),
\end{equation*}
\begin{equation*}
\frac{\partial {\mathbf{W}}_{ij}}{\partial t} = -\frac{1}{\Delta x}(\mathbf{F}_{i+1/2,j}(t) - \mathbf{F}_{i-1/2,j}(t)) - \frac{1}{\Delta y}(\mathbf{G}_{i,j+1/2}(t) - \mathbf{G}_{i,j-1/2}(t)) := \mathcal{L}_{ij}(W),
\end{equation*}
where $\mathbf{W}$ is the conservative variables and $\mathbf{F}$ is the corresponding flux.
The two-stage fourth-order time marching scheme~\cite{doi:10.1137/15M1052512} is used to solve the initial value
problem and is written as
\begin{equation}\label{equ2.13}
\begin{aligned}
& {{\mathbf{W}}^{*}}={{\mathbf{W}}^{n}}+\frac{1}{2}\Delta t\mathcal{L}\left( {{\mathbf{W}}^{n}} \right)+\frac{1}{8}\Delta {{t}^{2}}\frac{\partial }{\partial t}\mathcal{L}\left( {{\mathbf{W}}^{n}} \right) \\
& {{\mathbf{W}}^{n+1}}={{\mathbf{W}}^{n}}+\Delta t\mathcal{L}\left( {{\mathbf{W}}^{n}} \right)+\frac{1}{6}\Delta {{t}^{2}}\left( \frac{\partial }{\partial t}\mathcal{L}\left( {{\mathbf{W}}^{n}} \right)+2\frac{\partial }{\partial t}\mathcal{L}\left( {{\mathbf{W}}^{*}} \right) \right), \\
\end{aligned}
\end{equation}
where $\partial \mathcal{L}\left( \mathbf{W} \right)/\partial t$ is the time derivative of the spatial operator. The
proposition can be proved using the expansion
\begin{equation}\label{equ2.14}
\begin{aligned}
{{\mathbf{W}}^{n+1}}={{\mathbf{W}}^{n}}+\Delta t\mathcal{L}\left( {{\mathbf{W}}^{n}} \right)+\frac{1}{2}\Delta {{t}^{2}}\frac{\partial }{\partial t}\mathcal{L}\left( {{\mathbf{W}}^{n}} \right)+\frac{1}{6}\Delta {{t}^{3}}\frac{{{\partial }^{2}}}{\partial {{t}^{2}}}\mathcal{L}\left( {{\mathbf{W}}^{n}} \right)+\frac{1}{24}\Delta {{t}^{4}}\frac{{{\partial }^{3}}}{\partial {{t}^{3}}}\mathcal{L}\left( {{\mathbf{W}}^{n}} \right)+\mathcal{O}\left( \Delta {{t}^{5}} \right)
\end{aligned}
\end{equation}
With the time-dependent gas distribution function, the flux for the macroscopic flow variables can be calculated
by Eq.~\eqref{equ2.12}. For the flux in the time interval $\left[ {{t}_{n}},\ {{t}_{n}}+\Delta t \right]$,  the two-stage
gas-kinetic scheme needs the ${\mathbf{F}_{i\pm 1/2}}\left( \mathbf{W} \right)$ and ${{\partial }_{t}}{\mathbf{F}_{i\pm
1/2}}\left( \mathbf{W} \right)$ at both ${{t}_{n}}$ and ${{t}_{*}}={{t}_{n}}+\Delta t/2$. The algorithm of two-stage
gas-kinetic scheme is as follows.\par
Firstly, introduce the following notation,
\begin{equation}\label{equ2.15}
{{\mathbb{F}}_{i+1/2,j}}\left( {{\mathbf{W}}^{n}},\delta  \right)=\int_{{{t}_{n}}}^{{{t}_{n}}+\delta }{{{\mathbf{F}}_{i+1/2,j}}\left( {{\mathbf{W}}^{n}},t \right)=\int_{{{t}_{n}}}^{{{t}_{n}}+\delta }{\int{u{{\psi }_{\alpha }}f\left( {{x}_{i+1/2,j}},t,u,v,\xi  \right)}}}d\Xi dt
\end{equation}
With the reconstruction at ${{t}_{n}}$, the flux ${{\mathbb{F}}_{i+1/2,j}}\left( {{\mathbf{W}}^{n}},\Delta t \right)$,
${{\mathbb{F}}_{i+1/2,j}}\left( {{\mathbf{W}}^{n}},\Delta t/2 \right)$ in the time interval $\left[ {{t}_{n}},\ {{t}_{n}}+
\frac{\Delta t}{2} \right]$ can be evaluated by Eq.~\eqref{equ2.12}.
In the time interval $\left[ {{t}_{n}},\ {{t}_{n}}+\Delta t \right]$, the flux is expanded as the linear form as
\begin{equation}\label{equ2.16}
{\mathbf{F}_{i+1/2,j}}\left( {{\mathbf{W}}^{n}},{t} \right)=\mathbf{F}_{_{i+1/2,j}}^{n}+{\frac{\partial }{\partial t}}
\mathbf{F}_{_{i+1/2,j}}^{n}\left( t-{{t}_{n}} \right).
\end{equation}
The terms $\mathbf{F}_{_{i+1/2,j}}^{{}}\left( {{\mathbf{W}}^{n}},{{t}_{n}} \right)$ and $\partial t\mathbf{F}_{_{i+1/2,j}}^
{{}}\left( {{\mathbf{W}}^{n}},{{t}_{n}} \right)$ can be determined as
\begin{equation}\label{equ2.17}	
{\mathbf{F}_{i+1/2,j}}\left( {{\mathbf{W}}^{n}},{{t}_{n}} \right)\Delta t+\frac{1}{2}{\frac{\partial }{\partial t}}
{\mathbf{F}_{i+1/2,j}}\left( {{\mathbf{W}}^{n}},{{t}_{n}} \right)\Delta {{t}^{2}}={{\mathbb{F}}_{i+1/2,j}}\left( {{\mathbf{W}}^{n}},\Delta t \right),
\end{equation}
\begin{equation}\label{equ2.18}	
\frac{1}{2}{\mathbf{F}_{i+1/2,j}}\left( {{\mathbf{W}}^{n}},{{t}_{n}} \right)\Delta t+\frac{1}{8}{\frac{\partial }
{\partial t}}{\mathbf{F}_{i+1/2,j}}\left( {{\mathbf{W}}^{n}},{{t}_{n}} \right)\Delta {{t}^{2}}={{\mathbb{F}}_{i+1/2,j}}
\left( {{\mathbf{W}}^{n}},\Delta t/2 \right).
\end{equation}
The terms $\mathbf{F}_{_{i+1/2,j}}^{{}}\left( {{\mathbf{W}}^{n}},{{t}_{n}} \right)$ and ${\frac{\partial }{\partial t}}
\mathbf{F}_{_{i+1/2,j}}^{{}}\left( {{\mathbf{W}}^{n}},{{t}_{n}} \right)$ can be computed by solving this linear system as
\begin{equation}\label{equ2.19}
{\mathbf{F}_{i+1/2,j}}\left( {{\mathbf{W}}^{n}},{{t}_{n}} \right)=\left( 4{{\mathbb{F}}_{i+1/2,j}}\left( {{\mathbf{W}}^{n}},\Delta t/2 \right)-{{\mathbb{F}}_{i+1/2,j}}\left( {{\mathbf{W}}^{n}},\Delta t \right) \right)/\Delta t,
\end{equation}
\begin{equation}\label{equ2.20}
{\frac{\partial }{\partial t}}{\mathbf{F}_{i+1/2,j}}\left( {{\mathbf{W}}^{n}},{{t}_{n}} \right)=4\left( {{\mathbb{F}}_
{i+1/2,j}}\left( {{\mathbf{W}}^{n}},\Delta t \right)-2{{\mathbb{F}}_{i+1/2,j}}\left( {{\mathbf{W}}^{n}},\Delta t/2
\right) \right)/\Delta {{t}^{2}}.
\end{equation}
Secondly, update $\mathbf{W}_{i}^{*}$ at ${{t}_{*}}={{t}_{n}}+\Delta t/2$ by
\begin{equation}\label{equ2.21}
\begin{aligned}
& \mathbf{W}_{i}^{*}=\mathbf{W}_{i}^{n}-\frac{1}{\Delta x}\left[ {{\mathbb{F}}_{i+1/2,j}}\left( {{\mathbf{W}}^{n}},
\Delta t/2 \right)-{{\mathbb{F}}_{i-1/2,j}}\left( {{\mathbf{W}}^{n}},\Delta t/2 \right) \right].
\end{aligned}
\end{equation}
Then we can get ${\frac{\partial }{\partial t}}\mathbf{F}_{_{i+1/2,j}}^{{}}\left( {{\mathbf{W}}^{*}},{{t}_{*}} \right)$ in
the time interval $\left[ {{t}_{*}},\ {{t}_{*}}+\Delta t \right]$ to compute the middle stage by the same way.\par
Finally, The numerical fluxes $\mathcal{F}_{i+1/2,j}^{n}$ can be computed by
\begin{equation}\label{equ2.22}
\mathcal{F}_{i+1/2,j}^{n}={\mathbf{F}_{i+1/2,j}}\left( {{\mathbf{W}}^{n}},{{t}_{n}} \right)+\frac{\Delta t}{6}\left[
{\frac{\partial }{\partial t}}{\mathbf{F}_{i+1/2,j}}\left( {{\mathbf{W}}^{n}},{{t}_{n}} \right)+2{\frac{\partial }
{\partial t}}{\mathbf{F}_{i+1/2,j,j}}\left( {{\mathbf{W}}^{*}},{{t}_{*}} \right) \right],
\end{equation}
Similarly, $\mathbf{G}_{i,j+1/2}(t)$ and $\mathcal{G}_{i,j+1/2}^{n}$ can be constructed as well. \par
Update $\mathbf{W}_{ij}^{n+1}$ by
\begin{equation}\label{equ2.23}
\mathbf{W}_{ij}^{n+1}=\mathbf{W}_{ij}^{n}-\frac{\Delta t}{\Delta x}\left(\mathcal{F}_{i+1/2,j}^{n}-\mathcal{F}_
{i-1/2,j}^{n} \right)-\frac{\Delta t}{\Delta y}\left(\mathcal{G}_{i,j+1/2}^{n}-\mathcal{G}_{i,j-1/2}^{n} \right).
\end{equation}\par
To implement a two-stage fourth-order temporal discretization for the gas distribution function, a third-order
gas-kinetic solver is required. This approach necessitates the calculation of both the first and second derivatives
of the distribution function, which is typically achieved through its approximation using a quadratic function as
\begin{equation}\label{equ2.24}
f(t) = f(x_{i+1/2,j}, t, u, v, \xi) = f^n + f_t^n (t - t^n) + \frac{1}{2} f_{tt}^n (t - t^n)^2.
\end{equation}	
So the gas distribution function at the cell interface at \( t^{n+1} \) becomes
\begin{equation}\label{equ2.25}
f^{n+1} = f^n + \Delta t f_t^n + \frac{1}{6} \Delta t^2 \left( f_{tt}^n + 2 f_{tt}^* \right),
\end{equation}
where \( f^* \) is for the middle state at time \( t^* = t^n + \Delta t/2 \),
\begin{equation}\label{equ2.26}
f^* = f^n + \frac{1}{2} \Delta t f_t^n + \frac{1}{8} (\Delta t)^2 f_{tt}^n.
\end{equation}
By a simple derivation, it can be obtained that
\begin{equation}\label{equ2.27}
f^{n+1} = f^n + \Delta t f_t^n + \frac{\Delta t^2}{2} f_{tt}^n + \frac{\Delta t^3}{6} f_{ttt}^n + \frac{\Delta t^4}{24}
f_{tttt}^n + \mathcal{O}(\Delta t^5).
\end{equation}
According to the gas-distribution function at $ t = 0, \Delta t/2 $ and $ \Delta t $
\begin{equation}\label{equ2.28}
\begin{aligned}
&f^n = f(0), \\
&f^n + \frac{1}{2} f_t^n \Delta t + \frac{1}{8} f_{tt}^n \Delta t^2 = f(\Delta t/2), \\
&f^n + f_t^n \Delta t + f_{tt}^n \Delta t^2 = f(\Delta t),
\end{aligned}
\end{equation}	
the coefficients \( f^n, f_t^n \) and \( f_{tt}^n \) can be determined
\begin{equation}\label{equ2.29}
\begin{aligned}
&f^n = f(0), \\
&f_t^n = \frac{(4f(\Delta t/2) - 3f(0) - f(\Delta t))}{\Delta t}, \\
&f_{tt}^n = \frac{4(f(\Delta t) + f(0) - 2f(\Delta t/2))}{\Delta t^2}.
\end{aligned}
\end{equation}	
Therefore, the flow variables \( \mathbf{W}_{i+1/2,j}^{n+1} \) and \( \mathbf{W}_{i,j+1/2}^{n+1} \) can be explicitly
obtained, i.e.,
$\mathbf{W}_{i+1/2,j}^{n+1} = \int \mathbf{\psi} f_{i+1/2,j}^{n+1} d\Xi,$
from which the slope inside each cell can be updated as
\begin{equation}\label{equ2.30}
\begin{aligned}
(\mathbf{W}_x)_{ij}^{n+1} = \frac{\mathbf{W}_{i+1/2,j}^{n+1} - \mathbf{W}_{i-1/2,j}^{n+1}}{\Delta x},
(\mathbf{W}_y)_{ij}^{n+1} = \frac{\mathbf{W}_{i,j+1/2}^{n+1} - \mathbf{W}_{i,j-1/2}^{n+1}} {\Delta y}.
\end{aligned}
\end{equation}	
More details about compact gas-kinetic scheme can be found in ~\cite{JI2018446}.

\section{Adaptive stencil extension reconstruction with discontinuous feedback factor}
\label{sec3}
In ~\cite{zhang2024adaptivereconstructionmethodarbitrary}, an efficient class of adaptive stencil extension
reconstruction method based on the discontinuity feedback factor is introduced. By expanding the stencil, we can easily obtain arbitrary higher-order reconstruction method.
We extend it to adaptive compact stencil framework. Our primary focus is the development of an efficient
compact gas-kinetic scheme spatial reconstruction framework capable of achieving very high order accuracy,
with straightforward implementation in both one-dimensional and two-dimensional cases.
\subsection{Discontinuity feedback factor}
\label{sec3.1}
The discontinuity feedback factor is introduced by Ji~\cite{Ji09082021,Zhang2024}. It can solve shock wave and
rarefaction wave very well in three-dimensional hypersonic numerical simulation.
It is characterized by its high efficiency and accurate prediction of discontinuity, and is very robust.
Subsequently, Ji et al. applied the discontinuity feedback factor in structured mesh as an important tool
for high-order reconstruction. It is robust and based on DFF, designing 7-th order and 9-th order adaptive
high-order scheme~\cite{zhang2024adaptivereconstructionmethodarbitrary}. The discontinuity feedback factor
predicts by using the physical quantities on either side of the interface from the previous time step ${{t}^{n-1}}$.
Due to the evolution and motion of shock discontinuity, it is possible to determine whether the stencils
cross a discontinuity. Unlike some WENO scheme, it does not completely discard the stencil. Instead, it
uses the strength of the discontinuity analyzed by the discontinuity feedback factor to reduce the stencil
to first-order accuracy.\par
The discontinuity feedback factor is as follows. First, the discontinuity strength at the cell interface, denoted
as $\sigma _{i+1/2,j}^{n-1}$
 , is calculated as follows
 \begin{equation}\label{equ3.1}
\begin{aligned}
 {\sigma _{i+1/2,j}^{n-1}}=\operatorname{Avg}\left\{ \sum\limits_{m=1}^{M}{{{\sigma }_{i+1/2,jm}^{n-1}}} \right\}
\end{aligned}
\end{equation}	
where \(\sigma_{i+1/2,j}^{n-1} \geq 0\). And \(\sigma_{i+1/2,jm}^{n-1}\) is the discontinuity strength of the \(m\)-th
Gauss point at the interface, which has the form
\begin{equation}\label{equ3.2}
\begin{aligned}
\sigma_{j+1/2,jm}^{n-1} = \frac{|p^l - p^r|}{p^l} + \frac{|p^l - p^r|}{p^r} + \left( \text{Ma}_n^l - \text{Ma}_n^r \right)^2 + \left( \text{Ma}_t^l - \text{Ma}_t^r \right)^2,
\end{aligned}
\end{equation}
where \( p^k, k = l, r \) denote the left and right pressure of the Gauss point \( x_{j+1/2,jm} \),
\( \text{Ma}_n^l \), \( \text{Ma}_t^l \) denote the left Mach number defined by the normal and tangential velocity,
respectively. The \(\sigma_{i+1/2,j}^{n-1}\) corresponds to the strength of the discontinuity. When the
\(\sigma_{i+1/2,j}^{n-1} = 0\), the flow is smooth. To calculate the DFF \(\alpha_S^{n}\) as
\begin{equation}\label{equ3.3}
\begin{aligned}
A^{n-1} = \cdots + \sigma_{i-3/2,j}^{n-1} + \sigma_{i-1/2,j}^{n-1} + \sigma_{j+1/2,j}^{n-1} + \sigma_{j+3/2,j}^{n-1} + \cdots\\
\alpha_S^{n} =
\begin{cases}
1.0 & \text{if } A^{n-1} < \sigma_{\text{thres}}, \\
\frac{\sigma_{\text{thres}}}{\cdots + \sigma_{i-3/2,j}^{n-1} + \sigma_{i-1/2,j}^{n-1}+ \sigma_{i+1/2,j}^{n-1} + \sigma_{i+3/2,j}^{n-1} + \cdots} & \text{otherwise},
\end{cases}
\end{aligned}
\end{equation}
where \(\alpha_S^{n} \in (0,1]\) is the DFF of the non-compact stencil \(S\). When \(\alpha_S^{n} = 1\), which means the
stencil is smooth, when \(\alpha_S^{n} \to 0\), there are strong discontinuities in the stencil. However, due to the
possibility that compact stencils of different order may have the same size, such as both the 7th and 9th order stencils
utilizing information from five cells. It is necessary to distinguish them by providing the DFF for the cells which
provide the conservative variables and derivatives.
\begin{equation}\label{equ3.4}
\begin{aligned}
A^{n-1} = \cdots + \eta \sigma_{i-3/2,j}^{n-1} + \eta\sigma_{i-1/2,j}^{n-1} + \eta\sigma_{i+1/2,j}^{n-1} + \eta\sigma_{i+3/2,j}^{n-1} + \cdots,\eta=1,2\\
\alpha_S^{n} =
\begin{cases}
1.0 & \text{if } A^{n-1} < \sigma_{\text{thres}}, \\
\frac{\sigma_{\text{thres}}}{\cdots + \eta\sigma_{i-3/2,j}^{n-1} + \eta\sigma_{i-1/2,j}^{n-1} + \eta\sigma_{i+1/2,j}^{n-1} + \eta\sigma_{i+3/2,j}^{n-1} + \cdots} & \text{otherwise},
\end{cases}
\end{aligned}
\end{equation}
The $\sigma_{\text{thres}}$ is used to assess the strength of discontinuities within the stencil. For non-compact
stencils, it is generally set to 2.0. While for compact stencils, it is typically set to 1.0.
The role of the discontinuity feedback factor is to adaptively select stencils based on the magnitude of its value,
thereby enabling adaptive order reduction at discontinuity for the subsequent adaptive stencil extension. For
the polynomial of the reconstruction stencil, the discontinuity feedback factor of stencil is substituted such as,
\begin{equation}\label{equ3.5}
\begin{aligned}
& {{\mathbb{P}}^{r5}}(x)={{Q}_{0}}+\alpha _{{{\text{S}}^{r5}}}^{n}\left[ {{Q}_{x}}+{{Q}_{xx}}+{{Q}_{xxx}}+{{Q}_{xxxx}} \right],\alpha _{{{\text{S}}^{r5}}}^{n}\in \left[ 0,1 \right], \\
& \text{if} \alpha _{{{\text{S}}^{r5}}}^{n}=0,{{\mathbb{P}}^{r5}}(x)={{Q}_{0}}, \\
& \text{if} \alpha _{{{\text{S}}^{r5}}}^{n}=1,{{\mathbb{P}}^{r5}}(x)={{Q}_{0}}+{{Q}_{x}}+{{Q}_{xx}}+{{Q}_{xxx}}+{{Q}_{xxxx}}. \\
\end{aligned}
\end{equation}
Since DFF is a value between 0 and 1, when the discontinuity is very strong, DFF approaches 0, and the reconstruction of the
large stencil polynomial is reduced to first-order reconstruction due to the reduction of the slope. Similarly, when the
flow is smooth and DFF approaches 1, it ensures that the stencil maintains high-order smooth reconstruction.It can enhance
the robustness of the reconstruction.
\subsection{Adaptive stencil extension reconstruction}
\label{sec3.2}
We design the adaptive stencil extension reconstruction method based on the discontinuity feedback factor for the CGKS.
All stencils
satisfy the same condition. Here, we first introduce fifth-order HWENO-AO reconstruction.
\begin{equation}\label{equ3.6}
\begin{aligned}
\frac{1}{\Delta x} \int_{I_{i+j}} p_3(x) \, dx &= \overline{Q}_{i+j}, \quad j = -1, 0, 1, \\
\frac{1}{\Delta x} \int_{I_{i+j}} (p_3)_x(x) \, dx &= (\overline{Q}_x)_{i+j}. \quad j = -1, 0, 1,
\end{aligned}
\end{equation}
To reconstruct the left value \( Q_{i+1/2}^l \) at the cell interface \( x_{i+1/2} \), three sub-stencils are
selected
\[
S_0 = \{I_{i-1}, I_i\}, \quad S_1 = \{I_i, I_{i+1}\}, \quad S_2 = \{I_{i-1}, I_i, I_{i+1}\}.
\]
The Hermite quadratic reconstruction polynomials \( p_k^{rq}(x) \)(q is the order) corresponding to the
substencil \( S_k, k = 0, 1, 2 \) are constructed according to the following conditions
\begin{equation}\label{equ3.7}
\begin{aligned}
\frac{1}{\Delta x} \int_{I_{i-j}} p_0^{r3}(x) \, dx &= \overline{Q}_{i-j}, \quad j = 0, 1, \\
\frac{1}{\Delta x} \int_{I_{i-1}} (p_0^{r3})_x(x) \, dx &= (\overline{Q}_x)_{i-1}, \\
\frac{1}{\Delta x} \int_{I_{i+j}} p_1^{r3}(x) \, dx &= \overline{Q}_{i+j}, \quad j = 0, 1, \\
\frac{1}{\Delta x} \int_{I_{i+1}} (p_1^{r3})_x(x) \, dx &= (\overline{Q}_x)_{i+1}, \\
\frac{1}{\Delta x} \int_{I_{i+j}} p_2^{r3}(x) \, dx &= \overline{Q}_{i+j}, \quad j = -1, 0, 1.
\end{aligned}
\end{equation}
For the reconstructed polynomials, the point value at the cell interface \( x_{i+1/2} \) can be given in
terms of the cell averages value and the averaged spatial derivative as follows
\begin{equation}\label{equ3.8}
\begin{aligned}
&p_0^{r3}(x_{i+1/2}) = -\frac{7}{6} \overline{Q}_{i-1} + \frac{13}{6} \overline{Q}_i - \frac{2 \Delta x}{3} (\overline{Q}_x)_{i-1}, \\
&p_1^{r3}(x_{i+1/2}) = \frac{1}{6} \overline{Q}_i + \frac{5}{6} \overline{Q}_{i+1} - \frac{\Delta x}{3} (\overline{Q}_x)_{i+1}, \\
&p_2^{r3}(x_{i+1/2}) = -\frac{1}{6} \overline{Q}_{i-1} + \frac{5}{6} \overline{Q}_i + \frac{1}{3} \overline{Q}_{i+1},\\
&(p_0^{r3})_x(x_{i+1/2}) = \frac{- 2(\overline{Q}_x)_{i-1}(\Delta x) + 3 \overline{Q}_i - 3\overline{Q}_{i-1}}{\Delta x}, \\
&(p_1^{r3})_x(x_{i+1/2}) = \frac{- \overline{Q}_{i} + \overline{Q}_{i+1}}{\Delta x}, \\
&(p_2^{r3})_x(x_{i+1/2}) = \frac{- \overline{Q}_{i} + \overline{Q}_{i+1}}{\Delta x},\\
&(p_0^{r3})_{xx}(x_{i+1/2}) = -\frac{2((\overline{Q}_x)_{i-1}(\Delta x) - \overline{Q}_i + \overline{Q}_{i-1})}{\Delta x^2}, \\
&(p_1^{r3})_{xx}(x_{i+1/2}) = \frac{2((\overline{Q}_x)_{i+1}(\Delta x) + \overline{Q}_i - \overline{Q}_{i+1})}{\Delta x^2}, \\
&(p_2^{r3})_{xx}(x_{i+1/2}) = \frac{-2\overline{Q}_{i} + \overline{Q}_{i-1} + \overline{Q}_{i+1}}{\Delta x^2}.\\
\end{aligned}
\end{equation}
On the bigger stencil \( \mathbb{S} = \{S_0, S_1, S_2\} \), a fourth-order reconstruction polynomial \( p_3(x) \) are
constructed according to the following conditions
\begin{equation}\label{equ3.9}
\begin{aligned}
\frac{1}{\Delta x} \int_{I_{i+j}} p_3^{r5}(x) \, dx &= \overline{Q}_{i+j}, \quad j = -1, 0, 1, \\
\frac{1}{\Delta x} \int_{I_{i+j}} (p_3^{r5})_x(x) \, dx &= (\overline{Q}_x)_{i+j}. \quad j = -1, 0, 1,
\end{aligned}
\end{equation}
and the point value at the cell interface \( x_{i+1/2} \) can be written as
\begin{equation}\label{equ3.10}
\begin{aligned}
&p_3^{r5}(x_{i+1/2}) = -\frac{23}{120} \overline{Q}_{i-1} + \frac{19}{30} \overline{Q}_i + \frac{67}{120} \overline{Q}_{i+1} - \Delta x \left( \frac{3}{40} (\overline{Q}_x)_{i-1} + \frac{7}{40} (\overline{Q}_x)_{i+1} \right),\\
&(p_3^{r5})_x(x_{i+1/2}) = \frac{(\overline{Q}_x)_{i-1}(\Delta x) - 3(\overline{Q}_x)_{i+1}(\Delta x) -16 \overline{Q}_i +3 \overline{Q}_{i-1}+13 \overline{Q}_{i+1}}{8\Delta x}, \\
&(p_3^{r5})_{xx}(x_{i+1/2}) = \frac{3(\overline{Q}_x)_{i-1}(\Delta x) + 3(\overline{Q}_x)_{i+1}(\Delta x) -8\overline{Q}_{i} + 7\overline{Q}_{i-1}+ \overline{Q}_{i+1}}{4\Delta x^2}.
\end{aligned}
\end{equation}
Start from rewriting \( p_3^{r5}(x) \) as
\begin{equation}\label{equ3.11}
p_3^{r5}(x) = \gamma_3 \left( \frac{1}{\gamma_3} p_3^{r5}(x) - \sum_0^2 \frac{\gamma_k}{\gamma_3} p_k^{r3}(x) \right) + \sum_0^2 \gamma_k p_k^{r3}(x), \gamma_k \neq 0,
\end{equation}
where \( \gamma_k, l = 0, 1, 2, 3 \) are defined as linear weights.
\begin{equation}\label{equ3.12}
\begin{aligned}
\gamma_3 &= \gamma_{Hi}; \quad \gamma_0 = (1 - \gamma_{Hi})(1 - \gamma_{Lo})/2; \quad \gamma_1 = (1 - \gamma_{Hi}); \quad \gamma_2 = \gamma_0,
\end{aligned}
\end{equation}
which satisfy \( r_l > 0, l = 0, 1, 2, 3 \) and \( \sum_0^3 \gamma_k = 1 \), and
suggest \( \gamma_{Hi} \in [0.85, 0.95] \) and \( \gamma_{lo} \in [0.55, 0.95] \). Here we
choose \( \gamma_{Hi} = 0.85 \) and \( \gamma_{lo} = 0.6 \) in the numerical tests if no specification
values are provided~\cite{BALSARA2016780,CiCP-28-539}.
The \(\beta_k\) are the smoothness indicators which are defined as
\begin{equation}\label{equ3.13}
\beta_k = \sum_{q=1}^{q_k} \Delta x^{2q-1} \int_{x_{i-1/2}}^{x_{i+1/2}} \left( \frac{\mathrm{d}^q}{\mathrm{d}x^q} p_k(x) \right)^2 dx = \mathcal{O}(\Delta x^2),
\end{equation}
where \( q_k \) is the order of \( p_k(x) \).
To avoid the loss of order of accuracy at inflection points, the WENO-Z type non-linear weights~\cite{BORGES20083191} are
used as
\begin{equation}\label{equ3.14}
\omega_k = \gamma_k (1 + \frac{\tau^2}{(\beta_k + \epsilon)^2}),
\end{equation}
where the global smooth indicator \( \delta \) is defined as
\begin{equation}\label{equ3.15}
\tau = \frac{1}{3} (|\beta_3^{r5} - \beta_0^{r3}| + |\beta_3^{r5} - \beta_1^{r3}| + |\beta_3^{r5} - \beta_2^{r3}|) = \mathcal{O}(\Delta x^4).
\end{equation}
The normalized weights are given by
\begin{equation}\label{equ3.16}
\overline{\omega_k} = \frac{\omega_k}{\sum_0^3 \omega_k}.
\end{equation}
Then the final form of the reconstructed polynomial is
\begin{equation}\label{equ3.17}
P^{AO(5,3)}(x) = \overline{\omega_3} \left( \frac{1}{\gamma_3} p_3^{r5}(x) - \sum_0^2 \frac{\gamma_k}{\gamma_3} p_k^{r3}(x) \right) + \sum_0^2 \overline{\omega_k} p_k^{r3}(x).
\end{equation}
For adaptive stencil extension with discontinuity feedback reconstruction, first we present the results for the seventh-order linear compact stencil and the ninth-order compact stencil, as follows.
\begin{equation}\label{equ3.18}
\begin{aligned}
 p_{4}^{r7}({{x}_{i+1/2}})=&\frac{1}{1260}\left( -13{{{\bar{Q}}}_{i-2}}+717{{{\bar{Q}}}_{i}}+757{{{\bar{Q}}}_{i+1}} \right. \\
& \left. +22{{{\bar{Q}}}_{i+2}}-120{{({{{\bar{Q}}}_{x}})}_{i-1}}\Delta x-300{{({{{\bar{Q}}}_{x}})}_{i+1}}\Delta x \right). \\
{{(p_{4}^{r7})}_{x}}({{x}_{i+1/2}})=&\frac{1}{108\Delta x}\left( 24{{({{{\bar{Q}}}_{x}})}_{i-1}}(\Delta x)-48{{({{{\bar{Q}}}_{x}})}_{i+1}}(\Delta x)+2{{{\bar{Q}}}_{i-2}} \right. \\
& \left. +53{{{\bar{Q}}}_{i-1}}-243{{{\bar{Q}}}_{i}}+187{{{\bar{Q}}}_{i+1}}+{{{\bar{Q}}}_{i+2}} \right), \\
{{(p_{4}^{r7})}_{xx}}({{x}_{i+1/2}})=&\frac{1}{24\Delta {{x}^{2}}}\left( 24{{({{{\bar{Q}}}_{x}})}_{i-1}}(\Delta x)+36{{({{{\bar{Q}}}_{x}})}_{i+1}}(\Delta x)+3{{{\bar{Q}}}_{i-2}} \right. \\
& \left. +38{{{\bar{Q}}}_{i-1}}-30{{{\bar{Q}}}_{i}}-6{{{\bar{Q}}}_{i+1}}-5{{{\bar{Q}}}_{i+2}} \right). \\
p_{5}^{r9}({{x}_{i+1/2}})=&\frac{1}{15120}\left( -601{{{\bar{Q}}}_{i-2}}-2076{{{\bar{Q}}}_{i-1}}+7524{{{\bar{Q}}}_{i}}+9124{{{\bar{Q}}}_{i+1}} \right.+1149{{{\bar{Q}}}_{i+2}} \\
& \left. -150{{({{{\bar{Q}}}_{x}})}_{i-2}}\Delta x-1800{{({{{\bar{Q}}}_{x}})}_{i-1}}\Delta x-4920{{({{{\bar{Q}}}_{x}})}_{i+1}}\Delta x-270{{({{{\bar{Q}}}_{x}})}_{i+2}}\Delta x \right). \\
{{(p_{5}^{r9})}_{x}}({{x}_{i+1/2}})=&\frac{1}{2592\Delta x}\left( 66{{({{{\bar{Q}}}_{x}})}_{i-2}}(\Delta x)+984{{({{{\bar{Q}}}_{x}})}_{i-1}}(\Delta x)-1320{{({{{\bar{Q}}}_{x}})}_{i+1}}(\Delta x) \right. \\
& \left. -6{{({{{\bar{Q}}}_{x}})}_{i+2}}(\Delta x)+275{{{\bar{Q}}}_{i-2}}+1424{{{\bar{Q}}}_{i-1}}-6480{{{\bar{Q}}}_{i}}+4720{{{\bar{Q}}}_{i+1}}+61{{{\bar{Q}}}_{i+2}} \right), \\
{{(p_{5}^{r9})}_{xx}}({{x}_{i+1/2}})=&\frac{1}{2592\Delta {{x}^{2}}}\left( 330{{({{{\bar{Q}}}_{x}})}_{i-2}}(\Delta x)+3288{{({{{\bar{Q}}}_{x}})}_{i-1}}(\Delta x)+7080{{({{{\bar{Q}}}_{x}})}_{i+1}}(\Delta x) \right. \\
& \left. +642{{({{{\bar{Q}}}_{x}})}_{i+2}}(\Delta x)+1291{{{\bar{Q}}}_{i-2}}+2624{{{\bar{Q}}}_{i-1}}-432{{{\bar{Q}}}_{i}}-832{{{\bar{Q}}}_{i+1}}-2651{{{\bar{Q}}}_{i+2}} \right).
\end{aligned}
\end{equation}\par
(1)ASE-DFF(5,3)\par
The construction of the adaptive stencil extension with discontinuity feedback reconstruction scheme primarily
utilizes the discontinuity feedback factor of the stencil to switch to smoothing. If \(\alpha _{{{\text{S}}^{r5}}}^{n} < 1\), it indicates
the presence of discontinuities within the fifth-order compact stencil, and each sub-stencil uses its own DFF accordingly.
\begin{equation}\label{equ3.19}
\begin{aligned}
& \mathbb{P}_{k}^{rq}\left( {{x}_{i+1/2}} \right)={{Q}_{0}}+\alpha_{{{\text{S}_i^{rq}}}}^{n}\left( Q_{i+1/2}^{l}-{{Q}_{0}} \right), \\
& \left( {{\mathbb{P}}_{x}} \right)_{k}^{rq}\left( {{x}_{i+1/2}} \right)=\alpha_{{{\text{S}_i^{rq}}}}^{n}\left( \left( {{Q}_{x}} \right)_{i+1/2}^{l} \right), \\
& \left( {{\mathbb{P}}_{xx}} \right)_{k}^{rq}\left( {{x}_{i+1/2}} \right)=\alpha_{{{\text{S}_i^{rq}}}}^{n}\left( \left( {{Q}_{xx}} \right)_{i+1/2}^{l} \right). \\
\end{aligned}
\end{equation}
Then, the stencils are reassembled using the HWENO-AO reconstruction.
Otherwise, use fifth-order compact linear reconstruction.\par	
(2)ASE-DFF(7,5,3)\par
When \(\alpha _{{{\text{S}}^{r5}}}^{n} < 1\), it signifies the existence of discontinuities within the fifth-order
compact stencil, necessitating the application of the ASE-DFF (5, 3) method. Conversely,
if \(\alpha _{{{\text{S}}^{r5}}}^{n} = 1\), the stencil is deemed smooth. Following this, if \(\alpha _{{{\text{S}}^{r7}}}^{n} < 1\),
a fifth-order compact linear reconstruction based on the DFF is implemented.
Otherwise, a seventh-order compact linear reconstruction based on DFF is employed.\par
(3)ASE-DFF(9,7,5,3)\par
Firstly, we continue with the ACE-DFF (7, 5, 3) method. However, after evaluating the seventh-order
stencil, we proceed as follows: if \(\alpha _{{{\text{S}}^{r9}}}^{n} < 1\), indicating the presence of discontinuities
within the ninth-order stencil, we employ a seventh-order compact linear reconstruction based on DFF.
Otherwise, we utilize a ninth-order compact linear reconstruction based on DFF.\par
\begin{figure}[htp]	
\centering
\includegraphics[width=1.0\textwidth]{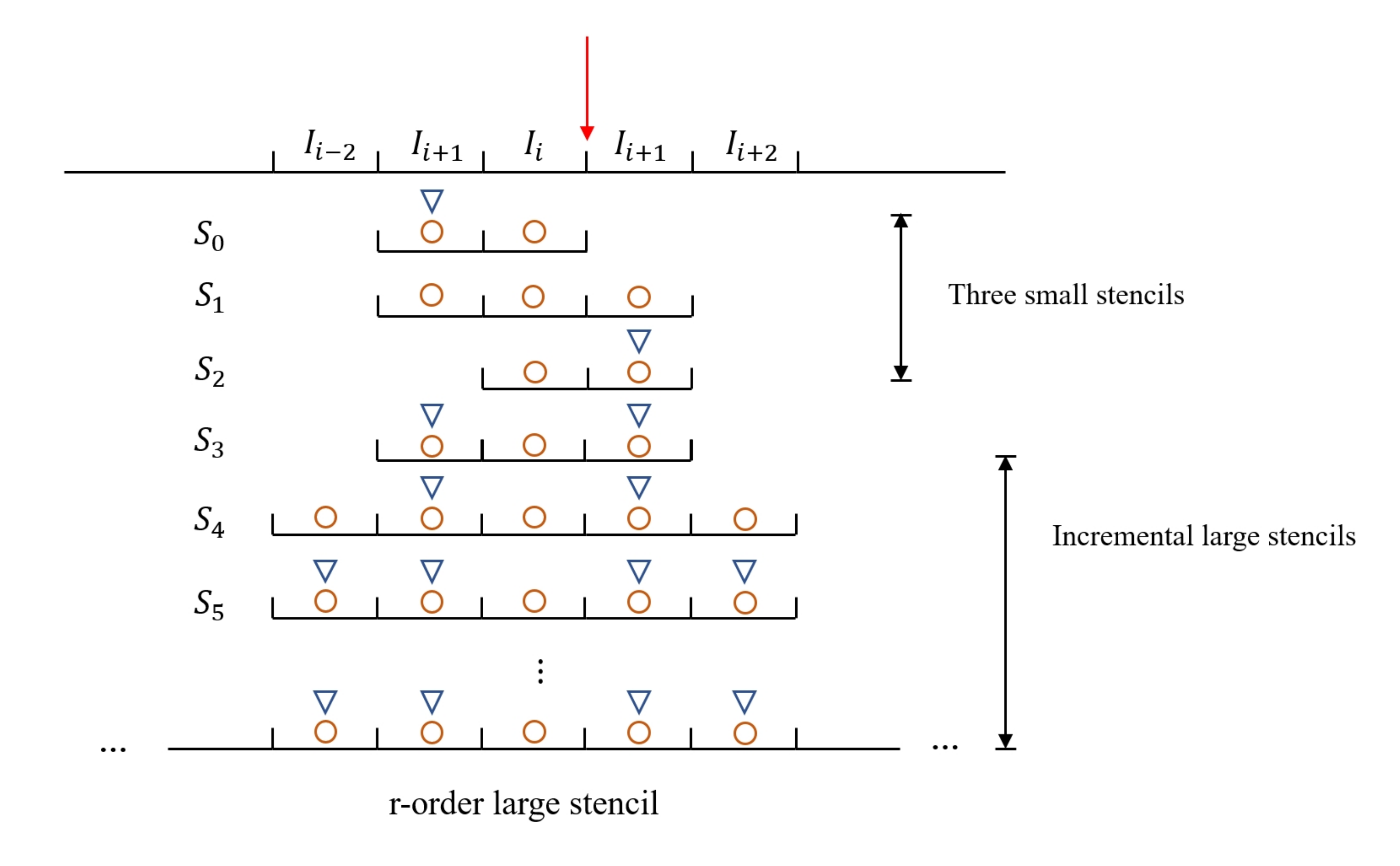}
\caption{\label{fig1Dstencil}
Candidate stencils with incremental width towards adaptive stencil extension reconstruction: the circles
represent cell averages, and the gradients represent cell averaged slopes.}
\end{figure}

The ASE-DFF algorithm for selecting stencils is shown in algorithm 1. And as shown in Fig.~\ref{fig1Dstencil},
we propose a new framework with incremental candidate stencils to construct the very-high-order
compact GKS scheme.\par
\begin{algorithm}
\caption{DFF-based adaptive stencil extension reconstruction method}
\begin{algorithmic}
\Require{Interfaces reconstructed values $\mathbf{W}^l, \mathbf{W}^r,\mathbf{W}_{x}^l,\mathbf{W}_{x}^r, \mathbf{W}_{xx}^l,\mathbf{W}_{xx}^r$.}
\Ensure{The needed stencil for the reconstruction.}
\State calculate the correspond DFF values $\alpha^{r5} \gets (\mathbf{W}^l, \mathbf{W}^r)$ by Eq.\eqref{equ3.4}.
\If{$\alpha^{r5} < 1$}
\State select the non-linear HWENO-AO with DFF reconstruction method;
\Else
\State calculate the correspond DFF values $\alpha^{r7} \gets (\mathbf{W}^l, \mathbf{W}^r)$ by Eq.\eqref{equ3.4}.
\If{$\alpha^{r7} < 1$}
\State select the linear 5th-order compact polynomial reconstruction method;
\Else
\State calculate the correspond DFF values $\alpha^{r9} \gets (\mathbf{W}^l, \mathbf{W}^r)$ by Eq.\eqref{equ3.4}.
\If{$\alpha^{r9} < 1$}
\State select the linear 7th-order compact polynomial reconstruction method;
\Else
\State select the linear 9th-order compact polynomial reconstruction method;
\EndIf
\EndIf
\EndIf
\end{algorithmic}
\end{algorithm}

\textbf{Remark 1.} \textit{ We assess  the computational complexity of r-th-order reconstruction for the conventional WENO reconstruction and the ASE-DFF reconstruction.
In WENO/HWENO formulations the reconstruction polynomial is assembled as the nonlinear convex combination of the
pointwise values obtained from $\mathcal{O}(r)$ distinct sub-stencils.  Each sub-stencil requires $\mathcal{O}(r)$ operations, and exactly $\mathcal{O}(r)$
such polynomials are combined. So the computational complexity of the polynomial calculation is $\mathcal{O}(r^2)$.
Smoothness indicators are evaluated on the same sub-stencils.  Each sub-stencil corresponds to a polynomial of $(r+1)/2$-th
order . Hence, according to Eq.\eqref{equ3.13}, up to $(r-1)/2$ successive derivatives must be computed.
This leads to an overall computational complexity of $\mathcal{O}(r^3)$ for the smoothness indicators evaluation.
Consequently, the total computational complexity of the classical WENO/HWENO reconstruction is $\mathcal{O}(r^3)$.
By contrast, the ASE-DFF reconstruction attains the r-th order by directly finishing the high-order
reconstruction with the corresponding DFF. The DFF itself can be inherited and no smoothness indicators are required.
The computational complexity of the ASE-DFF reconstruction is only $\mathcal{O}(r)$.
Therefore, the cost of both computation and memory for ASE-DFF method are relatively small.}

\subsection{Two dimensional reconstruction}
\label{sec3.3}
\begin{figure}[htp]	
\centering
\includegraphics[width=1.0\textwidth]{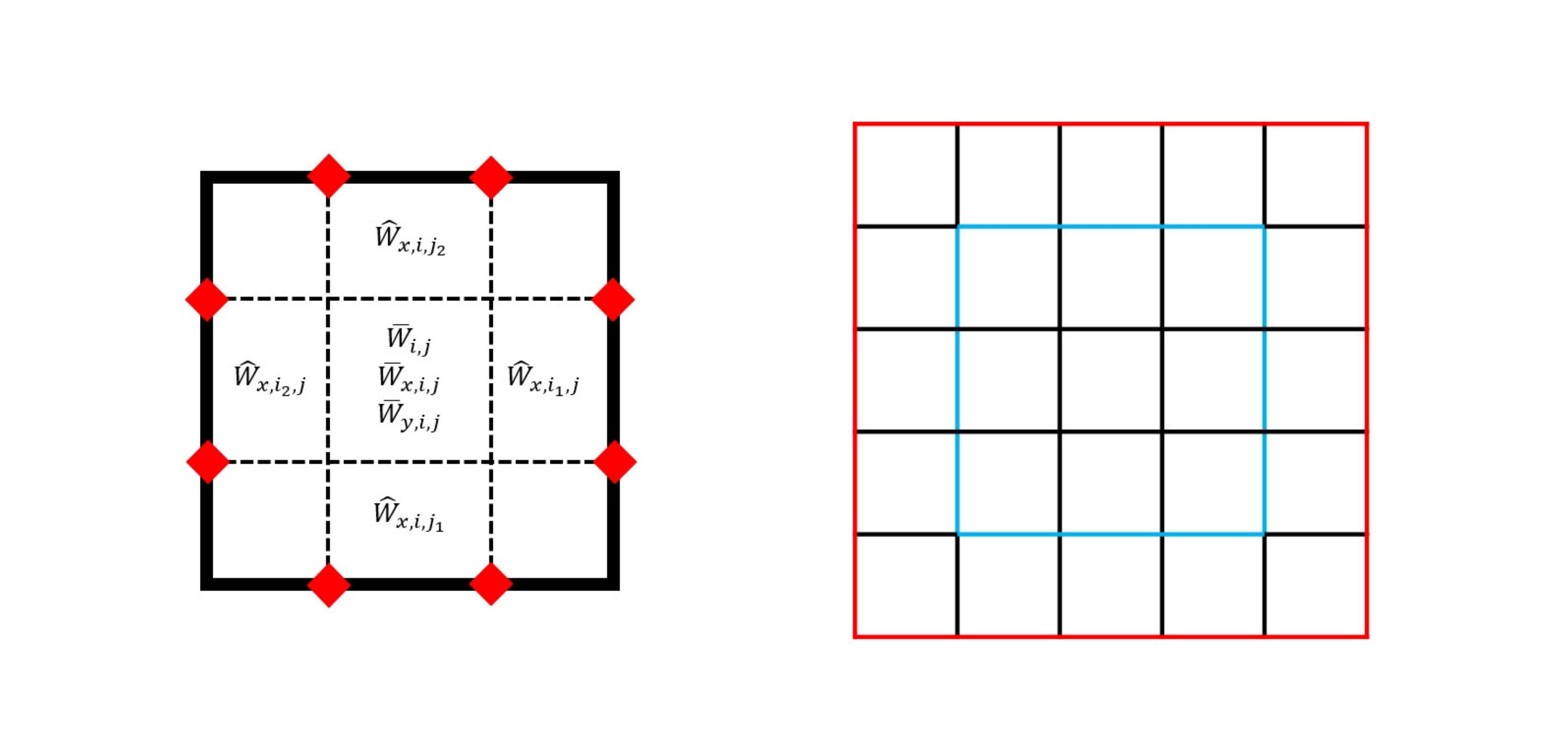}
\caption{\label{fig2Dstencil}
The schematic for 2-D ASE-DFF CGKS reconstruction. Left: Initial data for ASE-DFF CGKS reconstruction at Gaussian
points (in red color), such as the 5-th scheme. Right: Reconstruction stencils for ASE-DFF(5,3)-CGKS (blue box)
and, ASE-DFF(7,5,3)-CGKS and ASE-DFF(9,7,5,3)-CGKS  (red box). }
\end{figure}

Before introducing the reconstruction procedure, let's denote \(\overline{W}\) as cell averaged, \(\hat{W}\) as
line averaged, and \(W\) as pointwise values. Here \(W^{l,r}\) represent the reconstructed quantities on the
left and right sides, which correspond to the non-equilibrium initial part in GKS framework.
At \(t^n\) step, for cell \((i, j)\) the cell average
quantities \(\overline{W}_{i,j}^n\); \(\overline{W}_{x,i,j}^n\), \(\overline{W}_{y,i,j}^n\) are stored. For a
fifth-order scheme, two Gaussian points are required at each interface for numerical flux integration. For the
seventh-order scheme, three Gaussian points are needed, and for the ninth-order scheme, four Gaussian points are
required. Our target is to construct \(W^{l,r}_x, W^{l,r}_x, W^{l,r}_{xx}, W^{l,r}_y, W^{l,r}_{yy}, W^{l,r}_{xy}\) at each Gaussian point.
To obtain these quantities, the line averaged
slopes \(\hat{W}^n_{x,i,j_l}, \hat{W}^n_{y,i_l,j}\) are additionally evaluated, where \(l=1,...,m\) corresponds to
the number of Gaussian points required at each interface. For a better illustration, a schematic is plotted in
Fig.~\ref{fig2Dstencil} and the reconstruction procedure for the Gaussian point \(i + 1/2, j_l\) is summarized as follows. Here
the time level \(n\) is omitted.\par

Step 1. To obtain the line average values, i.e. \(\hat{W}_{i,j_l}\), we perform ASE-DFF(5,3) reconstruction in
tangential direction by using \(\overline{W}_{i,j-1}, \overline{W}_{i,j}, \overline{W}_{i,j+1}\),
and \(\overline{W}_{y,i,j-1}, \overline{W}_{y,i,j+1}\).For the ASE-DFF(7,5,3) reconstruction and
ASE-DFF(9,7,5,3) reconstruction, \(\overline{W}_{i,j-2}, \overline{W}_{i,j+2}\),
and \(\overline{W}_{y,i,j-2}, \overline{W}_{y,i,j+2}\) are required, additionally. \par

Step 2. With the reconstructed line average values
i.e. \(\hat{W}_{i-1,j_l}, \hat{W}_{i,j_l}, \hat{W}_{i+1,j_l}\) and
original \(\hat{W}_{x,i-1,j_l}, \hat{W}_{x,i+1,j_l}\), the one dimensional ASE-DFF(5,3) reconstruction is conducted
and \(W^r_{i-1/2,j_l}, W^l_{i+1/2,j_l}\) are obtained. And the derivatives \(W^r_{x,i-1/2,j_l}\),\(W^l_{x,i+1/2,j_l}\),\(W^r_{xx,i-1/2,j_l}\),\(W^l_{xx,i+1/2,j_l}\) are
constructed. The ASE-DFF(7,5,3) reconstruction and the ASE-DFF(9,7,5,3) reconstruction are similar with ASE-DFF(5,3),
and they also require \(\hat{W}_{i-2,j_l}, \hat{W}_{i+2,j_l}\) and
original \(\hat{W}_{x,i-2,j_l}, \hat{W}_{x,i+2,j_l}\).\par

Step 3. For the tangential derivatives, i.e. \(W^{l,r}_{y,i+1/2,j_l}, W^{l,r}_{yy,i+1/2,j_l}\), a
linear reconstruction with DFF is adopted by using \(W^{l,r}_{i+1/2,j_l}\),\(W^{l,r}_{i+1/2,(j-1)_l}\) and \(W^{l,r}_{i+1/2,(j+1)_l}\) at the Gaussian points.
And \(W^{l,r}_{xy,i-1/2,j_l}\) could be obtained in the same way with
corresponding \(W^{l,r}_{x,i-1/2,j_l}\),\(W^{l,r}_{x,i-1/2,(j-1)_l}\) and \(W^{l,r}_{x,i-1/2,(j+1)_l}\).

Step 4. To obtain the macro variables representing equilibrium state and derivatives for each Gaussian point  by using
Eq.~\eqref{equ2.8}.

Similar procedure can be performed to obtain all needed values at each Gaussian point. After gas evolution
process, the updated cell interface values are obtained, i.e. at
time \( t = t^*, W_{i\pm1/2,j}^*, W_{i,j\pm1/2}^* \), as well as the cell averaged slopes

\begin{equation}
\overline{W}^*_{x,i,j} = \frac{1}{\Delta x} \sum_{l=1}^m (W_{i+1/2,j_l}^* - W_{i-1/2,j_l}^*),
\end{equation}
\begin{equation}
\overline{W}^*_{y,i,j} = \frac{1}{\Delta y} \sum_{l=1}^m (W_{i_l,j+1/2}^* - W_{i_l,j-1/2}^*),
\end{equation}
according to the Gauss's theorem. The cell averaged values are computed through conservation laws
\begin{equation}
\overline{W}_{i,j}^* = \overline{W}_{i,j}^n - \frac{1}{\Delta x} \sum_{l=1}^m (F_{i+1/2,j_l}^* - F_{i-1/2,j_l}^*) - \frac{1}{\Delta y} \sum_{l=1}^m (G_{i_l,j+1/2}^* - G_{i_l,j-1/2}^*),
\end{equation}

where \( F \) and \( G \) are corresponding fluxes in x and y direction. Lastly, in the rectangular case, the
line averaged slopes are approximated by
\begin{equation}
\hat{W}^*_{x,i,j_l} = \frac{1}{\Delta x} (W_{i+1/2,j_l}^* - W_{i-1/2,j_l}^*),
\end{equation}
\begin{equation}
\hat{W}^*_{y,i_l,j} = \frac{1}{\Delta y} (W_{i_l,j+1/2}^* - W_{i_l,j-1/2}^*).
\end{equation}

\section{Numerical tests}
\label{sec4}
In this section, Numerical tests will be presented to validate the different compact GKS schemes. The ratio of specific heats takes $\gamma =1.4$. And for the inviscid flow, the collision time $\tau $ is
\begin{equation*}
\tau ={{c}_{1}}\Delta t+{{c}_{2}}\left| \frac{{{p}_{l}}-{{p}_{r}}}{{{p}_{l}}+{{p}_{r}}} \right|\Delta t,
\end{equation*}
where ${{p}_{l}}$ and ${{p}_{r}}$ denote the pressure on the left and right cell interface. Usually ${{c}_{1}}=0.05$ and ${{c}_{2}}=1$ are chosen in the classic HGKS. The pressure jump term in $\tau $ enhances shock thickness by adding artificial dissipation.\par
For the viscous flow~\cite{YANG2022110706,CiCP-28-539}, the collision time term related to the viscosity coefficient
is defined as
\begin{equation*}
\tau =\frac{\mu }{p}+{{c}_{2}}\left| \frac{{{p}_{l}}-{{p}_{r}}}{{{p}_{l}}+{{p}_{r}}} \right|\Delta t,
\end{equation*}
where $\mu $ is the dynamic viscous coefficient and $p$ is the pressure at the cell interface. In smooth viscous flow
region, it reduces to $\tau =\mu /p$.\par
The time step is determined by
\begin{equation*}
\Delta t={{C}_{CFL}}Min\left( \frac{\Delta x}{\left\| \mathbf{U} \right\|+{{a}_{s}}},\frac{{{\left( \Delta x \right)}
^{2}}}{4\nu } \right),
\end{equation*}
where $\left\| \mathbf{U} \right\|$ is the magnitude of velocities, ${{C}_{CFL}}$ is the CFL number, ${{a}_{s}}$ is the
sound speed and $\nu =\mu /\rho$ is the kinematic viscosity coefficient.\par

\subsection{Accuracy test in 1-D}
\label{sec4.1}
The advection of density perturbation is examined with initial conditions set as
\begin{equation*}
\rho \left( x \right)=1+0.2sin\left( \pi x \right),U\left( x \right)=1,p\left( x \right)=1,x\in \left[ 0,2 \right].
\end{equation*}
The test case employs periodic boundary conditions on both the left and right sides. The analytical solution for the
advection of density perturbation is given by
\begin{equation*}
\rho \left( x,t \right)=1+0.2sin\left( \pi \left( x-t \right) \right),U\left( x,t \right)=1,p\left( x,t \right)=1,x\in
\left[ 0,2 \right].
\end{equation*}
In this test, for the two-stage fourth-order time discretization and the r-th reconstruction compact GKS based on DFF,
the truncation error is $\mathcal{O}(\Delta {{x}^{r}}+\Delta {{t}^{4}})$. To test the r-th accuracy, the time step size
is taken
as $\Delta t=c\Delta {{x}^{r/4}}$. As the mesh is continuously refined, the results of the error and convergence
accuracy of are presented in Table \ref{1Daccuracy01}-\ref{1Daccuracy03}. Among them, ASE-DFF(5,3)-CGKS and
ASE-DFF(7,5,3)-CGKS achieved the expected error
and convergence accuracy of 5-th and 7-th order. ASE-DFF(9,7,5,3)-CGKS also achieved the expected 9-th order overall.
However, due to the insufficient time accuracy and reaching a sufficiently small machine error, the accuracy decreased
when the number of mesh points was sufficiently large.
\begin{table}[htp]
\footnotesize
\begin{center}
\vspace{-4mm} \caption{\label{1Daccuracy01} Accuracy test in 1-D for the advection of density perturbation by the ASE-DFF(5,3)-CGKS. $\Delta t=0.3\Delta {{x}^{5/4}}.$}
\def\temptablewidth{1.0\textwidth}
{\rule{\temptablewidth}{1pt}}
\begin{tabular*}{\temptablewidth}{@{\extracolsep{\fill}}c|cc|cc|cc}
mesh length & ${{L}^{1}}$error & Order & ${{L}^{2}}$error & Order & ${{L}^{\infty }}$error & Order\\
\hline
1/5 & 7.457899e-03 & ~ & 8.198459e-03 & ~ & 1.149022e-02 & ~ \\
1/10 & 2.525347e-04 & 4.88 & 2.877034e-04 & 4.83 & 4.049745e-04 & 4.83 \\
1/20 & 8.246889e-06 & 4.94 & 9.170583e-06 & 4.97 & 1.346789e-05 & 4.91 \\
1/40 & 2.594743e-07 & 4.99 & 2.878268e-07 & 4.99 & 4.238936e-07 & 4.99 \\
1/80 & 8.137760e-09 & 4.99 & 9.019865e-09 & 5.00 & 1.328944e-08 & 5.00 \\
\end{tabular*}
{\rule{\temptablewidth}{0.1pt}}
\end{center}
\end{table}
\begin{table}[htp]
\footnotesize
\begin{center}
\vspace{-4mm} \caption{\label{1Daccuracy02}Accuracy test in 1-D for the advection of density perturbation by the ASE-DFF(7,5,3)-CGKS. $\Delta t=0.3\Delta {{x}^{7/4}}.$}
\def\temptablewidth{1.0\textwidth}
{\rule{\temptablewidth}{1pt}}
\begin{tabular*}{\temptablewidth}{@{\extracolsep{\fill}}c|cc|cc|cc}
mesh length & ${{L}^{1}}$error & Order & ${{L}^{2}}$error & Order & ${{L}^{\infty }}$error & Order\\
\hline
1/5 & 1.131943e-03 & ~ & 1.280001e-03 & ~ & 1.750822e-03 & ~ \\
1/10 & 9.551121e-06 & 6.89 & 1.064429e-05 & 6.91 & 1.504539e-05 & 6.86 \\
1/20 & 7.577352e-08 & 6.98 & 8.389434e-08 & 6.99 & 1.230683e-07 & 6.93 \\
1/40 & 5.891611e-10 & 7.01 & 6.532323e-10 & 7.00 & 9.635079e-10 & 7.00 \\
1/80 & 4.575704e-12 & 7.01 & 5.072149e-12 & 7.01 & 7.505108e-12 & 7.00 \\
\end{tabular*}
{\rule{\temptablewidth}{0.1pt}}
\end{center}
\end{table}
\begin{table}[htp]
\footnotesize
\begin{center}
\vspace{-4mm} \caption{\label{1Daccuracy03}Accuracy test in 1-D for the advection of density perturbation by the ASE-DFF(9,7,5,3)-CGKS. $\Delta t=0.3\Delta {{x}^{9/4}}.$}
\def\temptablewidth{1.0\textwidth}
{\rule{\temptablewidth}{1pt}}
\begin{tabular*}{\temptablewidth}{@{\extracolsep{\fill}}c|cc|cc|cc}
mesh length & ${{L}^{1}}$error & Order & ${{L}^{2}}$error & Order & ${{L}^{\infty }}$error & Order\\
\hline
1/5 & 1.066673e-04 & ~ & 1.227691e-04 & ~ & 1.726543e-04 & ~ \\
1/10 & 2.196754e-07 & 8.92 & 2.410227e-07 & 8.99 & 3.486130e-07 & 8.95 \\
1/20 & 4.299771e-10 & 9.00 & 4.778505e-10 & 8.98 & 6.960887e-10 & 8.97 \\
1/40 & 8.676032e-13 & 8.95 & 9.636774e-13 & 8.95 & 1.413758e-12 & 8.94 \\
1/80 & 2.690764e-14 & 5.01 & 3.019528e-14 & 5.00 & 5.551115e-14 & 4.67 \\
\end{tabular*}
{\rule{\temptablewidth}{0.1pt}}
\end{center}
\end{table}	\par
	
\subsection{Shock-tube problem}
\label{sec4.2}
The standard shock tube problem is used to test the performance of the numerical scheme.
The initial condition of the Sod shock tube problem~\cite{SOD19781} is as follows
\begin{equation*}
\left( \rho ,u,p \right)=\left\{ \begin{aligned}
& \left( 1,\text{ }0,\text{ }1 \right),\text{ 0}\le x<0.5, \\
& \left( 0.125,\text{ }0,\text{ 0}\text{.1} \right),\text{ 0}\text{.5}\le x\le 1.0, \\
\end{aligned} \right.
\end{equation*}
The two ends have non-reflecting boundary conditions. The final simulation time is $t = 0.2$.\par
The initial condition for the Lax shock tube problem~\cite{cpa.3160070112} is as follows
\begin{equation*}
\left( \rho ,u,p \right)=\left\{ \begin{aligned}
& \left( 0.445,\text{ }0.698,\text{ }3.528 \right),\text{ 0}\le x<0.5, \\
& \left( 0.5,\text{ }0,\text{ 0}\text{.571} \right),\text{ 0}\text{.5}\le x\le 1.0, \\
\end{aligned} \right.
\end{equation*}
Similarly, as in the Sod problem, both ends have non-reflecting boundary conditions. The final time is $t = 0.14$.\par
\begin{figure}[htp]	
\centering
\includegraphics[width=0.4\textwidth]{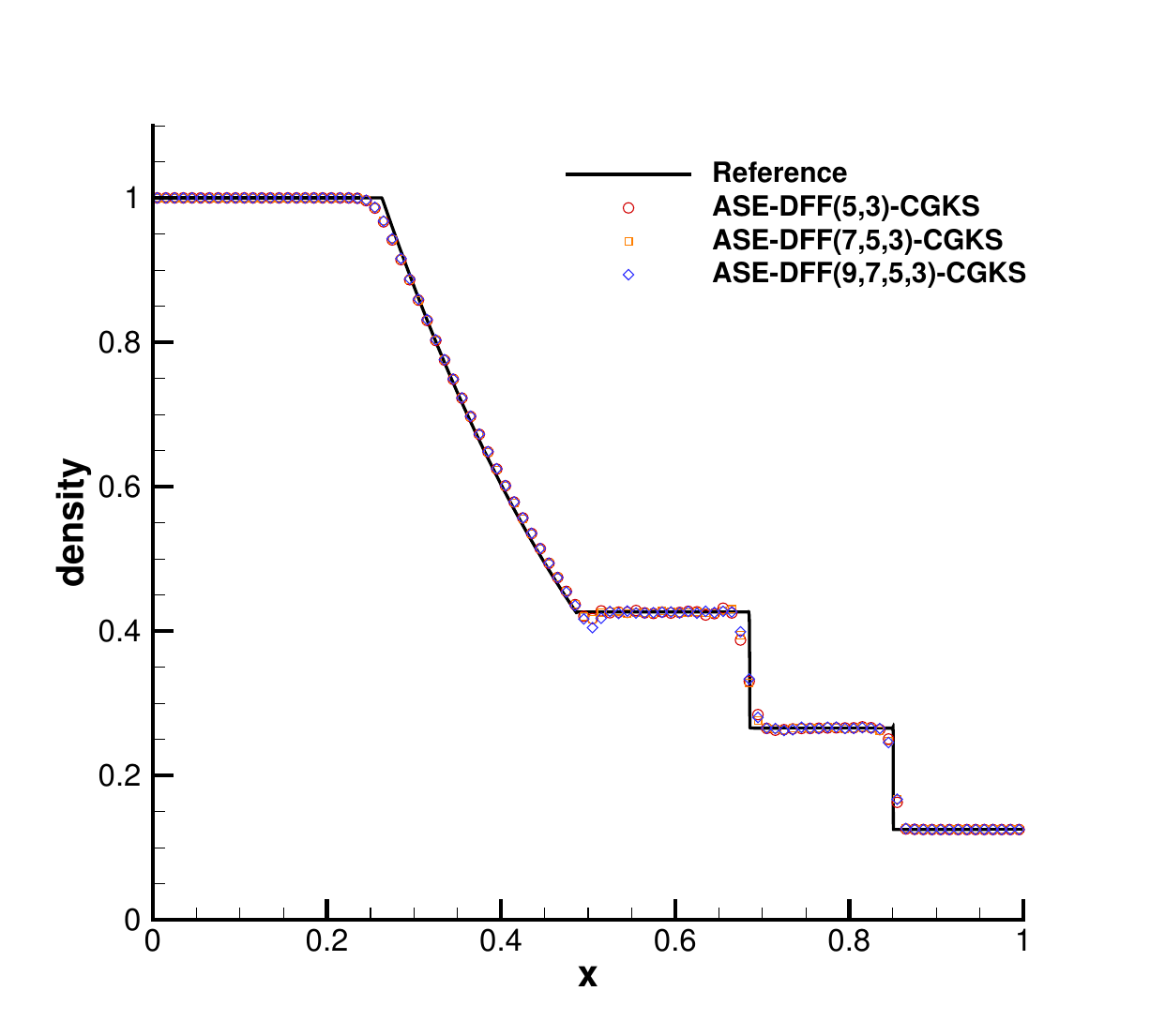}
\includegraphics[width=0.4\textwidth]{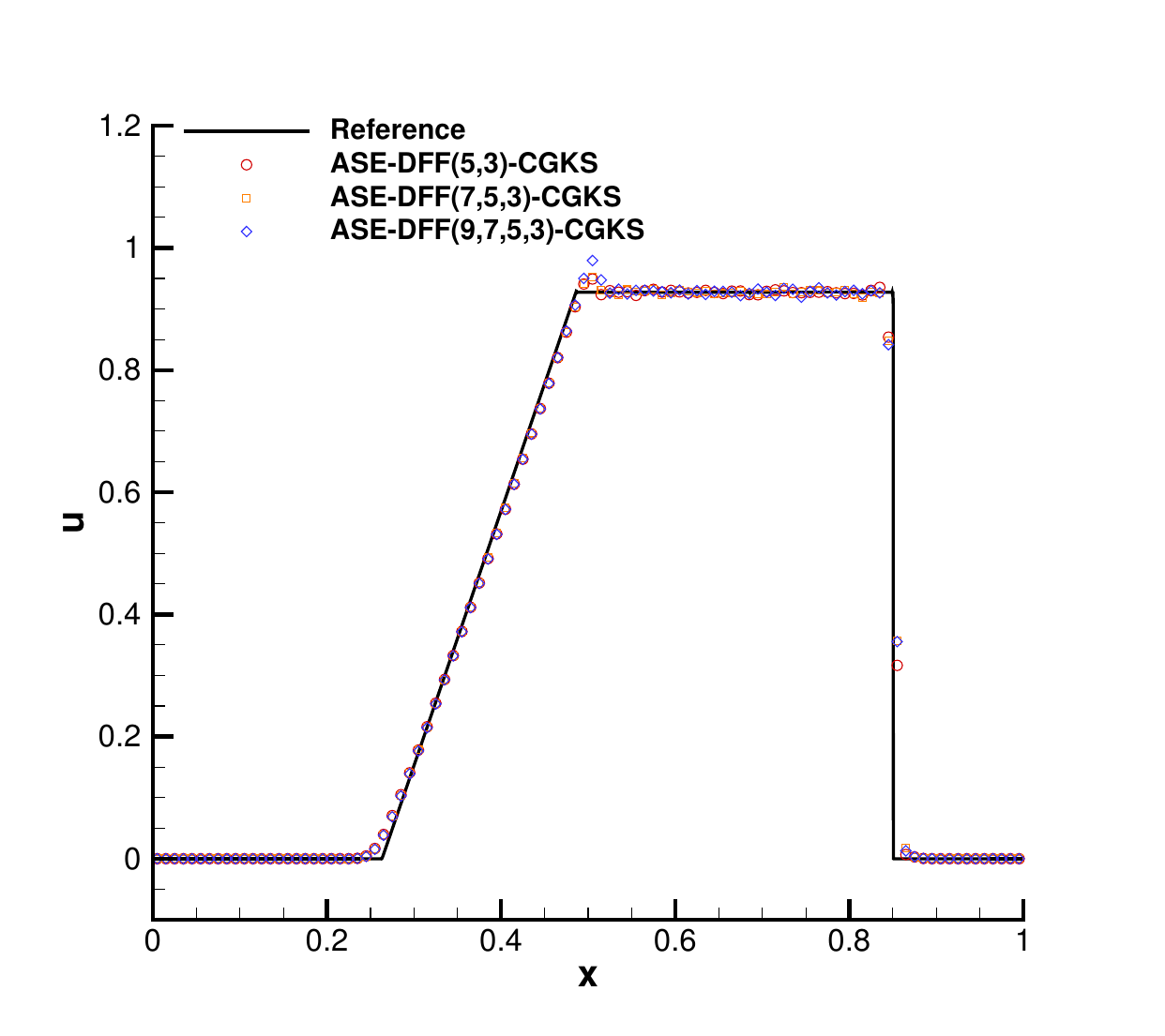}
\vspace{-4mm} \caption{\label{figSOD}
Sod problem: the density, velocity
distributions from ASE-DFF(5,3)-CGKS, ASE-DFF(7,5,3)-CGKS and ASE-DFF(9,7,5,3)-CGKS
with 100 cells. $CFL=0.5.$ $T=0.2.$ }
\end{figure}
\begin{figure}[htp]	
\centering
\includegraphics[width=0.4\textwidth]{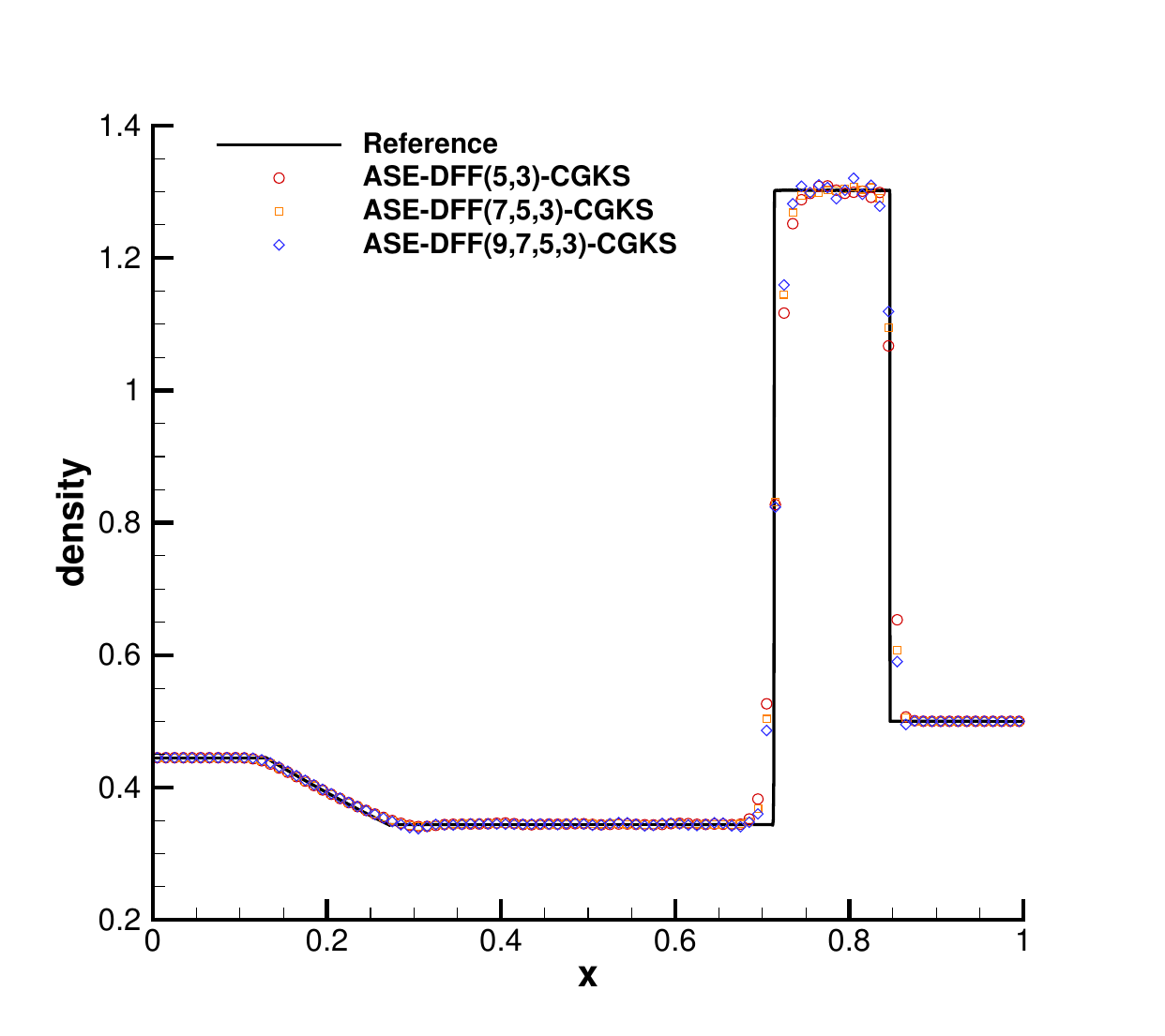}
\includegraphics[width=0.4\textwidth]{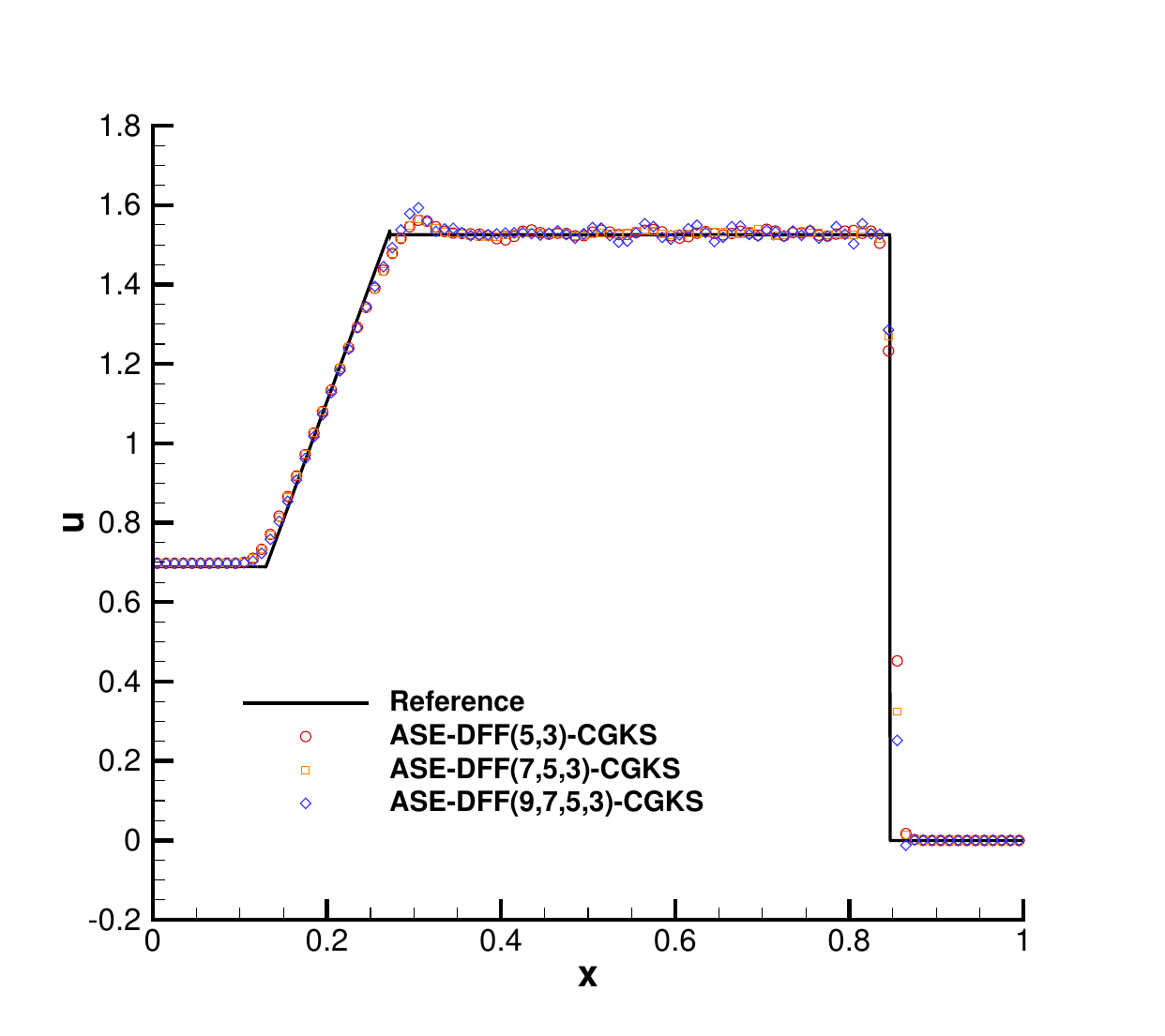}
\vspace{-4mm} \caption{\label{figLAX}
Lax problem: the density distributions and local enlargements from ASE-DFF(5,3)-CGKS, ASE-DFF(7,5,3)-CGKS and
ASE-DFF(9,7,5,3)-CGKS with 100 cells. $CFL=0.5.$ $T=0.14.$ }
\end{figure}

Fig.~\ref{figSOD} and Fig.~\ref{figLAX} show the results of three compact GKS schemes using 100 grid points respectively.
Compared with the reference result, the density and velocity distributions of the three compact GKS are in good agreement
with the reference solution. Moreover, the higher-order ASE-DFF(7,5,3)-CGKS and ASE-DFF(9,7,5,3)-CGKS schemes simultaneously
possess the ability to sharp capture discontinuity.

\subsection{Shock-density wave interaction}
\label{sec4.3}
The first case was proposed by Shu and Osher ~\cite{SHU198932}. The Shu-Osher shock wave acoustic interaction model
involves an interaction between a shock wave with a speed of 3 Mach and a disturbed density field. The calculation is
carried out within the domain [0, 10], using a total of 200 grid points. The final simulation time is $t = 1.8$. On the
left side, non-reflecting boundary condition is set. While on the right side, the fixed wave profile is extended on the
right. The initial conditions are as follows
\begin{equation*}
\left( \rho ,u,p \right)=\left\{ \begin{aligned}
& \left( 3.857134,2.629369,10.33333 \right),0\le x< 1, \\
& \left( 1+0.2\sin \left( 5(x-5) \right),0,1 \right),1\le x\le 10. \\
\end{aligned} \right.
\end{equation*}
As an extension of the Shu-Osher problem, Titarev and Toro ~\cite{SURESH199783} proposed a second case to test a severely
oscillatory
wave that interacts with shock discontinuity. The computational domain is [0,10], with N = 1000 uniformly
distributed grid cells. The initial conditions are as follows
\begin{equation*}
\left( \rho ,u,p \right)=\left\{ \begin{aligned}
& \left( 1.515695,0.523346,1.805 \right),0\le x< 0.5, \\
& \left( 1+0.1\sin \left( 20\pi(x-5) \right),0,1 \right),0.5\le x\le 10. \\
\end{aligned} \right.
\end{equation*}
The final evolution time is $t = 5.0$. The boundary conditions are the same as those in
the Shu-osher problem.\par
\begin{figure}[htp]	
\centering
\includegraphics[width=0.4\textwidth]{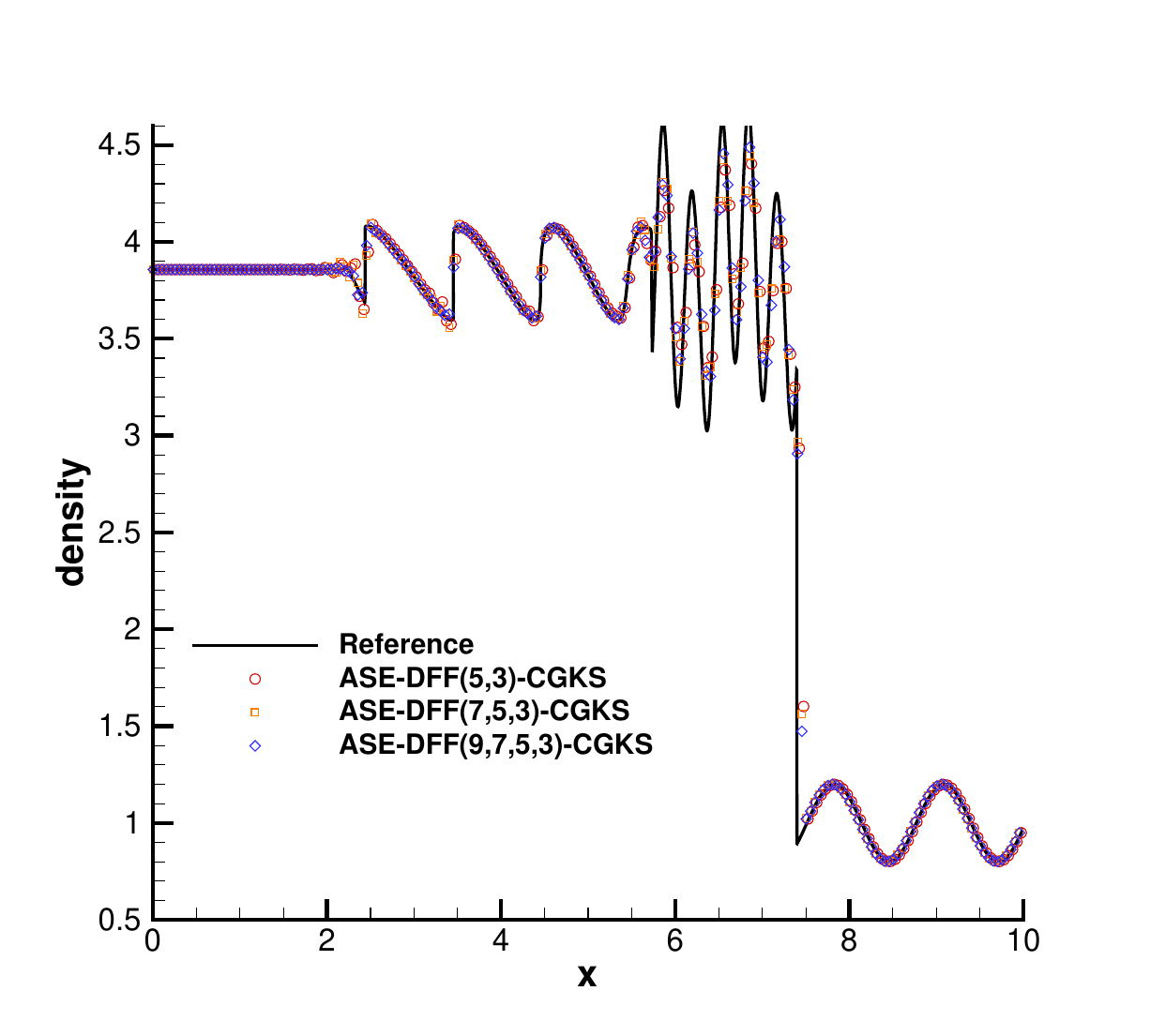}
\includegraphics[width=0.4\textwidth]{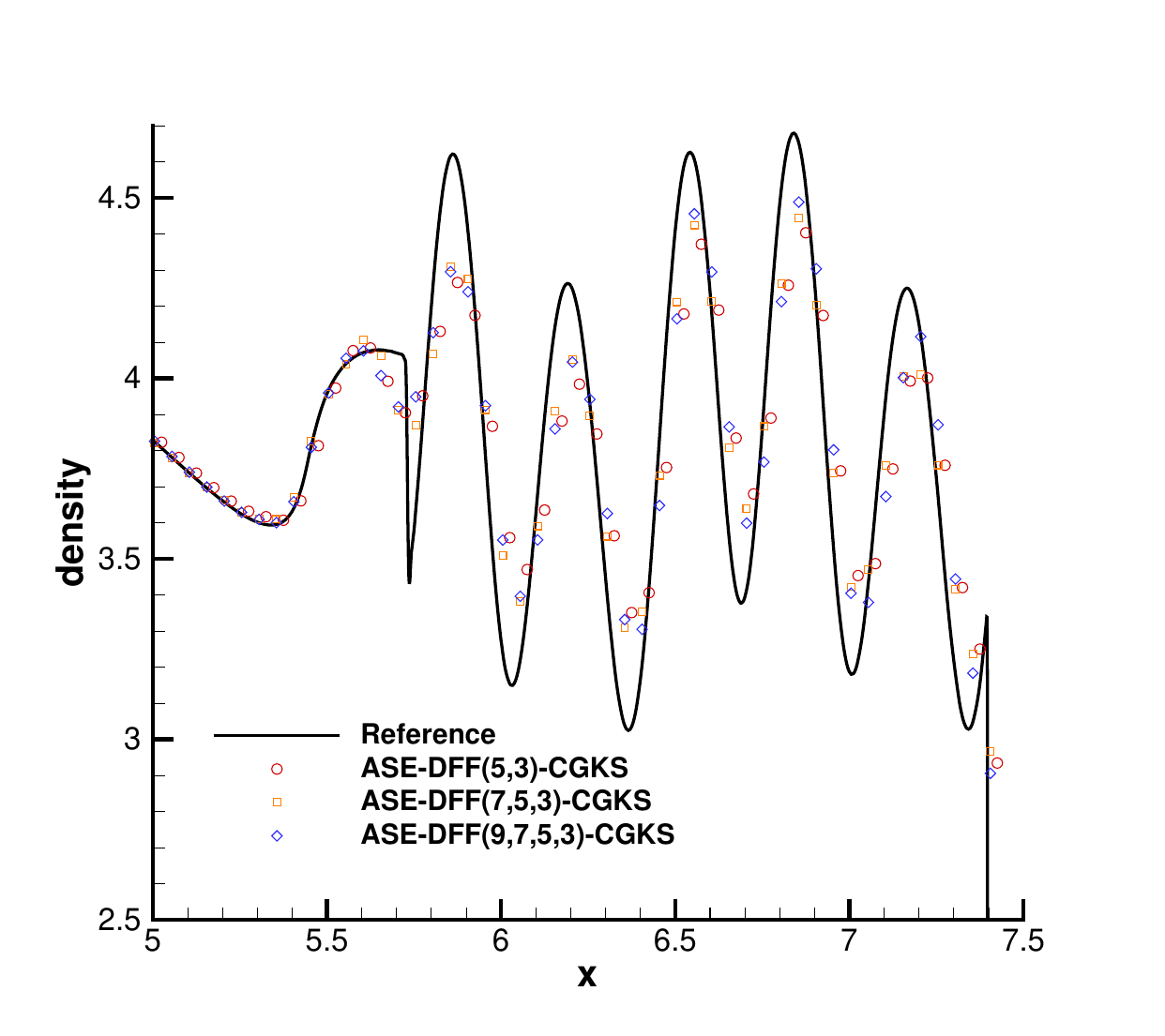}
\vspace{-4mm} \caption{\label{figshu}
Shu-Osher problem: the density distributions and local enlargements for different
schemes with 200 cells. $CFL=0.5.$ $T=1.8.$ }
\end{figure}
\begin{figure}[htp]	
\centering
\includegraphics[width=0.4\textwidth]{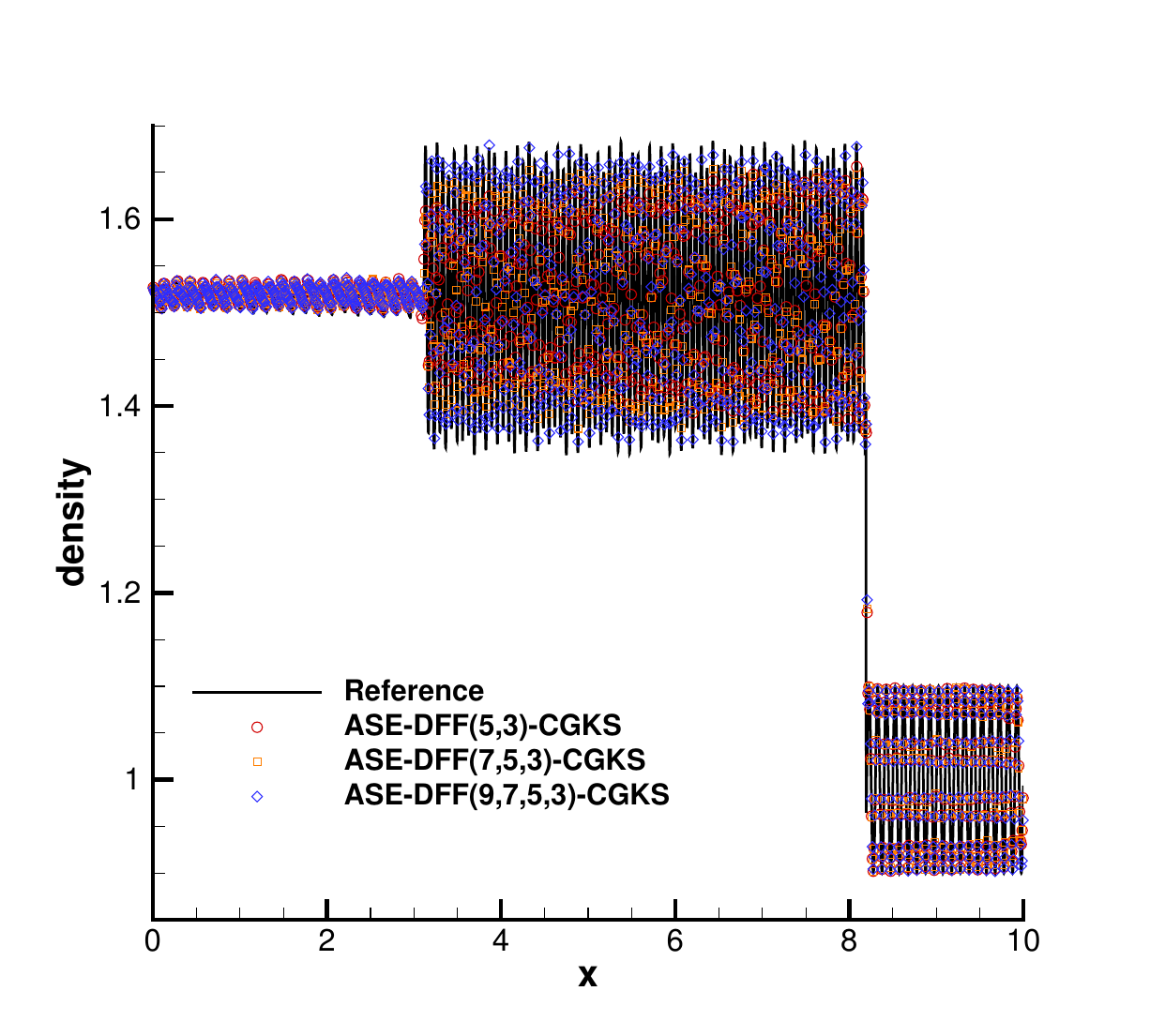}
\includegraphics[width=0.4\textwidth]{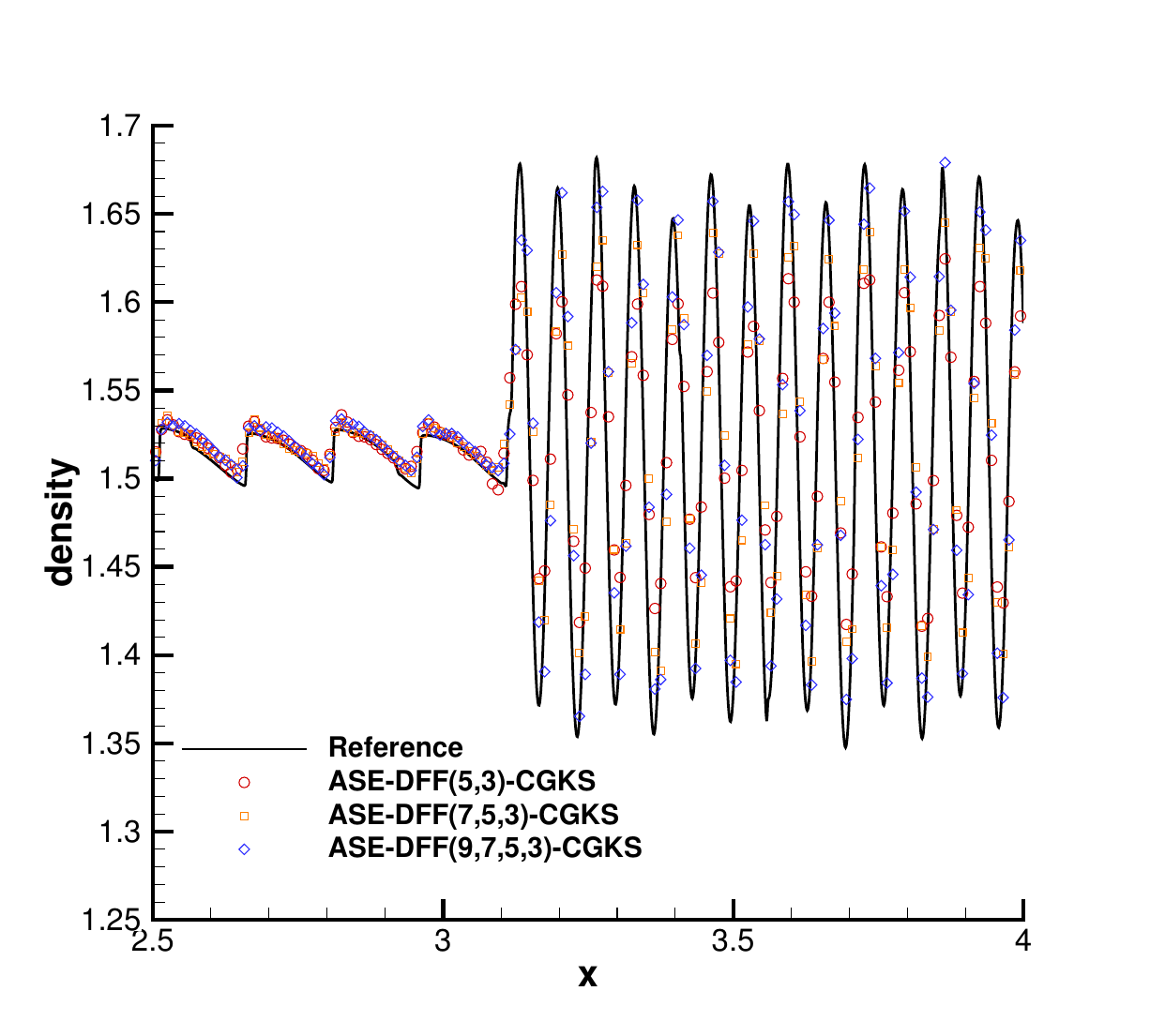}
\vspace{-4mm} \caption{\label{figtoro}
Titarev-Toro problem: the density distributions and local enlargements for different
schemes with 1000 cells. $CFL=0.5.$ $T=5.0.$ }
\end{figure}

Fig.~\ref{figshu} and \ref{figtoro} present the density distribution and enlarged images for the achieved computational
time.
In Fig.~\ref{figshu}, ASE-DFF(9,7,5,3)-CGKS and ASE-DFF(7,5,3)-CGKS capture the amplitude of the density wave more
effectively
and have higher resolution compared to ASE-DFF(5,3)-CGKS. While in Fig.~\ref{figtoro}, the compact CGKS demonstrates
extremely superior performance. Compared to the reference solution obtained by using 10,000 points WENO5-Z-GKS.
All three CGKS
using only 1,000 points have very high resolution. Moreover, the 9th-order
ASE-DFF(9,7,5,3)-CGKS even captures the same peak as the reference solution at some positions. This proves that
ASE-DFF-CGKS
has excellent shock-capturing and wave-resolution capabilities.\\

\subsection{Interacting blast waves}
\label{sec4.4}
\begin{figure}[htp]	
\centering
\includegraphics[width=0.4\textwidth]{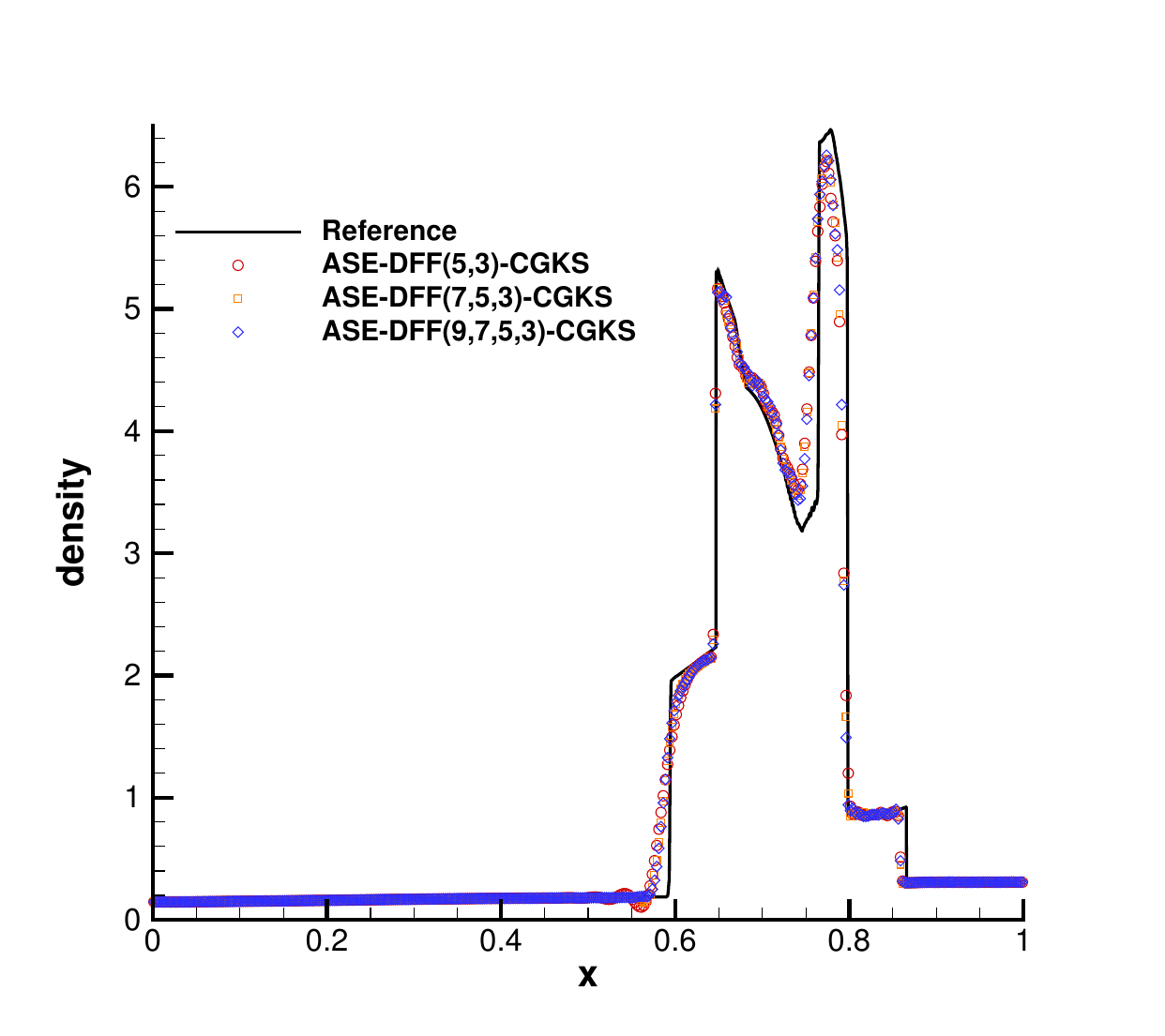}
\includegraphics[width=0.4\textwidth]{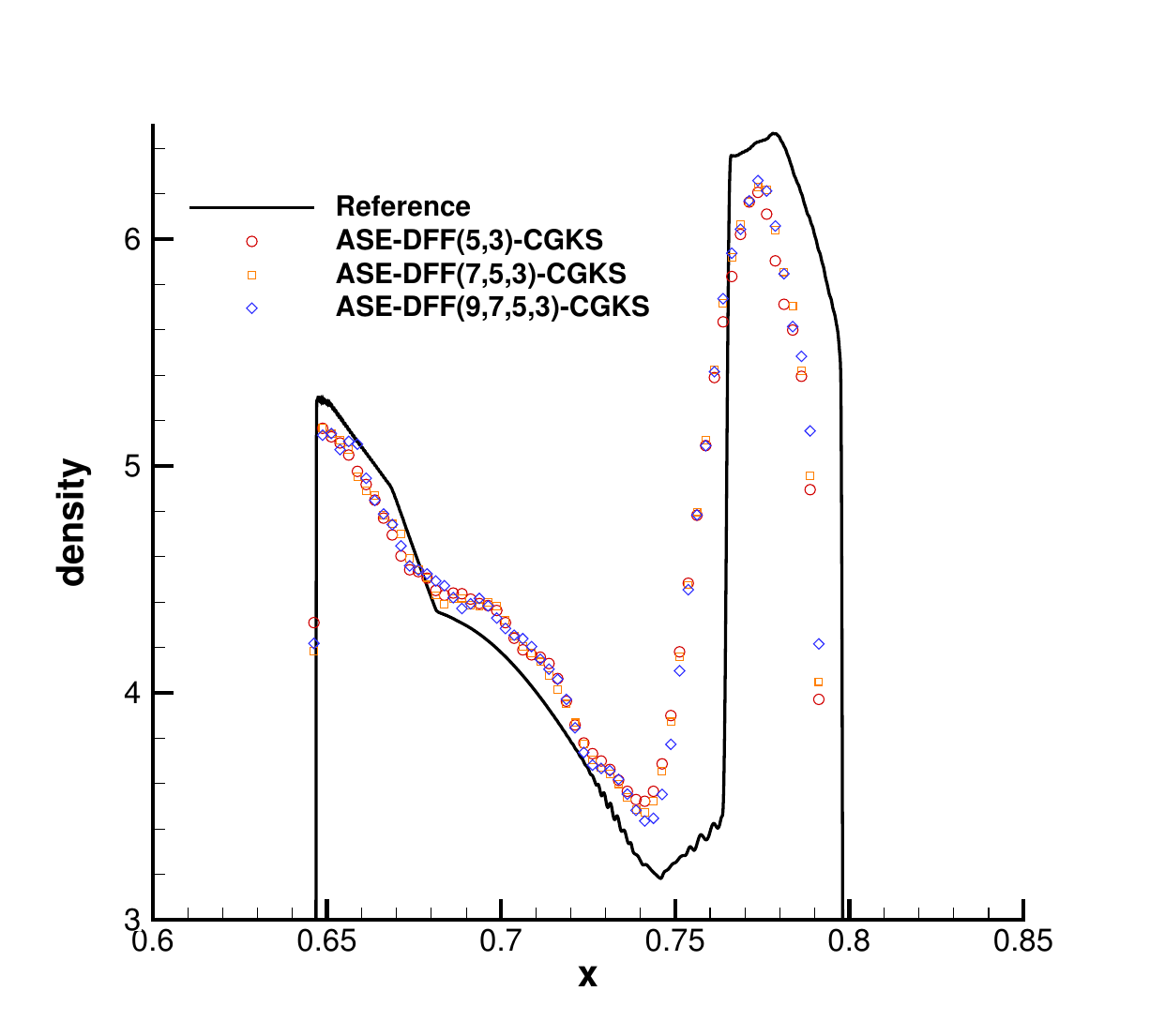}
\vspace{-4mm} \caption{\label{figblast}
Blast wave problem: the density distributions for different schemes with 400 cells.  $CFL=0.5.$ $T=0.038.$ }
\end{figure}

The blast wave problem is taken from Woodward-Colella blast~\cite{WOODWARD1984115}. The initial conditions
for the blast wave problem are given as follows
\begin{equation*}
\left( \rho ,u,p \right)=\left\{ \begin{aligned}
& \left( 1,0,1000 \right),0\le x<0.1, \\
& \left( 1,0,0.01 \right),0.1\le x<0.9, \\
& \left( 1,0,100 \right),0.9\le x\le 1.0. \\
\end{aligned} \right.
\end{equation*}
The computational domain employed 400 uniform grid points, and reflective boundary conditions are applied
at both ends. The conditions used in the calculation are $CFL = 0.5$. Fig.~\ref{figblast} shows the density
distribution curve at $t = 0.038$. The calculation results of ASE-DFF(5,3)-CGKS, ASE-DFF(7,5,3)-CGKS, and
ASE-DFF(9,7,5,3)-CGKS are in good agreement with the reference solution. The greatest
advantage brought by the discontinuity feedback factor is the improvement in the robustness of the high-order scheme.
\subsection{Low density and low pressure problems}
\label{sec4.5}
To further test the robustness of the schemes, two 1D test problems involving vacuum or strong discontinuity are
considered,
namely the double rarefaction wave problem~\cite{HU2013169} and the Le Blanc problem~\cite{LOUBERE2005105}. For the
double rarefaction wave problem, the density and pressure in the
central region approach zero. The initial conditions are
\begin{equation*}
\left( \rho ,u,p \right)=\left\{ \begin{aligned}
& \left( 1.0,\text{ }-2,\text{}0.1 \right),\text{ 0}\le x<0.5, \\
& \left( 1.0,\text{ }2, \text{0.1} \right),\text{ 0}\text{.5}\le x\le 1.0, \\
\end{aligned} \right.
\end{equation*}
The final simulation time is $t = 0.1$ and the number of grid points is $N = 400$. \par
For the Le Blanc problem, the ratio of initial density to pressure is relatively high. Therefore, an extremely intense
expansion wave will be generated in the high-pressure region. The initial conditions are
\begin{equation*}
\left( \rho ,u,p \right)=\left\{ \begin{aligned}
& \left( 1.0,0,\frac{2}{3}\times {{10}^{-1}} \right),\text{ 0}\le x<3.0, \\
& \left( {{10}^{-3}},0, \frac{2}{3}\times {{10}^{-10}} \right),\text{3.0}\le x<9.0, \\
\end{aligned} \right.
\end{equation*}
The $\gamma =\frac{5}{3}$.The grid resolution is set to $N = 800$, and the final simulation time is set to $t = 6.0$.\par
\begin{figure}[htp]	
\centering
\includegraphics[width=0.4\textwidth]{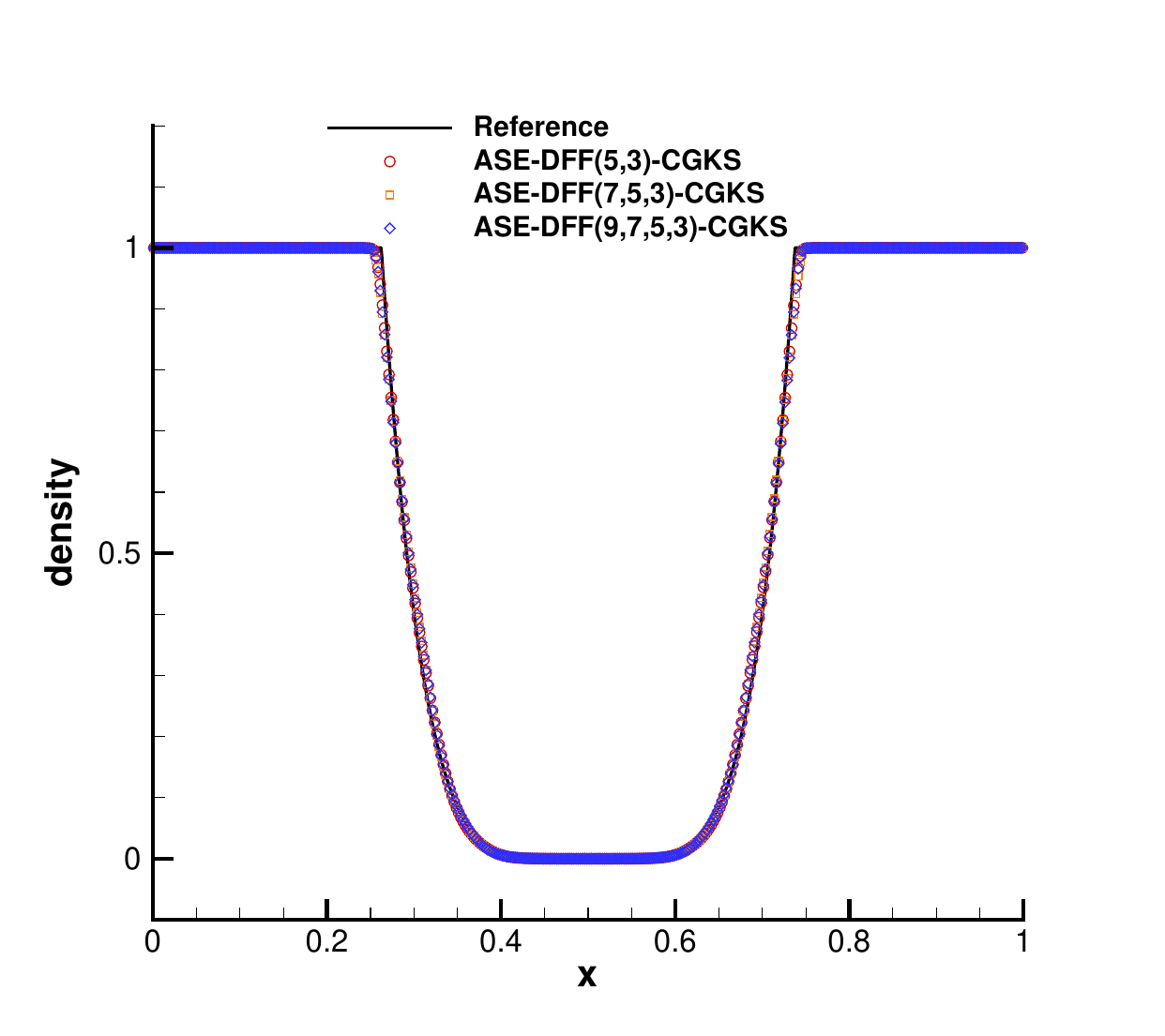}
\includegraphics[width=0.4\textwidth]{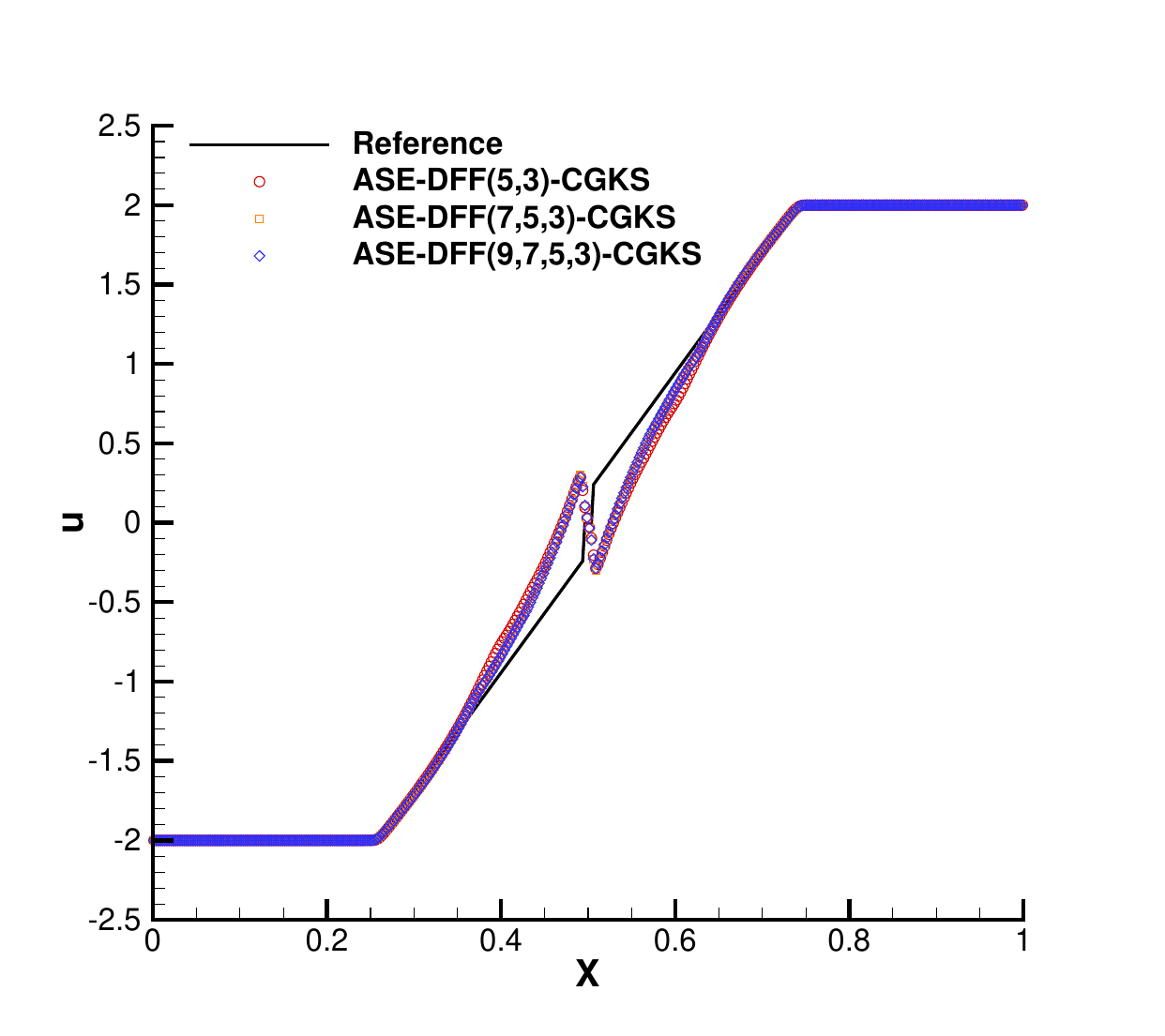}
\includegraphics[width=0.4\textwidth]{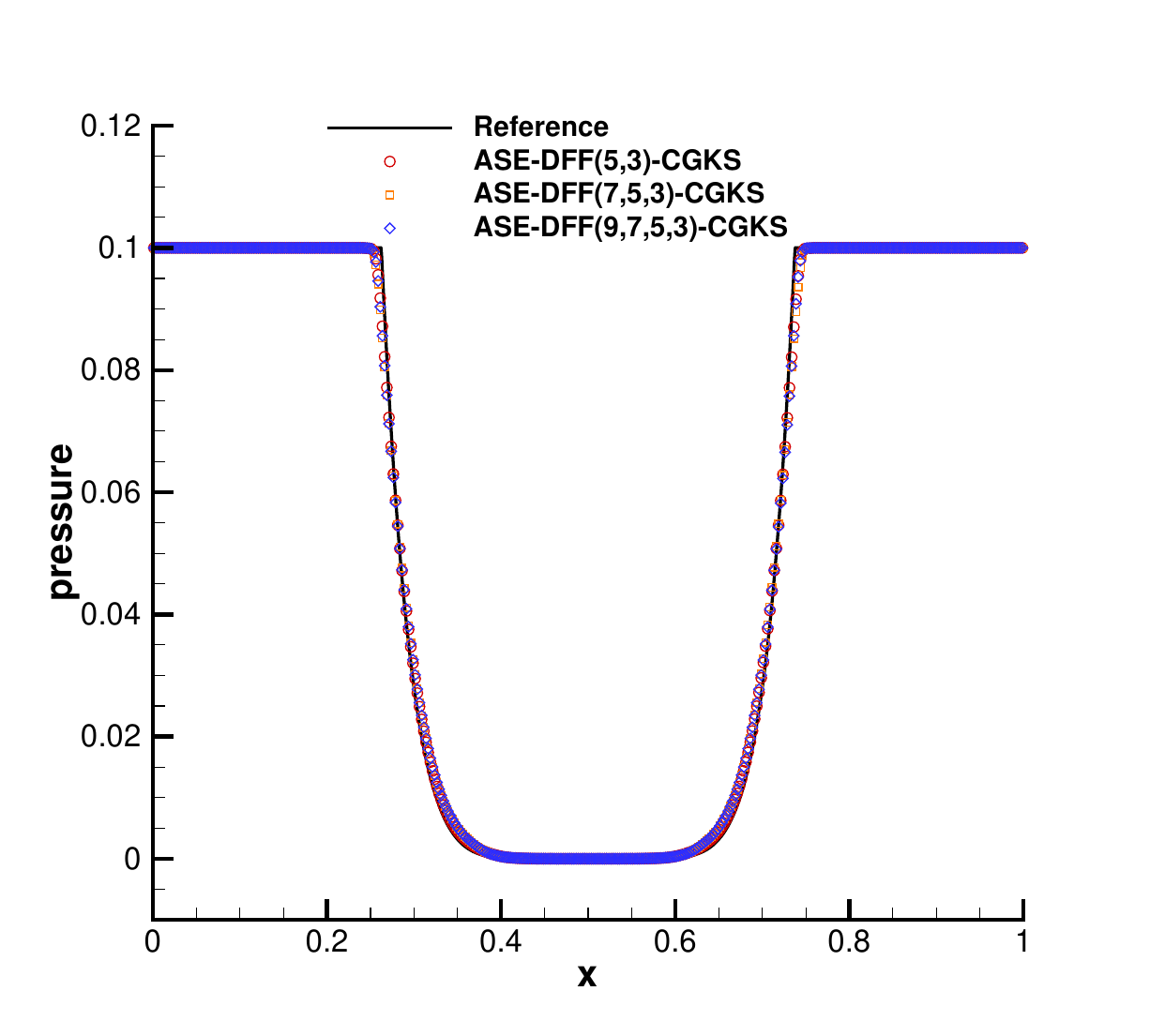}
\vspace{-4mm} \caption{\label{fig123}
Double rarefaction problem: the density, velocity and pressure distributions for different
schemes with 400 cells.  $CFL=0.5.$ $T=0.1.$ }
\end{figure}
\begin{figure}[htp]	
\centering
\includegraphics[width=0.4\textwidth]{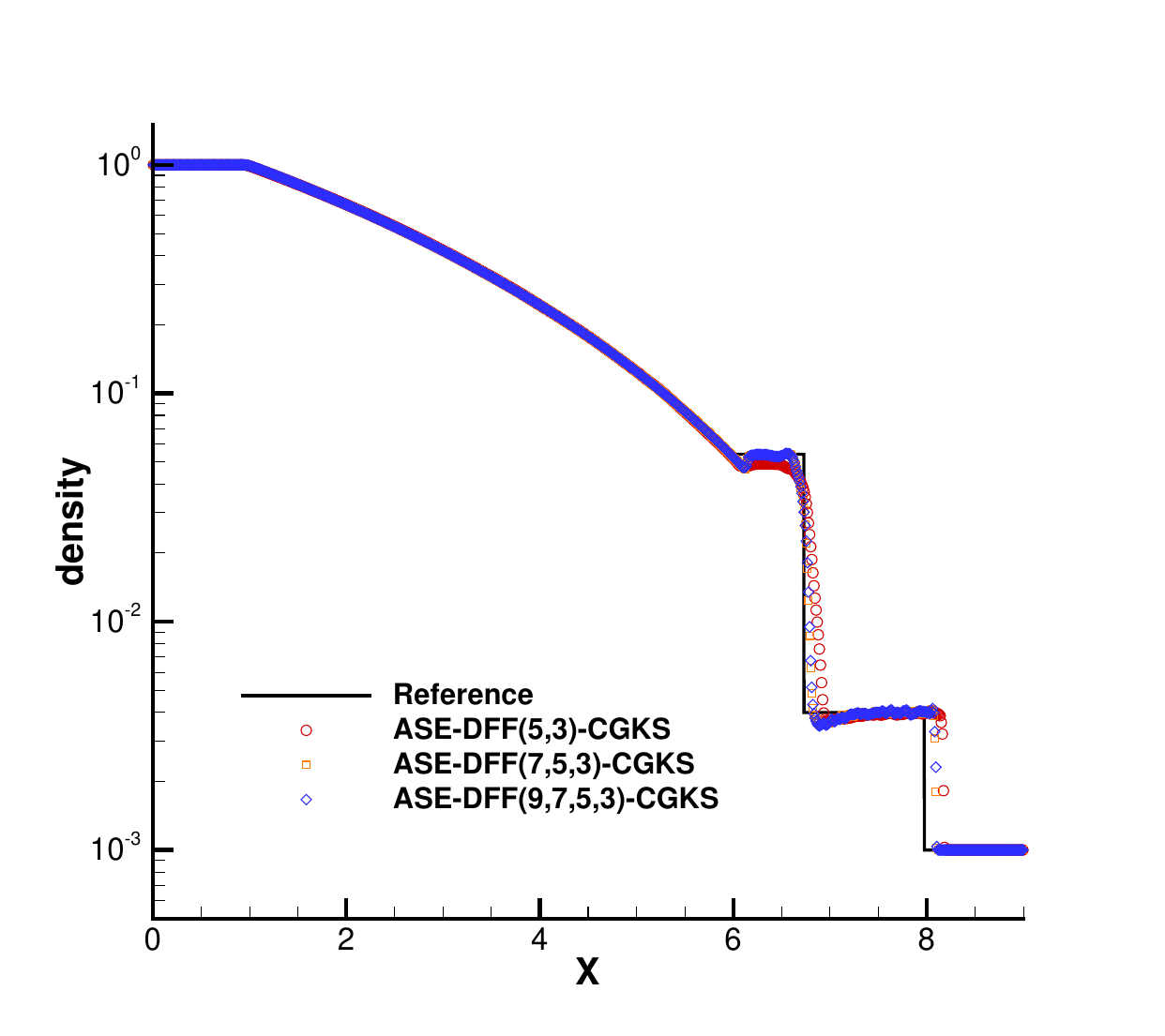}
\includegraphics[width=0.4\textwidth]{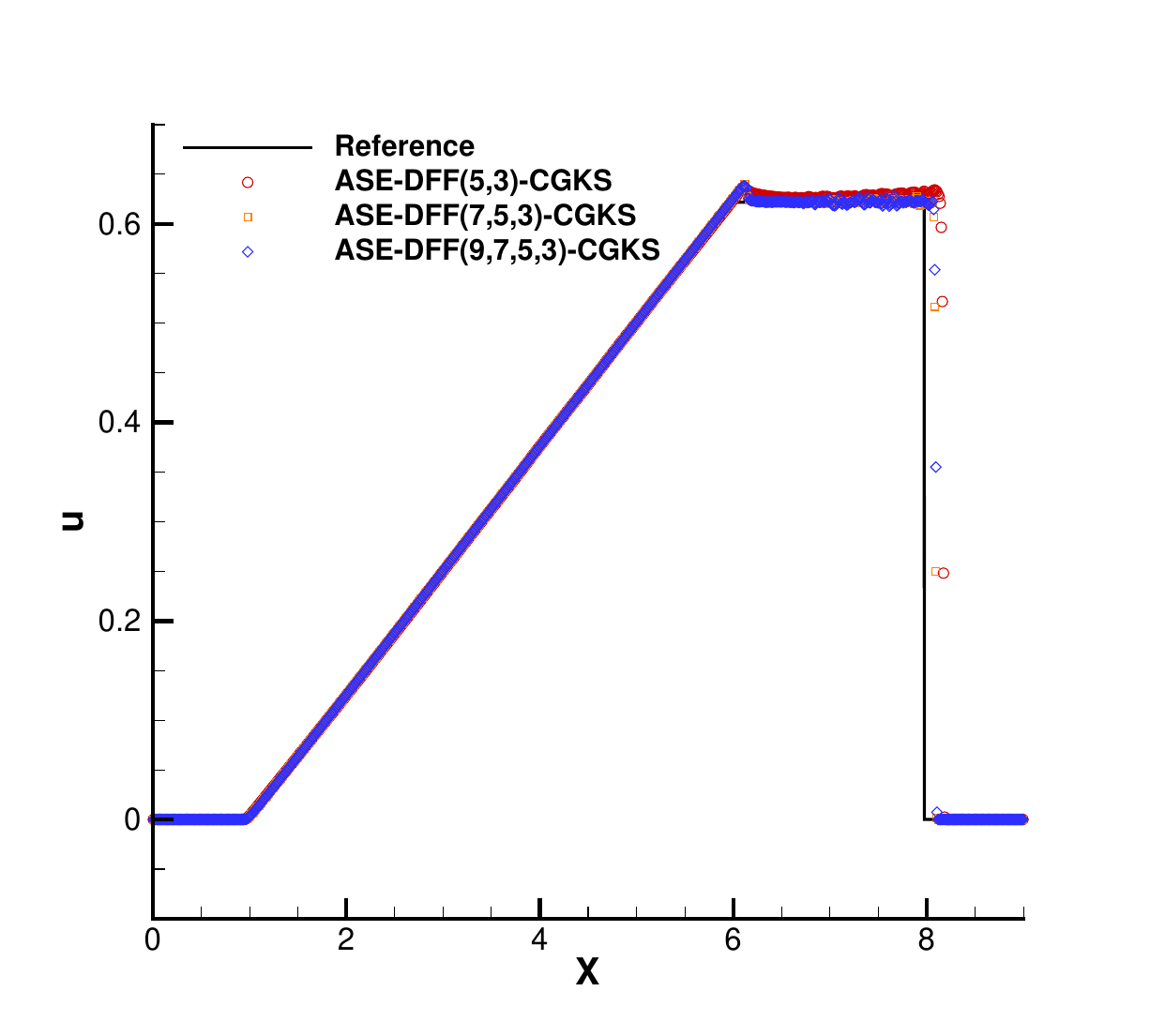}
\includegraphics[width=0.4\textwidth]{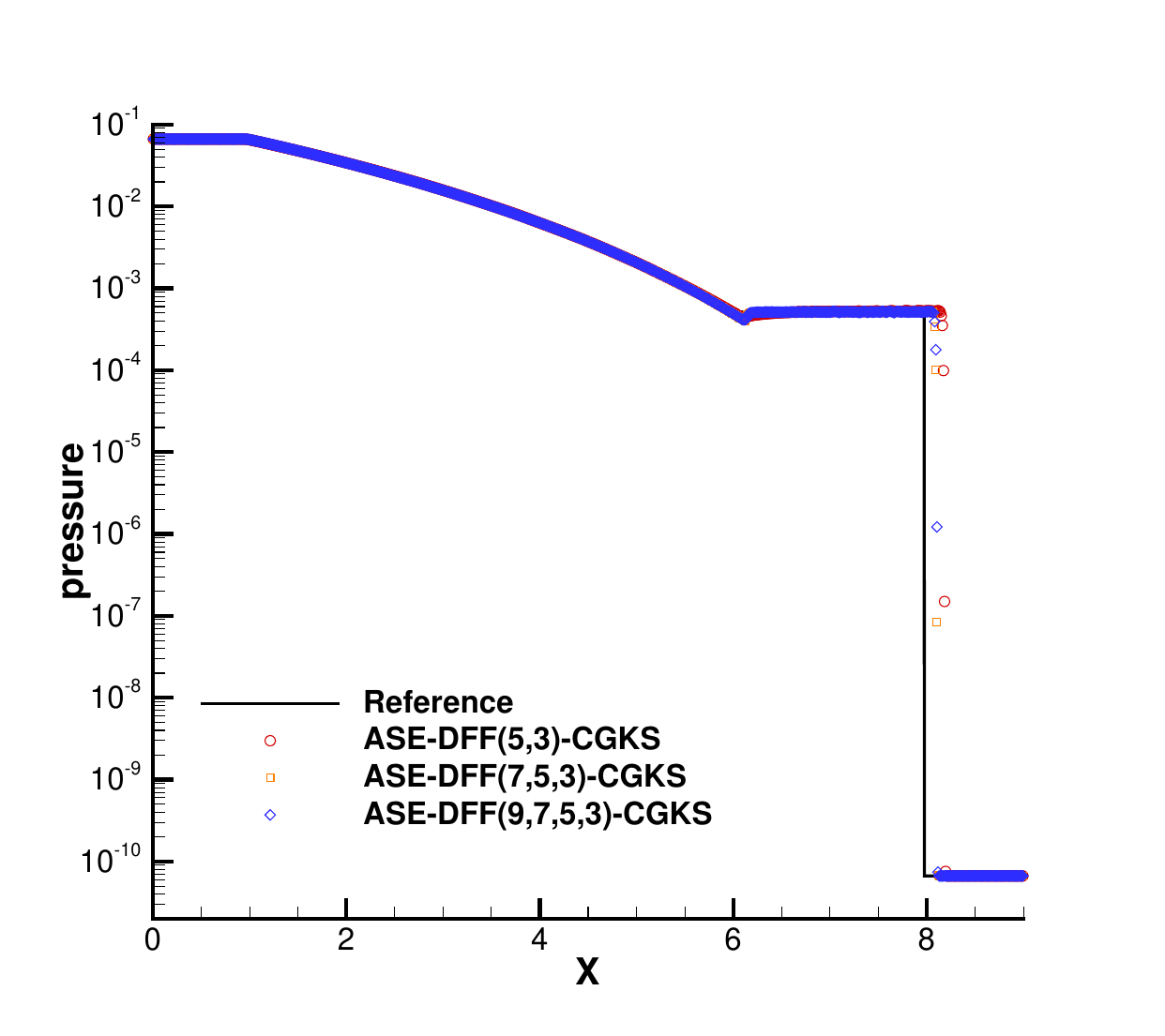}
\vspace{-4mm} \caption{\label{figLB}
Le Blanc problem:  the density, velocity and pressure distributions for different
schemes with 800 cells.  $\gamma =\frac{5}{3}.$ $CFL=0.5.$ $T=6.0.$  }
\end{figure}

In Fig.~\ref{fig123}-\ref{figLB}, the density, velocity and pressure distribution of two
numerical tests are shown for ASE-DFF(5,3)-CGKS, ASE-DFF(7,5,3)-CGKS and ASE-DFF(9,7,5,3)-CGKS.
The results of the three schemes are in good agreement with the reference solutions. In the results of the
double rarefaction wave problem and the
Le Blanc problem, ASE-DFF(7,5,3)-CGKS and ASE-DFF(9,7,5,3)-CGKS have higher resolution compared to the 5-th order
ASE-DFF(5,3)-CGKS. Overall, the discontinuity feedback factor is good at simulating some tests involving strong
shock waves and strong rarefaction waves. Furthermore, only ASE-DFF-CGKS was able to pass the Le Blanc problem.
The traditional WENO-GKS or HWENO-CGKS gas kinetic schemes fail to pass the Le Blanc problem test.Therefore,
the compact GKS using the discontinuity feedback factor can
maintain good robustness in extreme case.

\subsection{Accuracy test in 2-D}
\label{sec4.6}
Two-dimensional accuracy test uses the advection of the two-dimensional advection of density perturbation. For
inviscid flow, the collision time $\tau = 0$. The computational region is
$\left[ 0,2 \right] \times \left[ 0,2 \right]$. A uniform mesh of $N \times N$ is used and the boundary
conditions in both directions are periodic. The initial conditions are
\begin{equation*}
\rho \left( x,y \right)=1+0.2sin\left( \pi \left( x+y \right) \right),U\left( x,y \right)=\left( 1,1 \right),p\left(
x,y \right)=1,\left( x,y \right)\in \left[ 0,2 \right]\times \left[ 0,2 \right].
\end{equation*}
The analytic solution of the 2-D advection of density perturbation is
\begin{equation*}
\rho \left( x,y,t \right)=1+0.2sin\left( \pi \left( x+y-t \right) \right),U\left( x,y,t \right)=1,p\left( x,y,t
\right)=1,\left( x,y \right)\in \left[ 0,2 \right]\times \left[ 0,2 \right].
\end{equation*}
The time step size is taken as $\Delta t=0.3\Delta {{x}^{r/4}}$ to test the r-th accuracy.
The results of error and order for ASE-DFF(5,3)-CGKS, ASE-DFF(7,5,3)-CGKS, and ASE-DFF(9,7,5,3)-CGKS are tested
and presented in the Table \ref{2Daccuracy01} -\ref{2Daccuracy03} respectively. For the 5-th, 7-th and
9-th order compact GKS schemes, 2, 3 and 4 Gaussian points are used respectively on the interface.
All three schemes can achieve the expected accuracy. Among them, under the same number of grids, it can be
clearly seen that as the scheme accuracy increases, the error can be reduced to a smaller level. However, in
terms of convergence accuracy, it is affected by the number of Gaussian points and time accuracy. Among them,
ASE-DFF(5,3)-CGKS still maintains a accuracy close to 5-th order. ASE-DFF(7,5,3)-CGKS can well
achieve 7-th order accuracy. ASE-DFF(9,7,5,3)-CGKS gradually decreases from a accuracy close to 9-th order.
\begin{table}[htp]
\footnotesize
\begin{center}
\vspace{-4mm} \caption{\label{2Daccuracy01} Accuracy test in 2-D for the advection of density perturbation by the ASE-DFF(5,3)-CGKS. $\Delta t=0.3\Delta {{x}^{5/4}}.$}
\def\temptablewidth{1.0\textwidth}
{\rule{\temptablewidth}{1pt}}
\begin{tabular*}{\temptablewidth}{@{\extracolsep{\fill}}c|cc|cc|cc}
mesh length & ${{L}^{1}}$error & Order & ${{L}^{2}}$error & Order & ${{L}^{\infty }}$error & Order\\
\hline
1/5 & 1.208082e-02 & ~ & 1.343530e-02 & ~ & 1.871657e-02 & ~ \\
1/10 & 4.845383e-04 & 4.64 & 5.374590e-04 & 4.64 & 7.612098e-04 & 4.62 \\
1/20 & 1.627974e-05 & 4.90 & 1.802367e-05 & 4.90 & 2.574375e-05 & 4.89 \\
1/40 & 5.583361e-07 & 4.87 & 6.214484e-07 & 4.86 & 8.889290e-07 & 4.86 \\
1/80 & 2.196746e-08 & 4.67 & 2.444747e-08 & 4.67 & 3.492228e-08 & 4.67 \\
\end{tabular*}
{\rule{\temptablewidth}{0.1pt}}
\end{center}
\end{table}
\begin{table}[htp]
\footnotesize
\begin{center}
\vspace{-4mm} \caption{\label{2Daccuracy02}Accuracy test in 2-D for the advection of density perturbation by the ASE-DFF(7,5,3)-CGKS. $\Delta t=0.3\Delta {{x}^{7/4}}.$}
\def\temptablewidth{1.0\textwidth}
{\rule{\temptablewidth}{1pt}}
\begin{tabular*}{\temptablewidth}{@{\extracolsep{\fill}}c|cc|cc|cc}
mesh length & ${{L}^{1}}$error & Order & ${{L}^{2}}$error & Order & ${{L}^{\infty }}$error & Order\\
\hline
1/5 & 2.013854e-03 & ~ & 2.311136e-03 & ~ & 3.128605e-03 & ~ \\
1/10 & 1.856911e-05 & 6.76 & 2.071570e-05 & 6.80 & 2.955261e-05 & 6.73 \\
1/20 & 1.467971e-07 & 6.98 & 1.634051e-07 & 6.99 & 2.381770e-07 & 6.96 \\
1/40 & 1.139811e-09 & 7.01 & 1.263502e-09 & 7.01 & 1.844327e-09 & 7.01 \\
1/80 & 8.794032e-12 & 7.02 & 9.745087e-12 & 7.02 & 1.431633e-11 & 7.01 \\
\end{tabular*}
{\rule{\temptablewidth}{0.1pt}}
\end{center}
\end{table}
\begin{table}[htp]
\footnotesize
\begin{center}
\vspace{-4mm} \caption{\label{2Daccuracy03}Accuracy test in 2-D for the advection of density perturbation by the ASE-DFF(9,7,5,3)-CGKS. $\Delta t=0.3\Delta {{x}^{9/4}}.$}
\def\temptablewidth{1.0\textwidth}
{\rule{\temptablewidth}{1pt}}
\begin{tabular*}{\temptablewidth}{@{\extracolsep{\fill}}c|cc|cc|cc}
mesh length & ${{L}^{1}}$error & Order & ${{L}^{2}}$error & Order & ${{L}^{\infty }}$error & Order\\
\hline
1/5 & 2.122972e-04 & ~ & 2.317399e-04 & ~ & 3.197545e-04 & ~ \\
1/10 & 4.674029e-07 & 8.83 & 5.177266e-07 & 8.81 & 7.303062e-07 & 8.77 \\
1/20 & 1.590262e-09 & 8.20 & 1.761415e-09 & 8.20 & 2.495147e-09 & 8.19 \\
1/40 & 1.668123e-11 & 6.57 & 1.851308e-11 & 6.57 & 2.623513e-11 & 6.57 \\
1/80 & 3.290070e-13 & 5.66 & 3.668069e-13 & 5.66 & 6.803447e-13 & 5.27 \\
\end{tabular*}
{\rule{\temptablewidth}{0.1pt}}
\end{center}
\end{table}	\par
\subsection{Two-dimensional Riemann problems}
\label{sec4.7}
\noindent{\sl{(a) Configuration 2}}\par
\begin{figure}[htp]	
\centering
\includegraphics[width=0.4\textwidth]{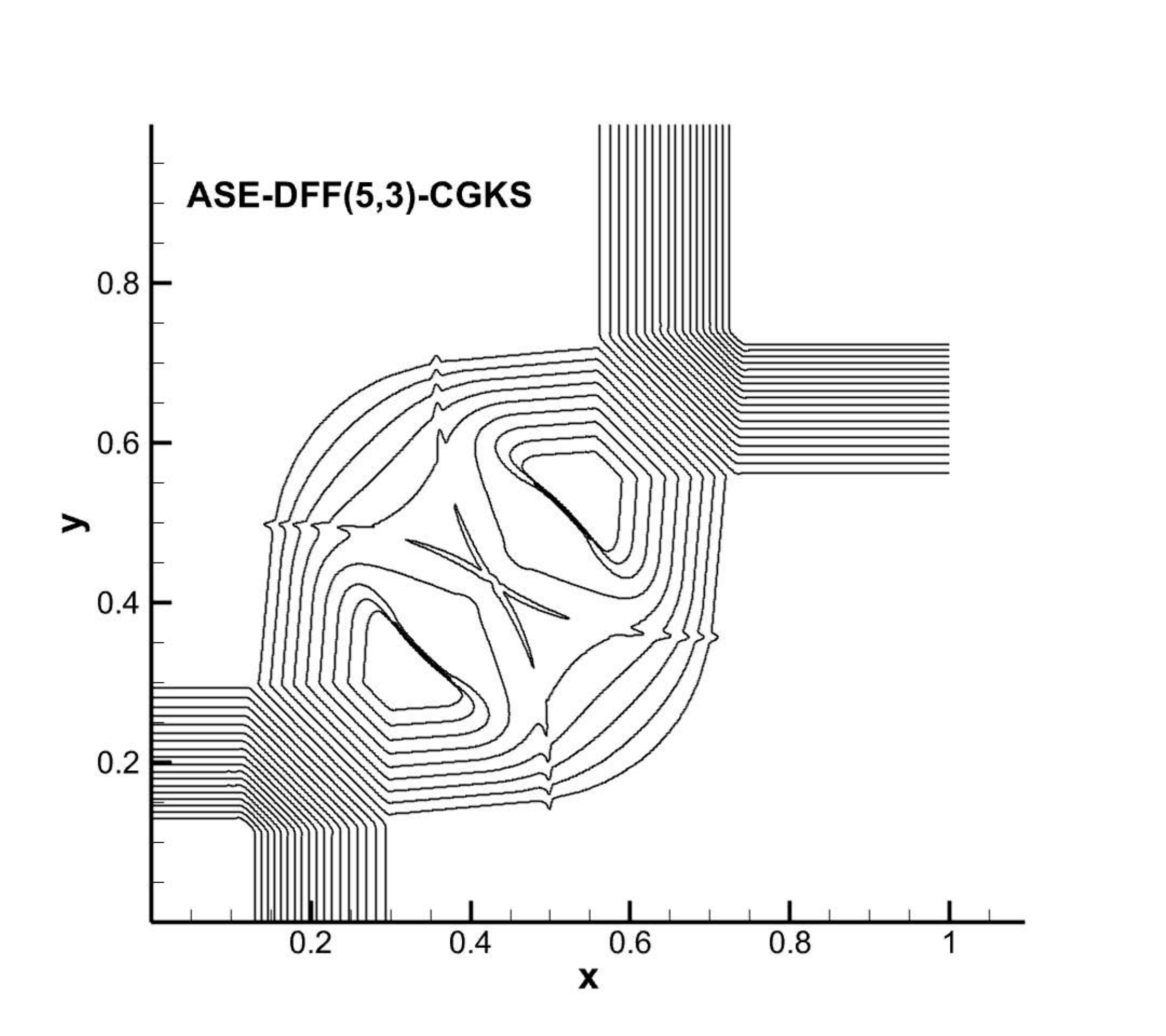}
\includegraphics[width=0.4\textwidth]{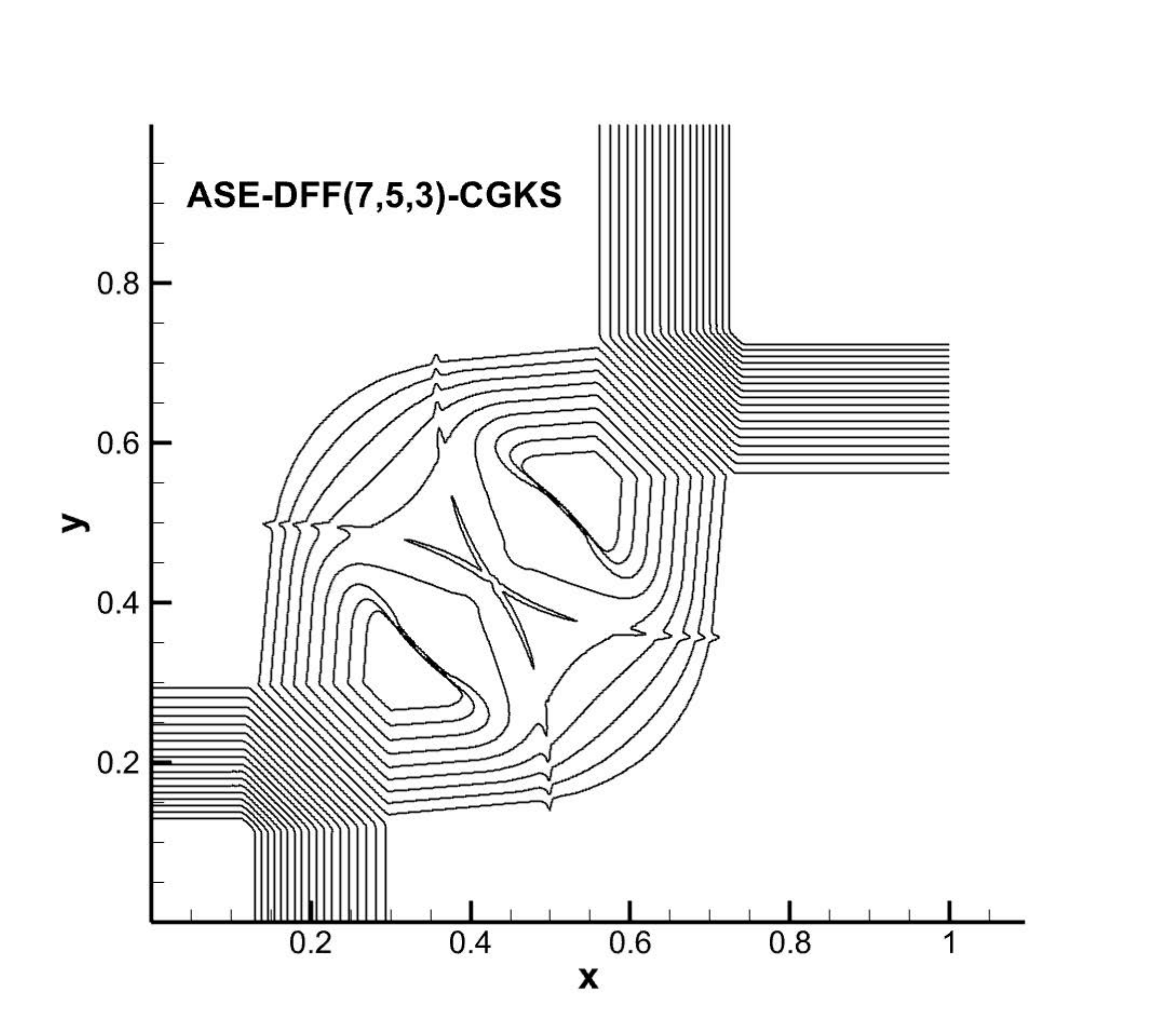}
\includegraphics[width=0.4\textwidth]{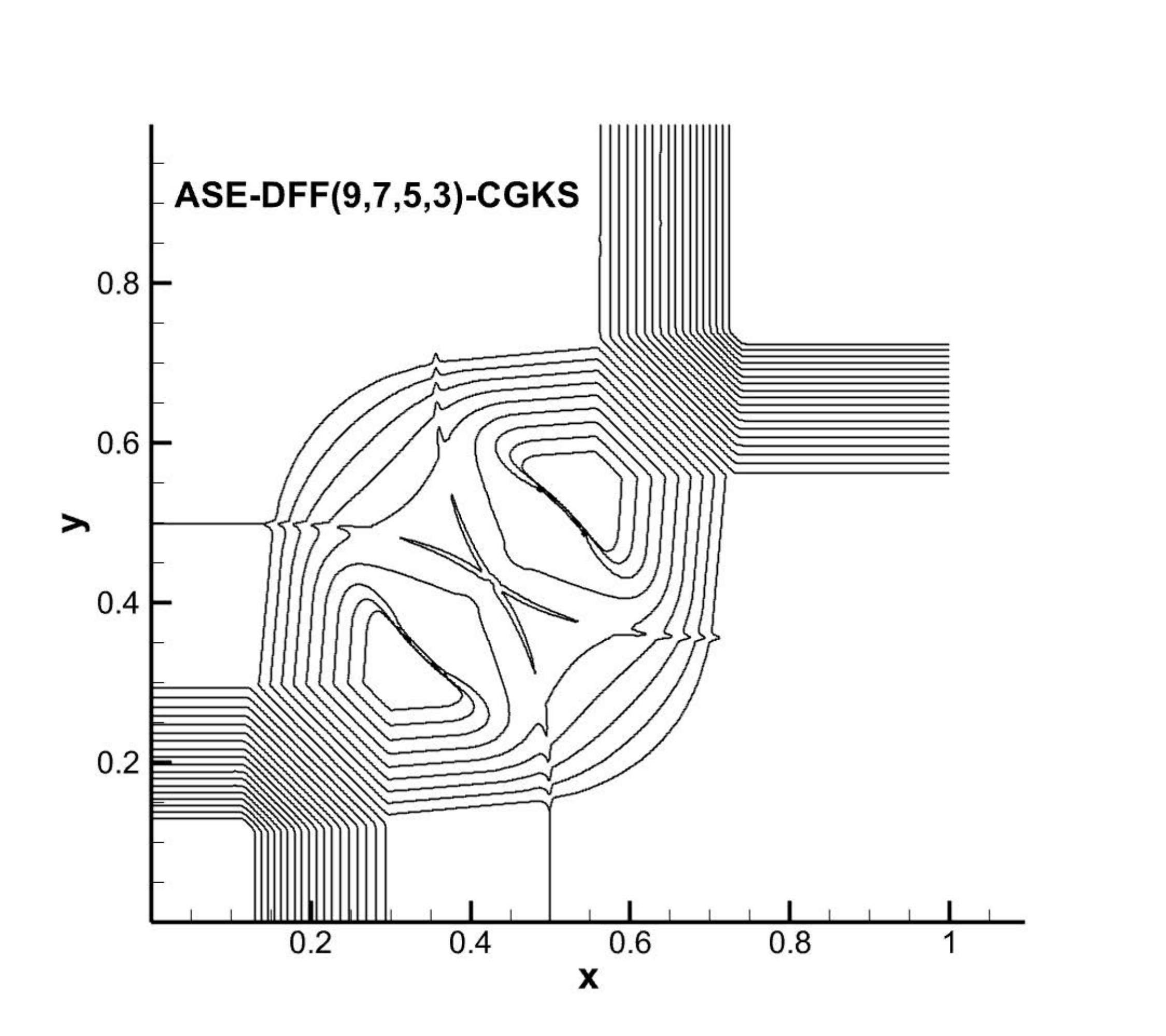}
\vspace{-4mm} \caption{\label{figliman02}
Two-dimensional Riemann problems: the density distributions for
Configuration 2. $CFL=0.5.$ $T=0.2.$ Mesh:$500\times 500.$}
\end{figure}

The Configuration 2 in ~\cite{doi:10.1137/S1064827595291819} is tested. For the case of Configuration 2
~\cite{doi:10.1137/S1064827595291819}, the initial conditions with four
rarefaction waves are given as
\begin{equation*}
\left( \rho ,u,v,p \right)=\left\{ \begin{aligned}
& \left( 1.0,-0.7259,-0.7259,1.0 \right),x<0.5,y<0.5, \\
& \left( 0.5197, 0.0,-0.7259, 0.4 \right),x\ge 0.5,y<0.5, \\
& \left( 1.0, 0.0, 0.0, 1.0 \right),x\ge 0.5,y\ge 0.5, \\
& \left( 0.5197, -0.7259, 0.0, 0.4 \right),x<0.5,y\ge 0.5. \\
\end{aligned} \right.
\end{equation*}
The mesh is $500\times 500.$ in the test. Although the discontinuity is weak,
The result schemes of the new ASE-DFF(5,3)-CGKS,
ASE-DFF(7,5,3)-CGKS and ASE-DFF(9,7,5,3)-CGKS at $t=0.2$ are shown in Fig.~\ref{figliman02}. Since the discontinuity
feedback factor is particularly good at solving rarefaction wave. The new ASE-DFF(5,3)-CGKS,
ASE-DFF(7,5,3)-CGKS and ASE-DFF(9,7,5,3)-CGKS have good robustness.\\
\noindent{\sl{(b) Configuration 3}}\par
\begin{figure}[htp]	
\centering
\includegraphics[width=0.4\textwidth]{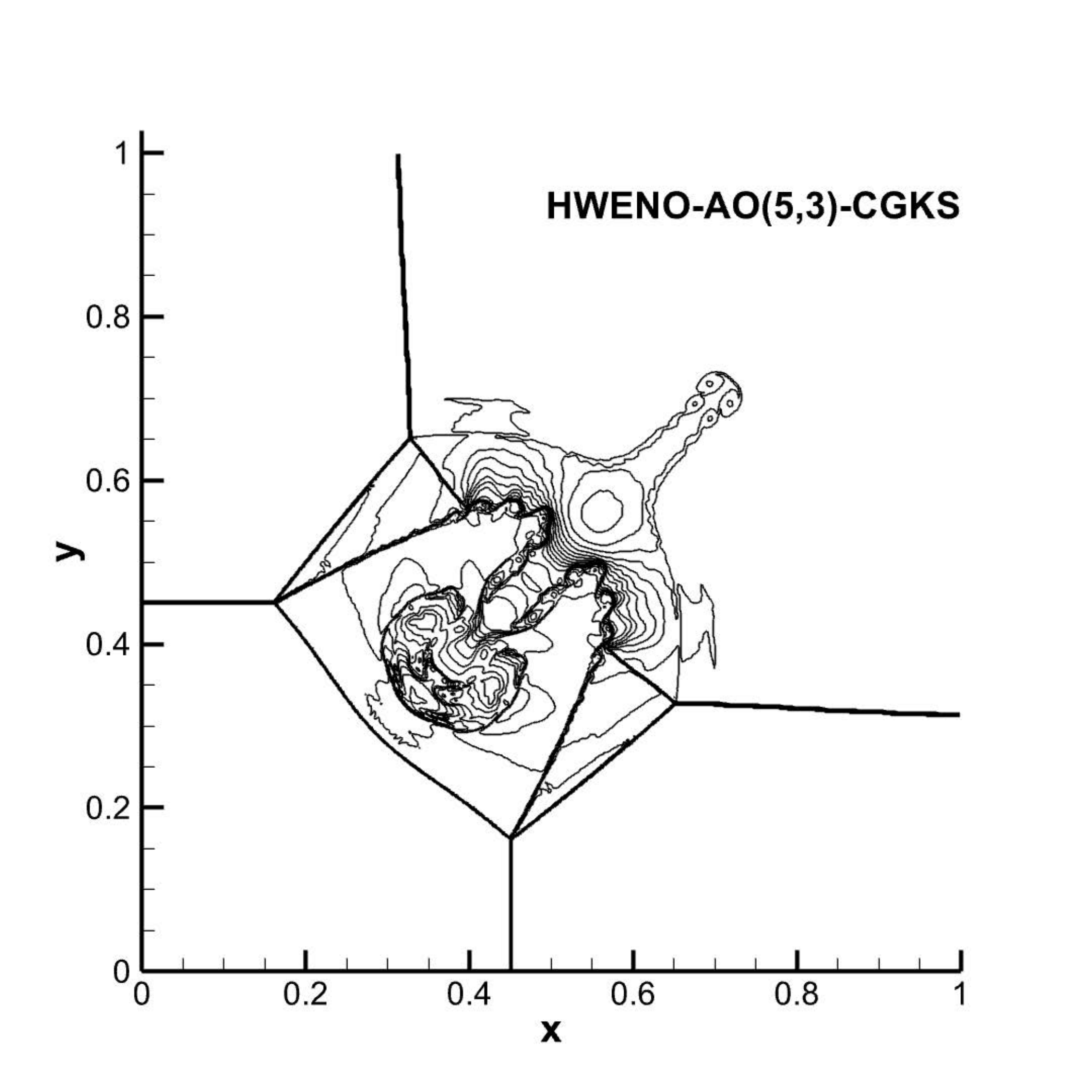}
\includegraphics[width=0.4\textwidth]{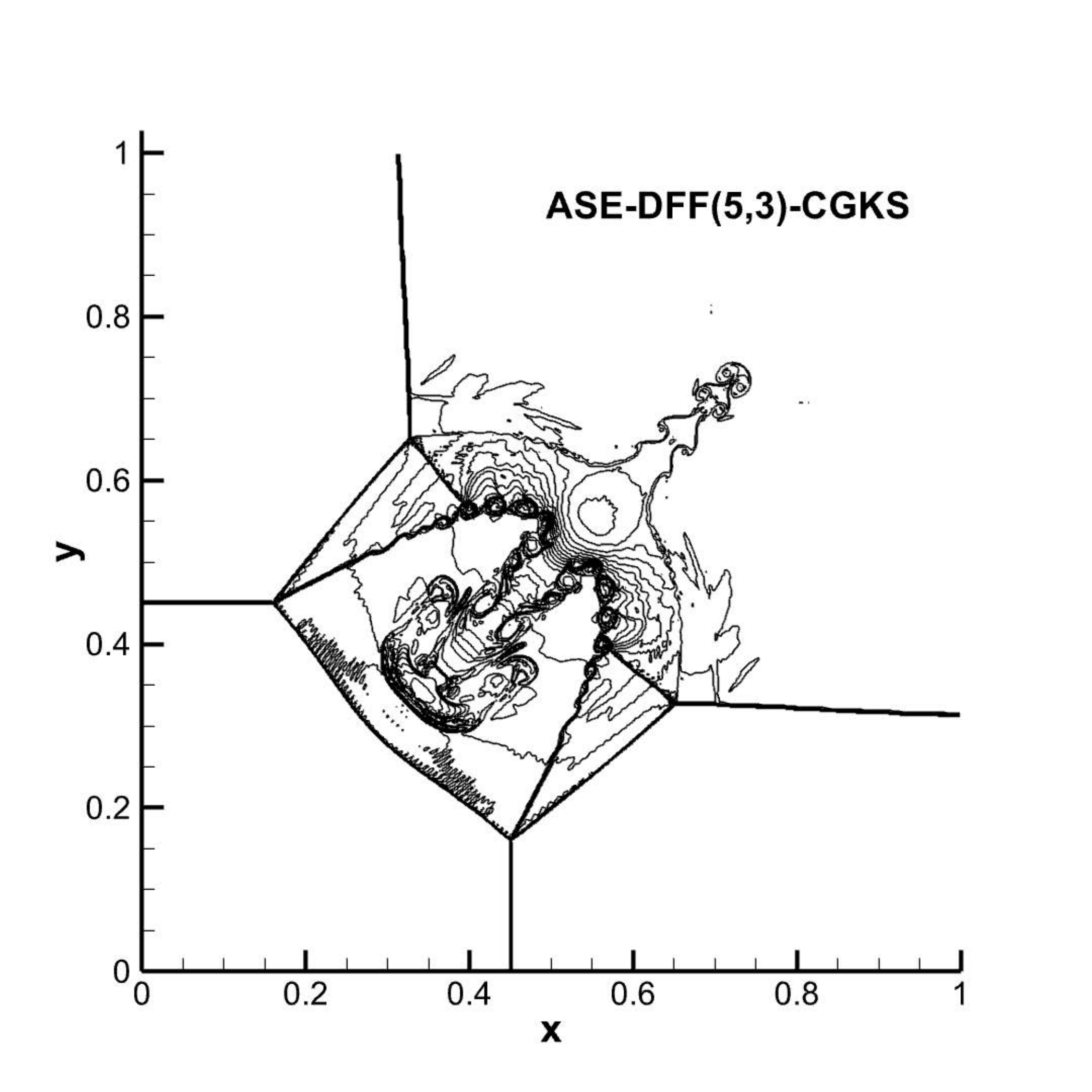}
\includegraphics[width=0.4\textwidth]{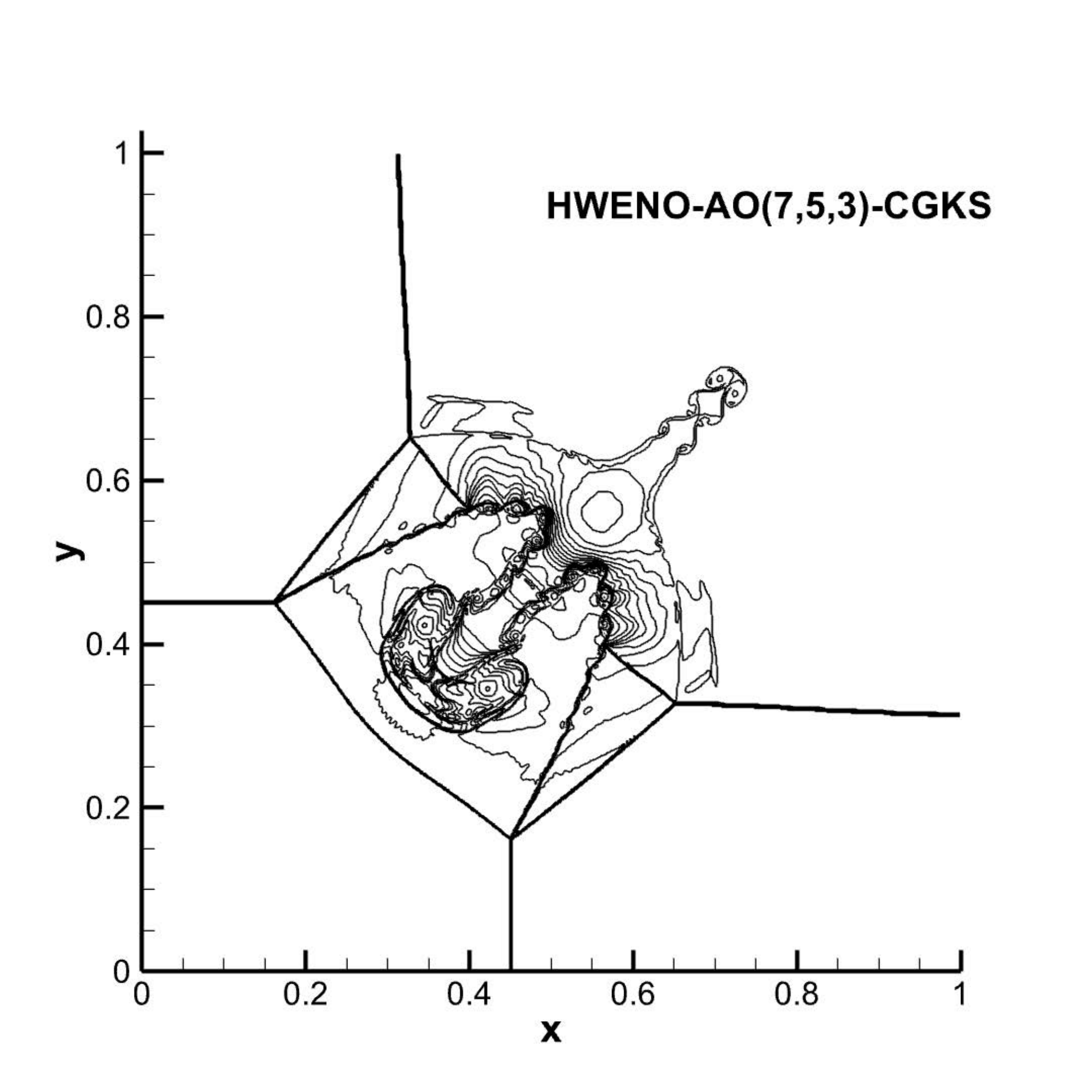}
\includegraphics[width=0.4\textwidth]{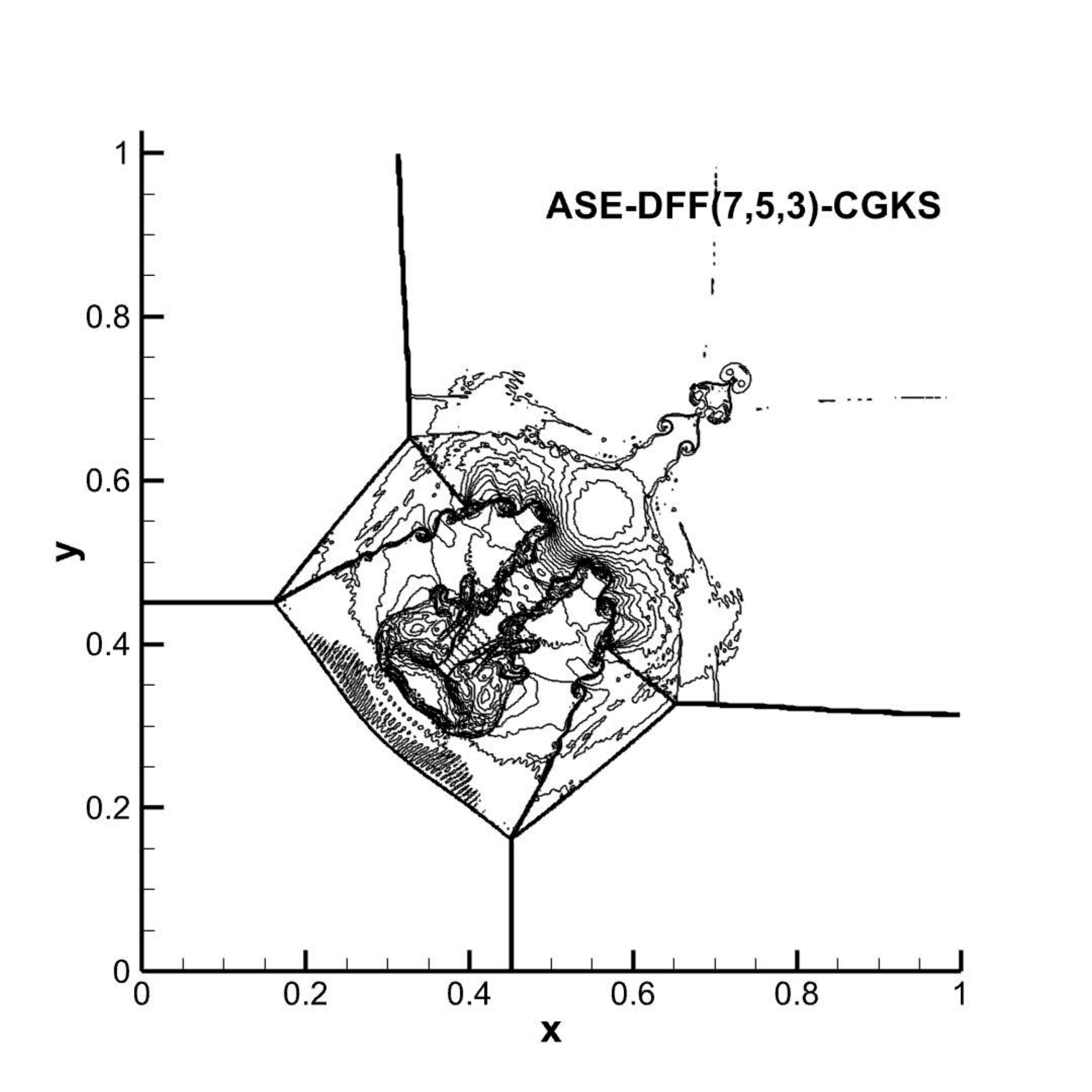}
\includegraphics[width=0.4\textwidth]{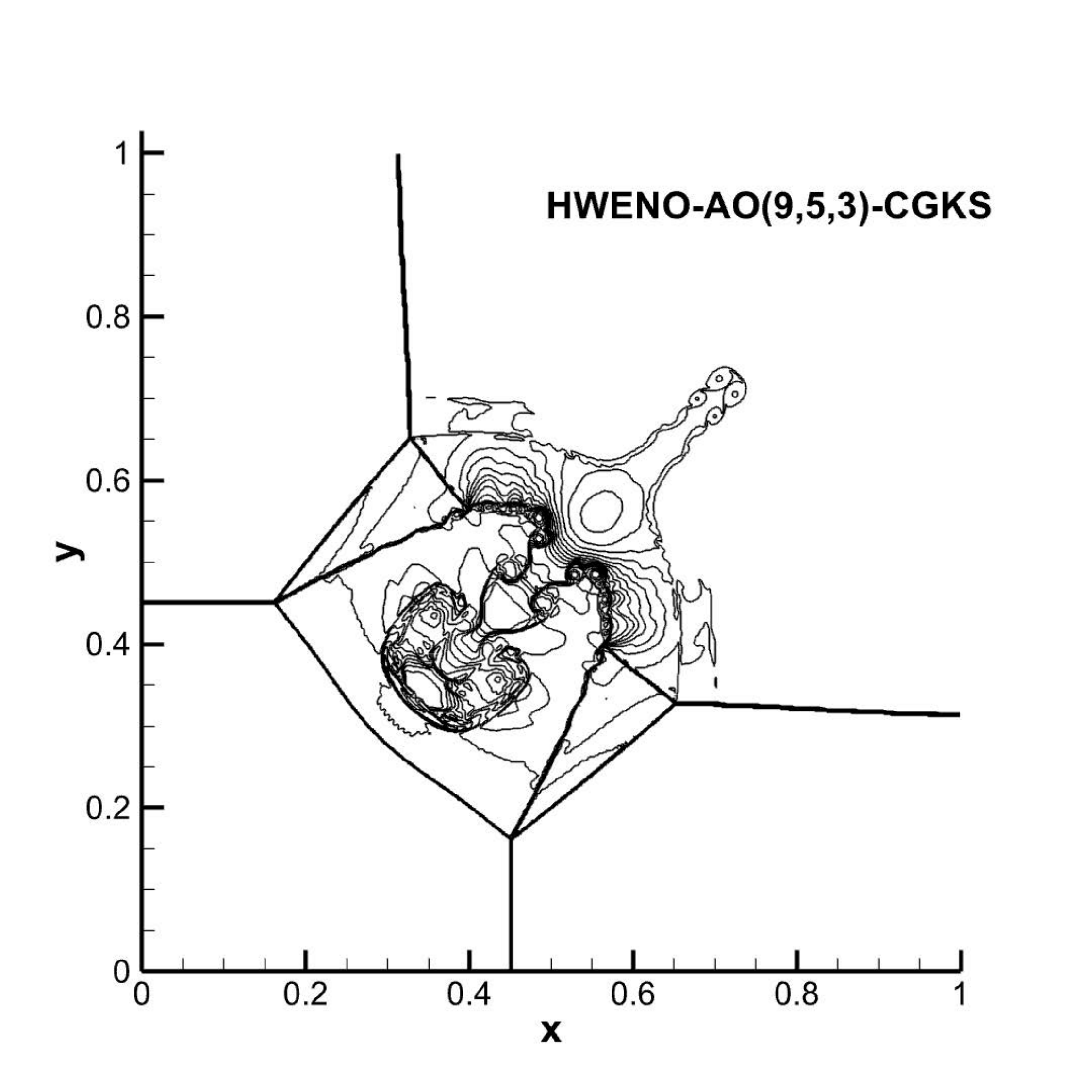}
\includegraphics[width=0.4\textwidth]{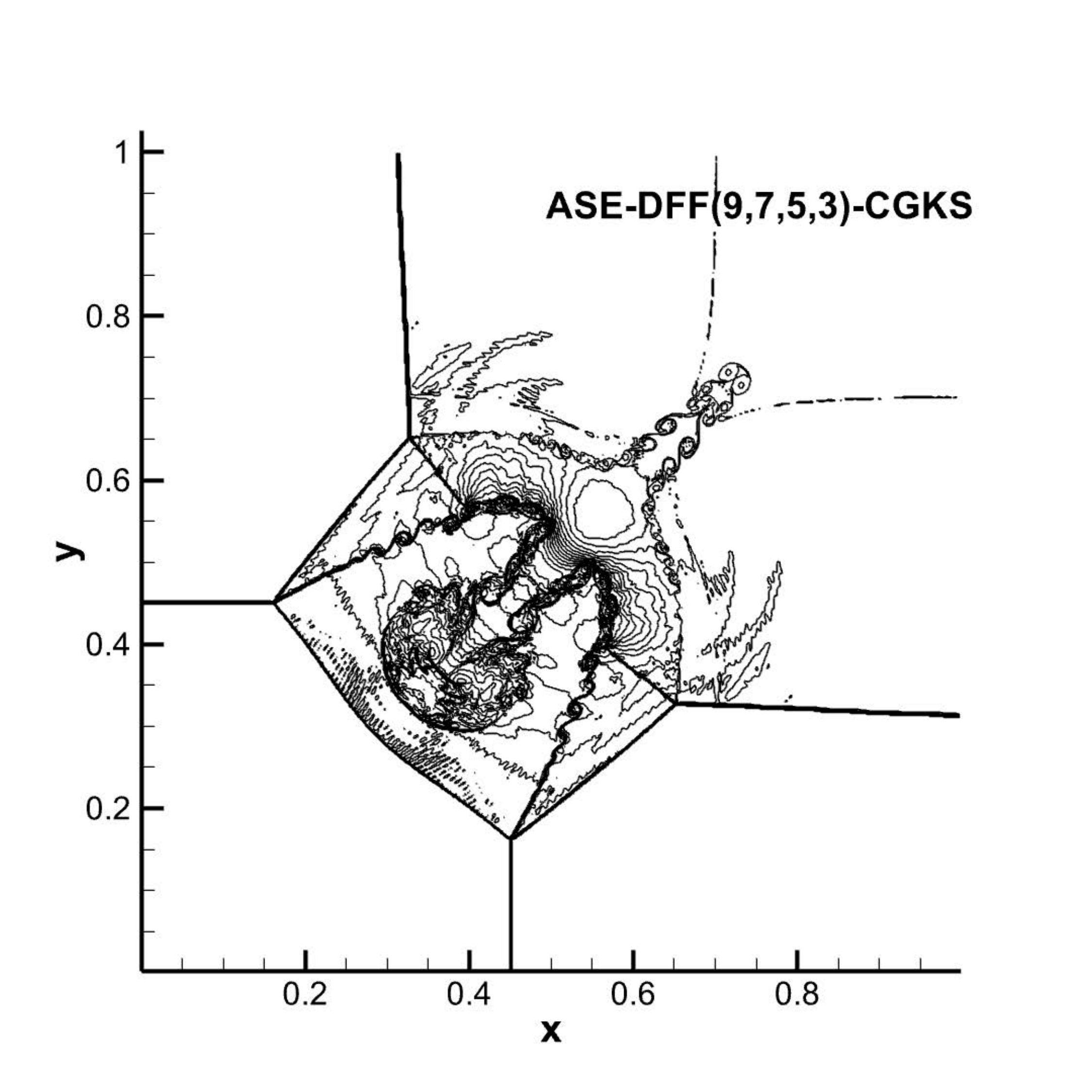}
\vspace{-4mm} \caption{\label{figliman03}
Two-dimensional Riemann problems: the density distributions for
Configuration 3. $CFL=0.5.$ $T=0.6.$ Mesh:$500\times 500.$}
\end{figure}

The initial conditions of the Configuration 3 are the shock-shock interaction and shock-vortex interaction
and given as~\cite{doi:10.1137/S1064827595291819}.
\begin{equation*}
\left( \rho ,u,v,p \right)=\left\{ \begin{aligned}
& \left( 0.138,1.206,1.206,0.029 \right),x<0.7,y<0.7, \\
& \left( 0.5323,0,1.206,0.3 \right),x\ge 0.7,y<0.7, \\
& \left( 1.5,0,0,1.5 \right),x\ge 0.7,y\ge 0.7, \\
& \left( 0.5323,1.206,0,0.3 \right),x<0.7,y\ge 0.7. \\
\end{aligned} \right.
\end{equation*}
The mesh is $500\times 500.$ The results of the Configuration 3 at $t=0.6$ for the ASE-DFF(5,3)-CGKS,
ASE-DFF(7,5,3)-CGKS, and ASE-DFF(9,7,5,3)-CGKS
are presented in Fig.~\ref{figliman03}.The proof shows that the adaptive high-order CGKS based on
discontinuity feedback factor demonstrates extremely high resolution. Furthermore, the ASE-DFF-CGKS has better resolution than the HWENO-CGKS.
The ASE-DFF-CGKS have revealed numerous small-scale structures. As the order increases, the performance continuously improves.\par

\subsection{Double Mach reflection problem}
\label{sec4.8}
Designed by Woodward and Colella~\cite{WOODWARD1984115} the inviscid double Mach reflection problem is chosen for
testing the robustness of high-order methods.
The computational domain is $\left[ 0,\ 4 \right]\times \left[ 0,\ 1 \right]$ with a slip boundary
condition applied on the bottom of the domain starting from $x=1/6$. The mesh is $960\times 240.$The simulation end time
is t = 0.2 The initial conditions are
\begin{equation*}
\left( \rho ,u,v,p \right)=\left\{ \begin{aligned}
& \left( 1.4,0,0,1 \right),\text{ }if\text{ }y<1.732\left( x-0.1667 \right), \\
& \left( 8,4.125\sqrt{3},-4.125,116.5 \right),\text{ otherwise}\text{.} \\
\end{aligned} \right.
\end{equation*}
Initially, a right-moving Mach 10 shock wave is positioned at \( x = 0.1667 \) with an incident
angle of \( 60^\circ \) to the \( x \)-axis. The post-shock conditions are imposed from \( x = 0 \)
to \( x = 0.1667 \), while a reflecting wall condition is enforced from
\( x = 0.1667 \) to \( x = 4 \) along the bottom boundary. The post-shock conditions are set
for the remainder of the bottom boundary. At the top boundary, the flow variables are specified
to describe the exact motion of the Mach 10 shock.\par
\begin{figure}[htp]	
\centering
\includegraphics[width=0.4\textwidth]{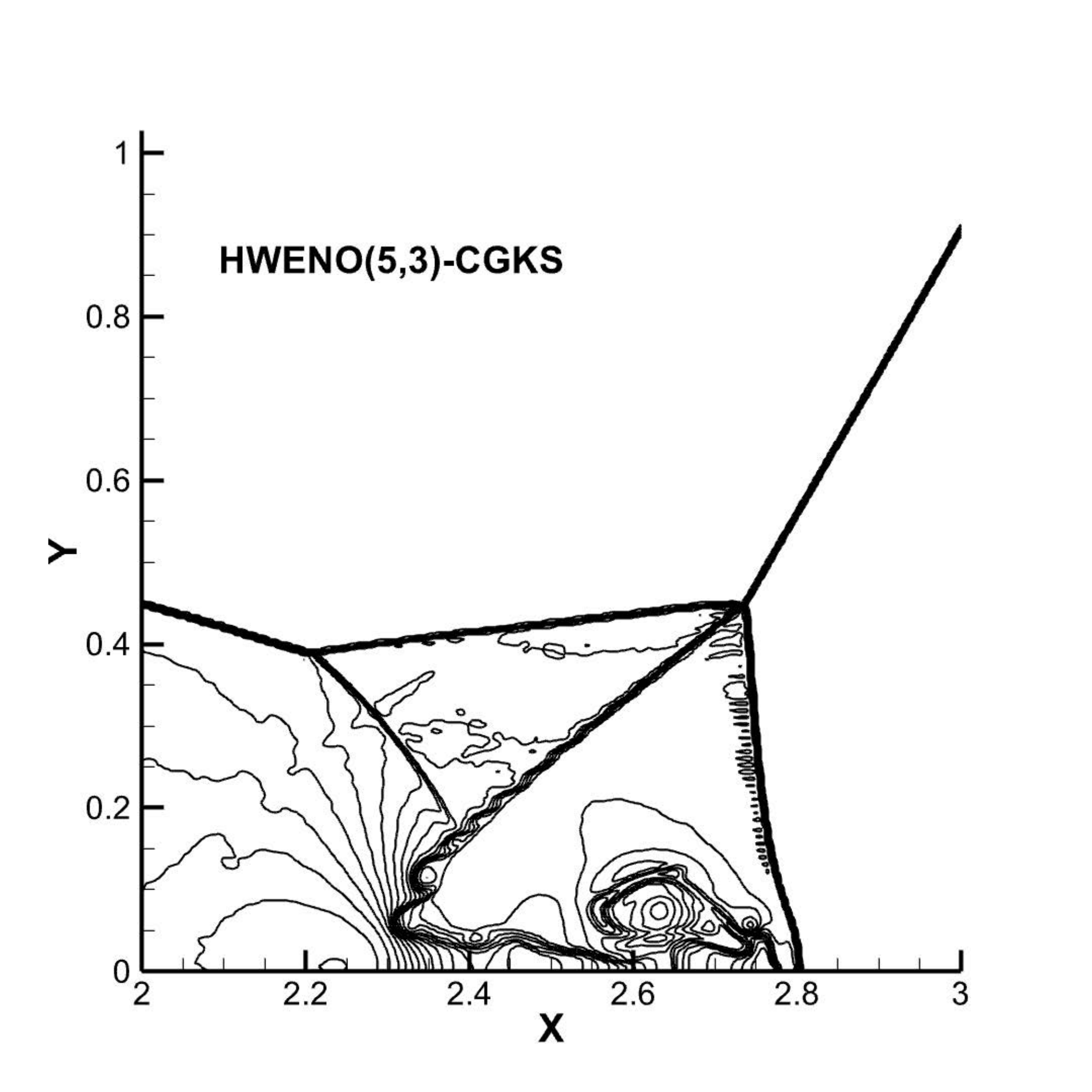}
\includegraphics[width=0.4\textwidth]{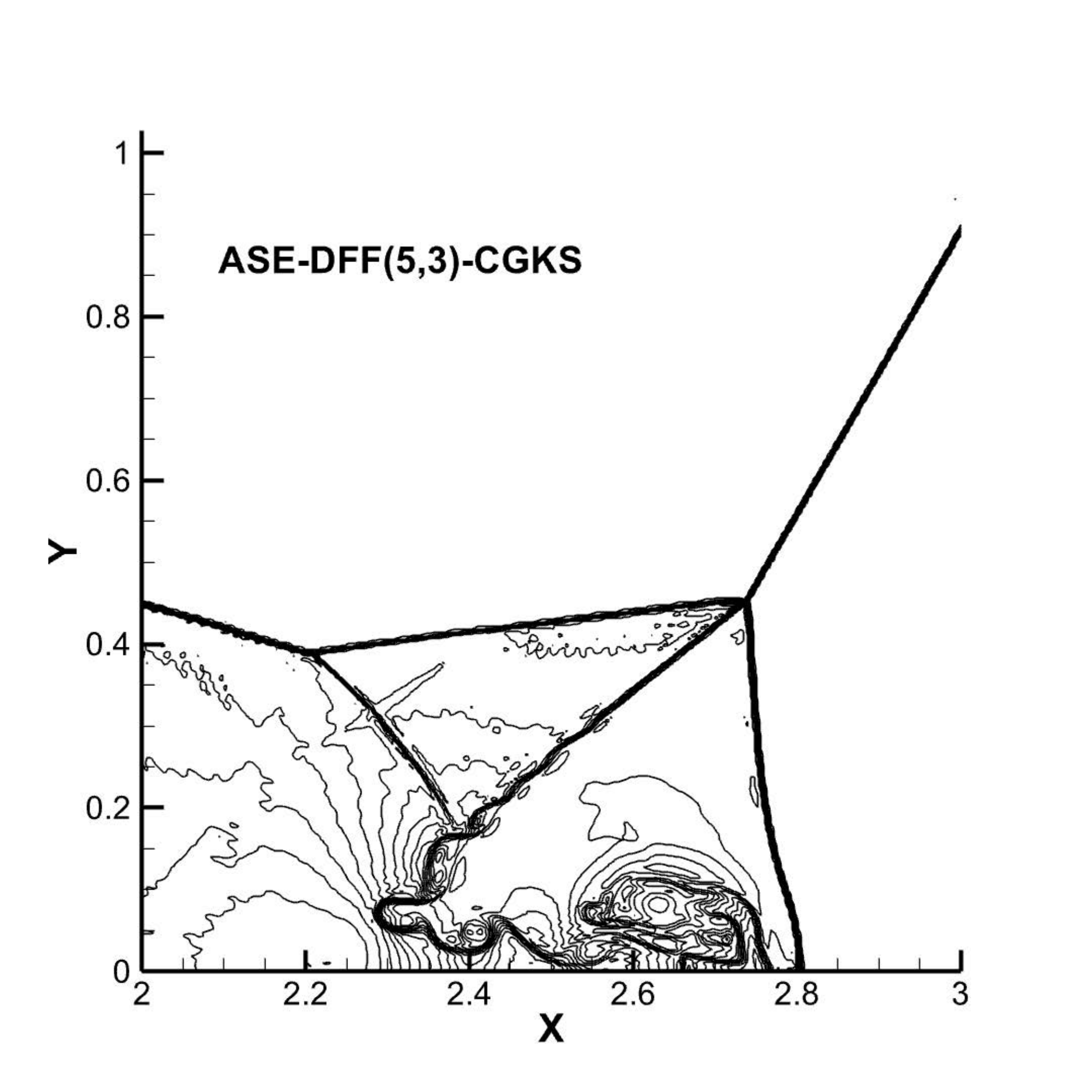}
\includegraphics[width=0.4\textwidth]{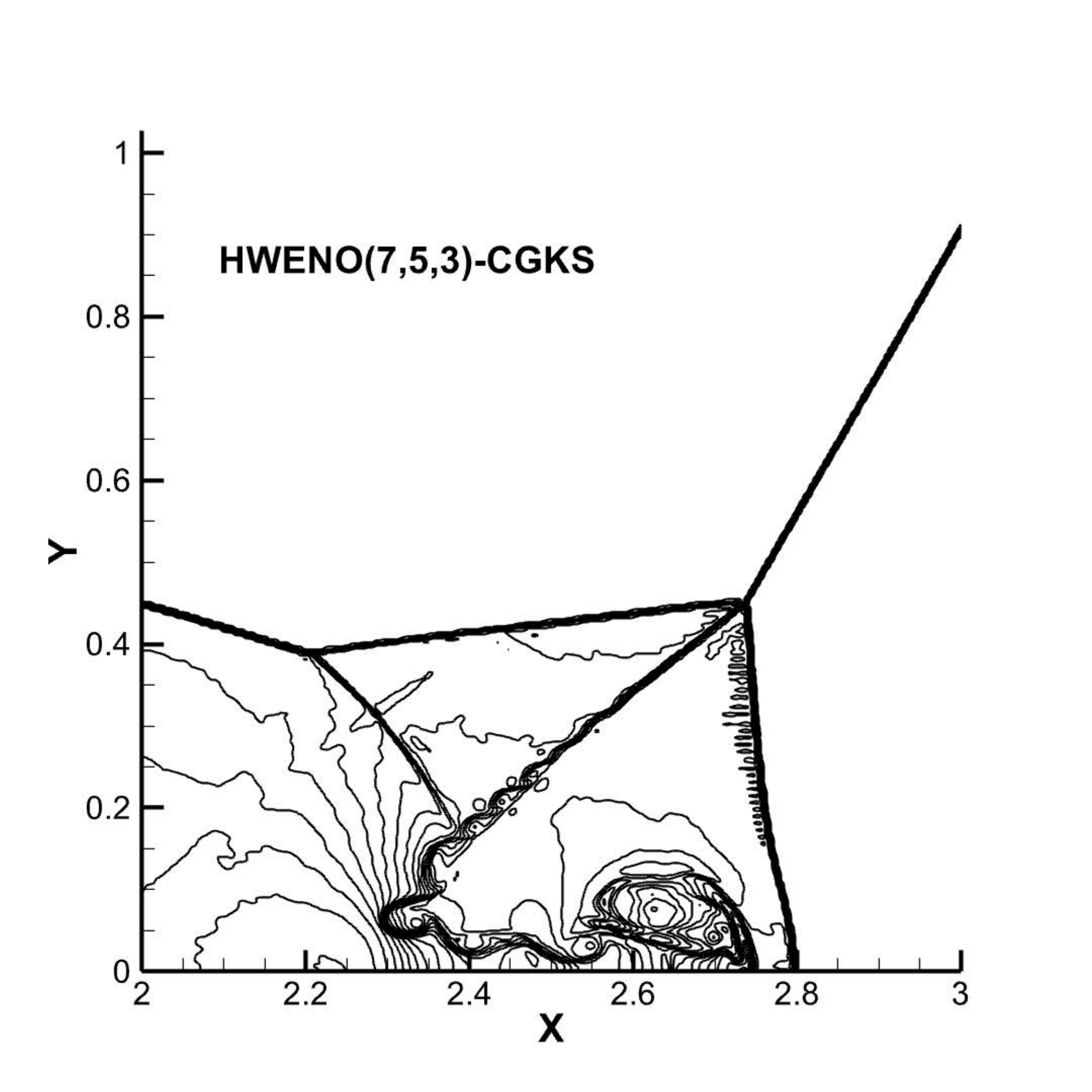}
\includegraphics[width=0.4\textwidth]{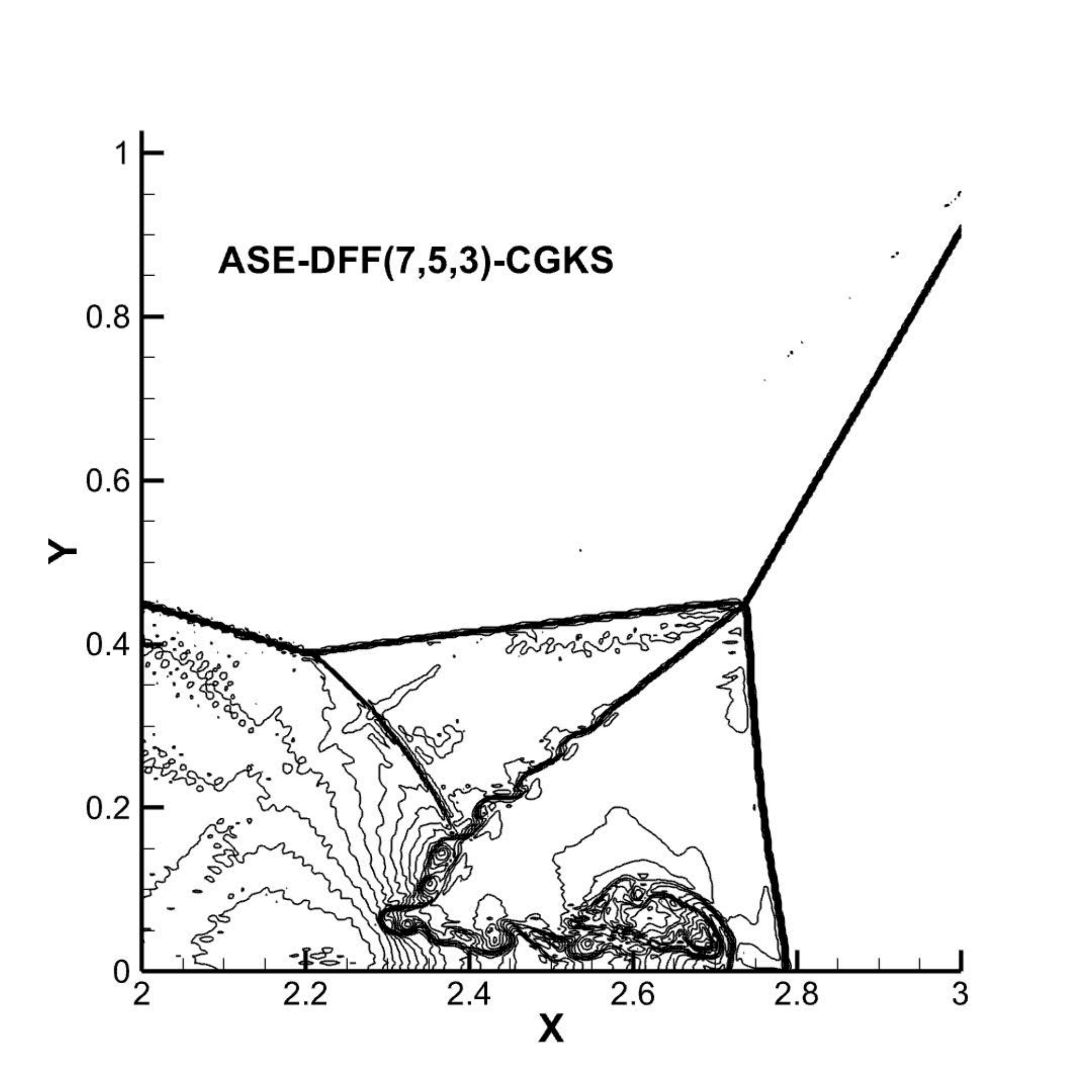}
\includegraphics[width=0.4\textwidth]{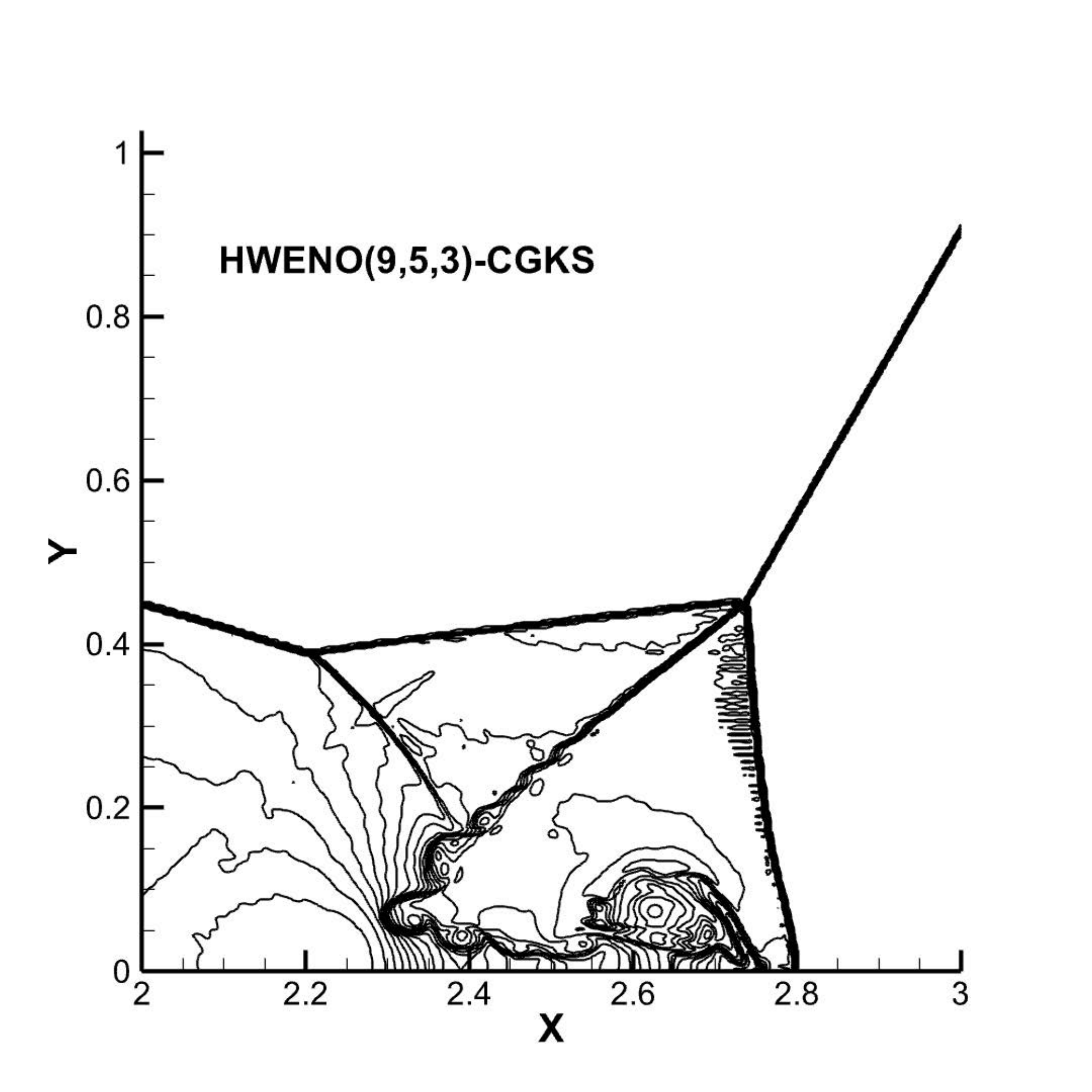}
\includegraphics[width=0.4\textwidth]{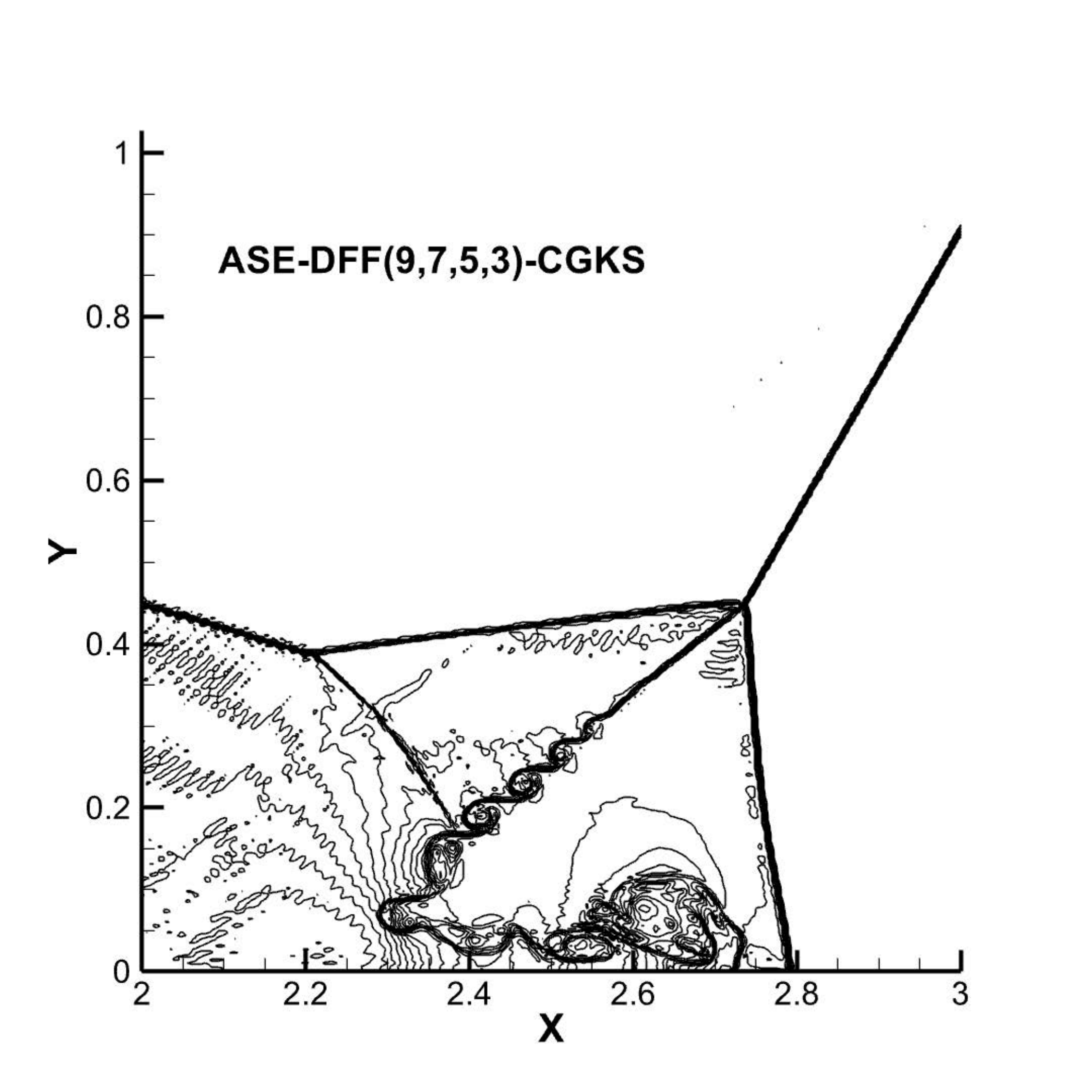}
\vspace{-4mm} \caption{\label{figdouble}
The density distributions for Double Mach reflection problem. Mesh: $960\times 240.$ $CFL=0.5.$ $T=0.2.$}
\end{figure}

The density distributions are shown in Fig.~\ref{figdouble}. In the double Mach reflection problem,
it is proved that all three compact CGKS schemes have excellent robustness and shock-capturing
capability. At the same time, ASE-DFF(5,3)-CGKS, ASE-DFF(7,5,3)-CGKS and ASE-DFF(9,7,5,3)-CGKS all
have lower dissipation and better resolution than the HWENO-CGKS. They can capture a larger number of small scale flow structures. However,
as higher-order compact linear stencils are added to the reconstruction, some numerical oscillations
can also be observed to increase.

\subsection{Viscous shock tubes problem}
\label{sec4.9}
This viscous shock tube problem~\cite{DARU2009664,KIM2005527} was investigated to valid the capability of
scheme for low Reynolds number viscous flow with the $Ma=2.37$ strong shock. In a two-dimensional
unit box $\left[ 0,\ 1 \right]\times \left[ 0,\ 0.5 \right]$, a membrane located at $x=0.5$ separates two
different states of the gas and the dimensionless initial states are
\begin{equation*}
\left( \rho ,u,p \right)=\left\{ \begin{aligned}
& \left( 120,0,120/\gamma  \right),0<x<0.5, \\
& \left( 1.2,0,1.2/\gamma  \right),0.5\le x<1, \\
\end{aligned} \right.
\end{equation*}
where $\gamma =1.4$, Prandtl number $\Pr =1.0$ and Reynolds number $\operatorname{Re} =200$. The
symmetric boundary condition is imposed on the top $x\in \left[ 0,\ 1 \right]$, $y=0.5$ and non-slip
adiabatic conditions applied on the other three solid wall boundaries. The mesh is $500\times 250.$The output time
is $t=1.0$.
The drastic change in velocity above the bottom wall of this problem introduced intense shear stress,
eventually leading to a complex two-dimensional shock/shear/boundary layer interaction flow. \par
\begin{figure}[htp]	
\centering
\includegraphics[width=0.4\textwidth]{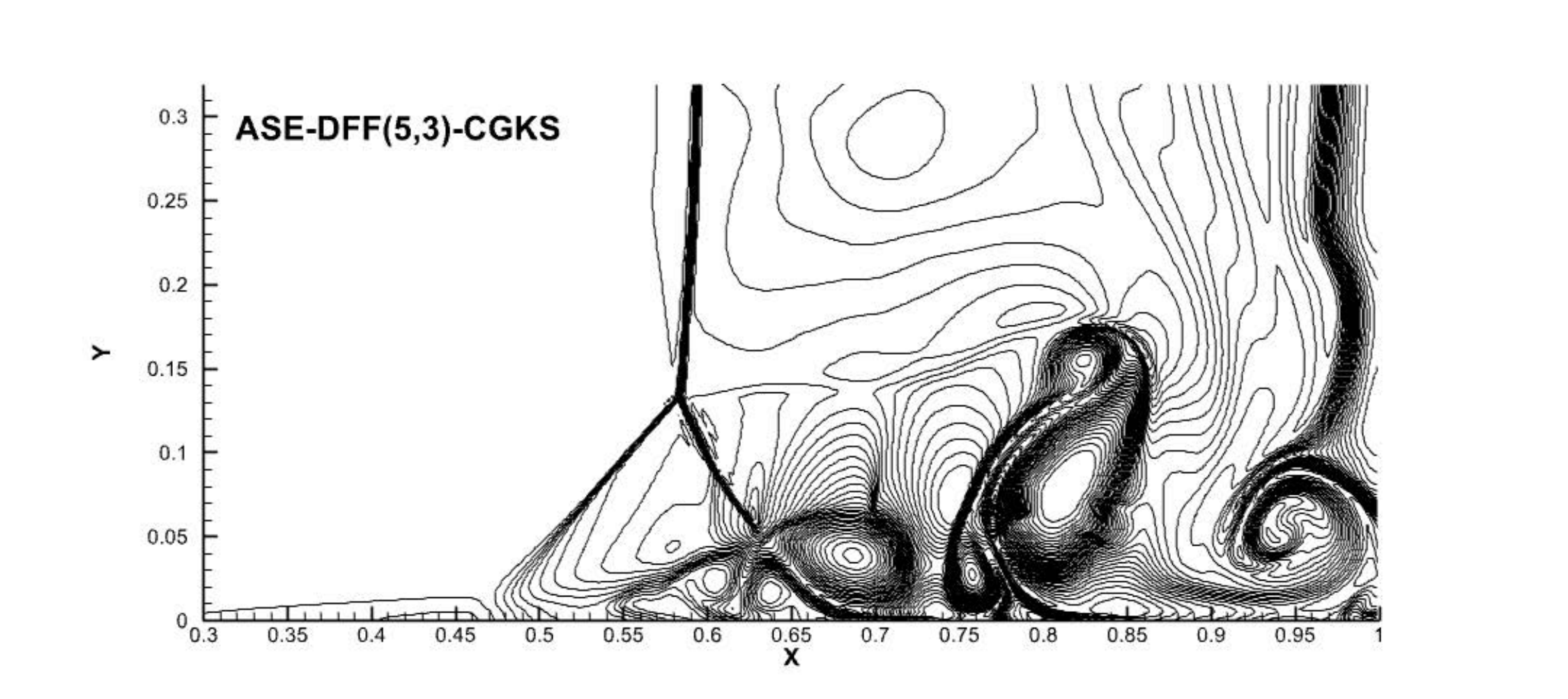}
\includegraphics[width=0.4\textwidth]{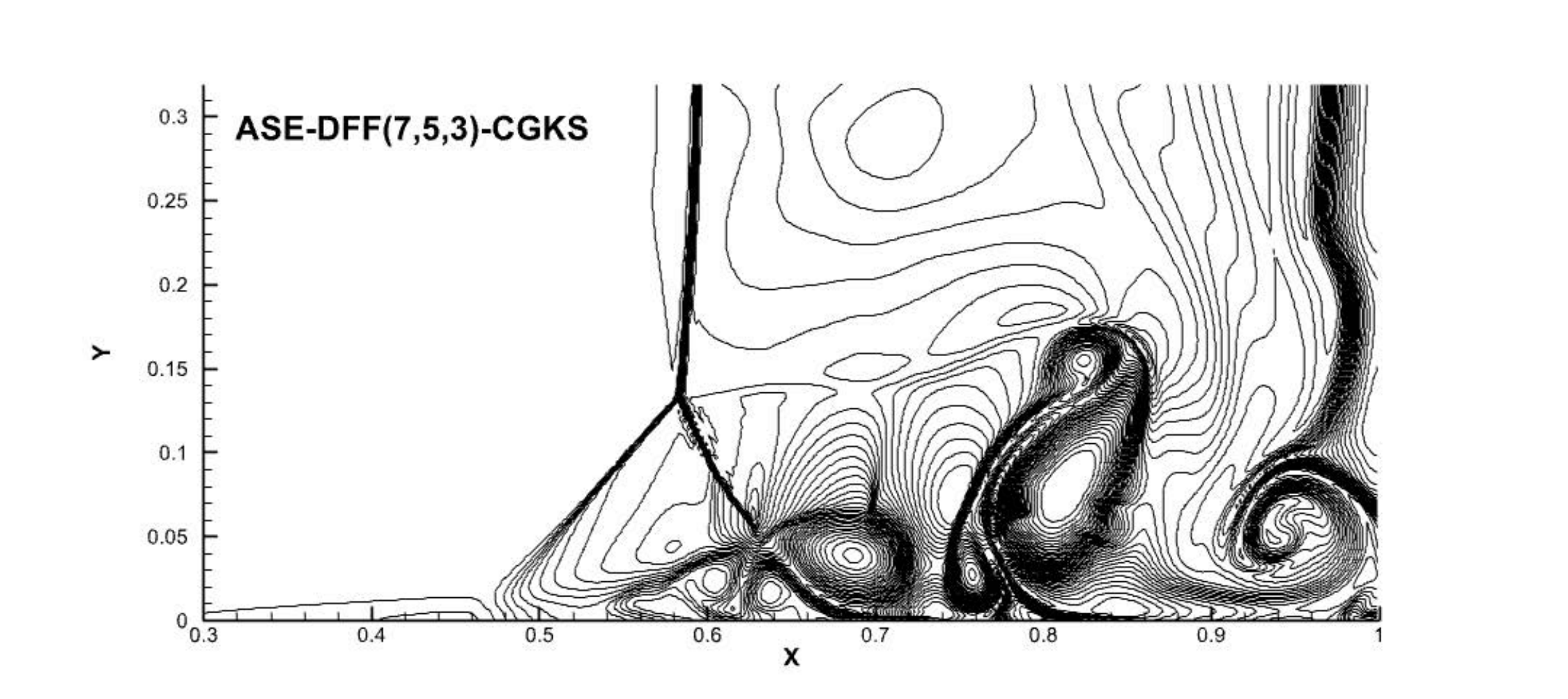}
\includegraphics[width=0.4\textwidth]{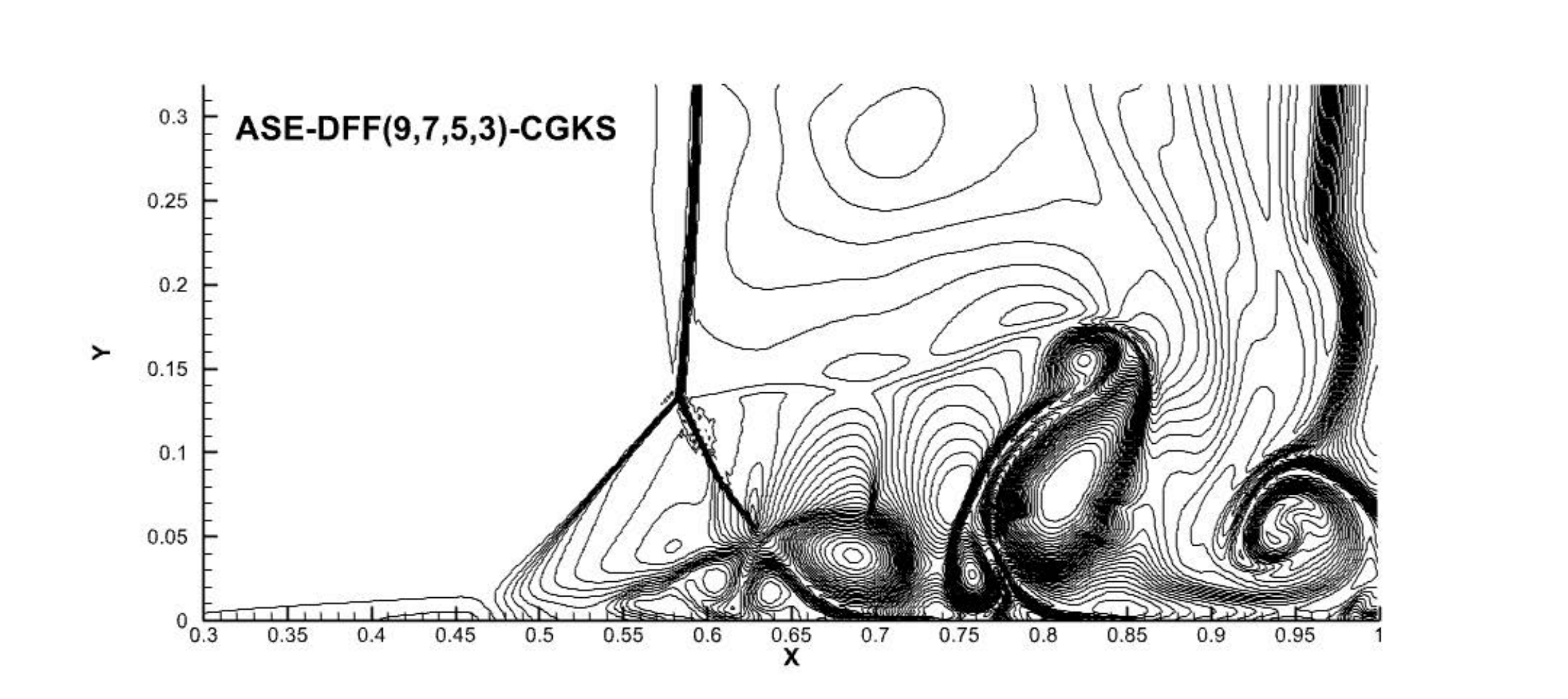}
\vspace{-4mm} \caption{\label{figviscous200}
Viscous shock tube problem with $Re =200$ by ASE-DFF(5,3)-CGKS, ASE-DFF(7,5,3)-CGKS and
ASE-DFF(9,7,5,3)-CGKS scheme: density distribution. For all cases, the CFL number is 0.2. The mesh
number is $500\times250$; and for the bottom two figures, the mesh number is $1000 \times500$.}
\end{figure}
\begin{figure}[htp]	
\centering
\includegraphics[width=1.0\textwidth]{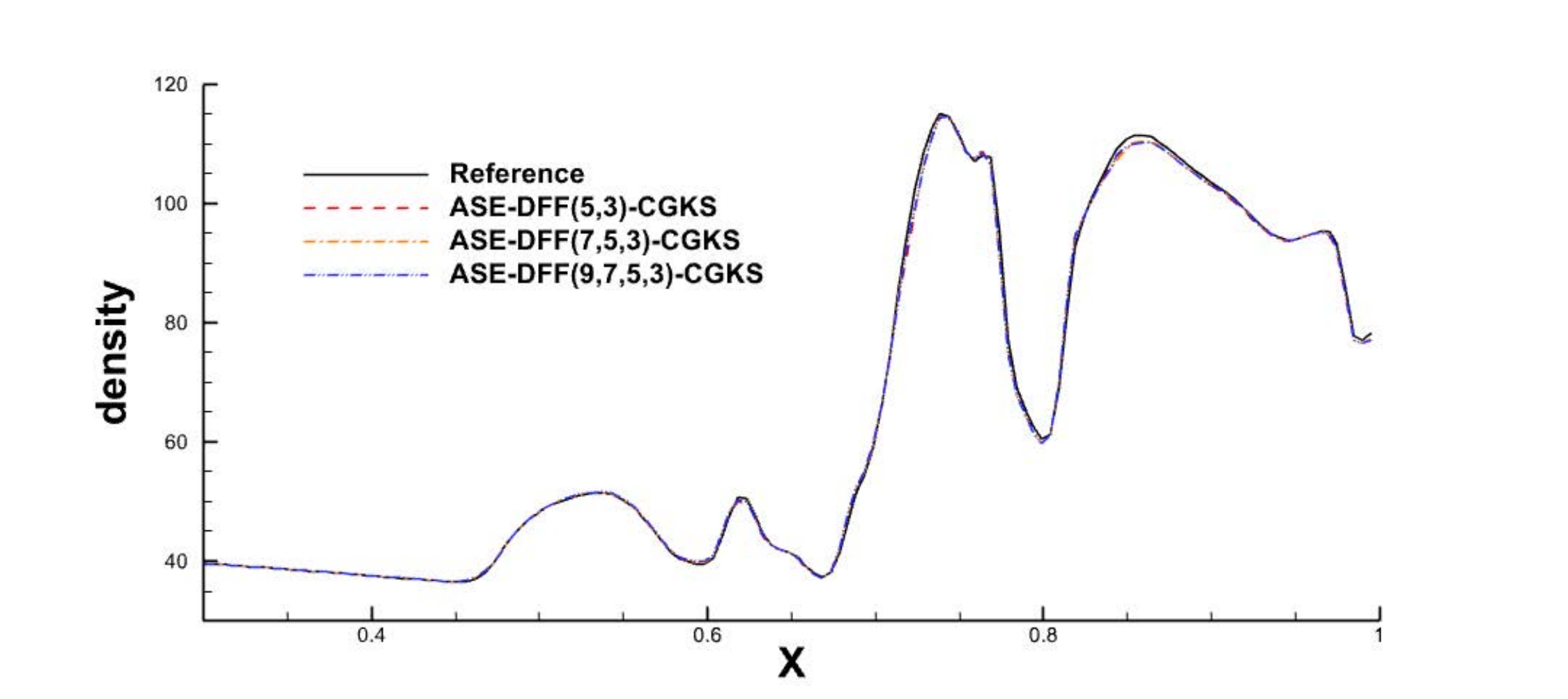}
\vspace{-4mm} \caption{\label{figviscous200rho}
Viscous shock tube problem of $Re =200$ by ASE-DFF(5,3)-CGKS, ASE-DFF(7,5,3)-CGKS and
ASE-DFF(9,7,5,3)-CGKS scheme: density profile along the bottom wall $(y =0)$. For all cases, the CFL
number is 0.2.}
\end{figure}

For the viscous shock tube, The density distributions for the ASE-DFF(5,3)-CGKS, ASE-DFF(7,5,3)-CGKS and
ASE-DFF(9,7,5,3)-CGKS
are plotted in Fig.~\ref{figviscous200}. The density profiles along the bottom wall are shown in
Fig.~\ref{figviscous200rho}. All three CGKS schemes exhibit excellent robustness in this problem involving
the interaction of viscous shock waves. The density distributions at the bottom of all three schemes also
compare well with the reference solution.

\subsection{High Mach number astrophysical jet}
\label{sec4.10}
The high Mach number astrophysical jet~\cite{ZHANG20108918} has two cases.
The initial conditions are as follows:
\begin{equation*}
(\rho, u, v, p, \gamma) = (0.5, 0, 0, 0.4127, \frac{5}{3}).
\end{equation*}
For the first case where \( Ma = 80 \), the computational domain
is \(\left[ 0,\ 2 \right]\times \left[ 0,\ 1 \right]\).
For the boundary conditions, if the y-coordinate of the left boundary is within
the range of \(\left[ 0.45,\ 0.55 \right]\),
then \((\rho, u, v, p) = (5, 30, 0, 0.4127)\). Otherwise, it is \((\rho, u, v, p) = (0.5, 0, 0, 0.4127)\).
The grid resolution is \(448 \times 224\), and the calculation time is \(t = 0.07\). For the second case
where
\( Ma = 2000 \), the calculation domain is \(\left[ 0,\ 1 \right]\times \left[ 0,\ 0.5 \right]\).
For the boundary conditions, if the y-coordinate on the left boundary is within the range \(\left[ 0.2,\ 0.3 \right]\),
then \((\rho, u, v, p) = (5, 800, 0, 0.4127)\). Otherwise, it is \((\rho, u, v, p) = (0.5, 0, 0, 0.4127)\). The
grid resolution is \(800 \times 400\). And the calculation time is \(t = 1.0 \times 10^{-3}\). For both of these cases,
outflow conditions are applied to the right, top, and bottom boundaries, with the specific
heats \(\gamma = \frac{5}{3}\).\par
\begin{figure}[htp]	
\centering

\includegraphics[width=0.4\textwidth]{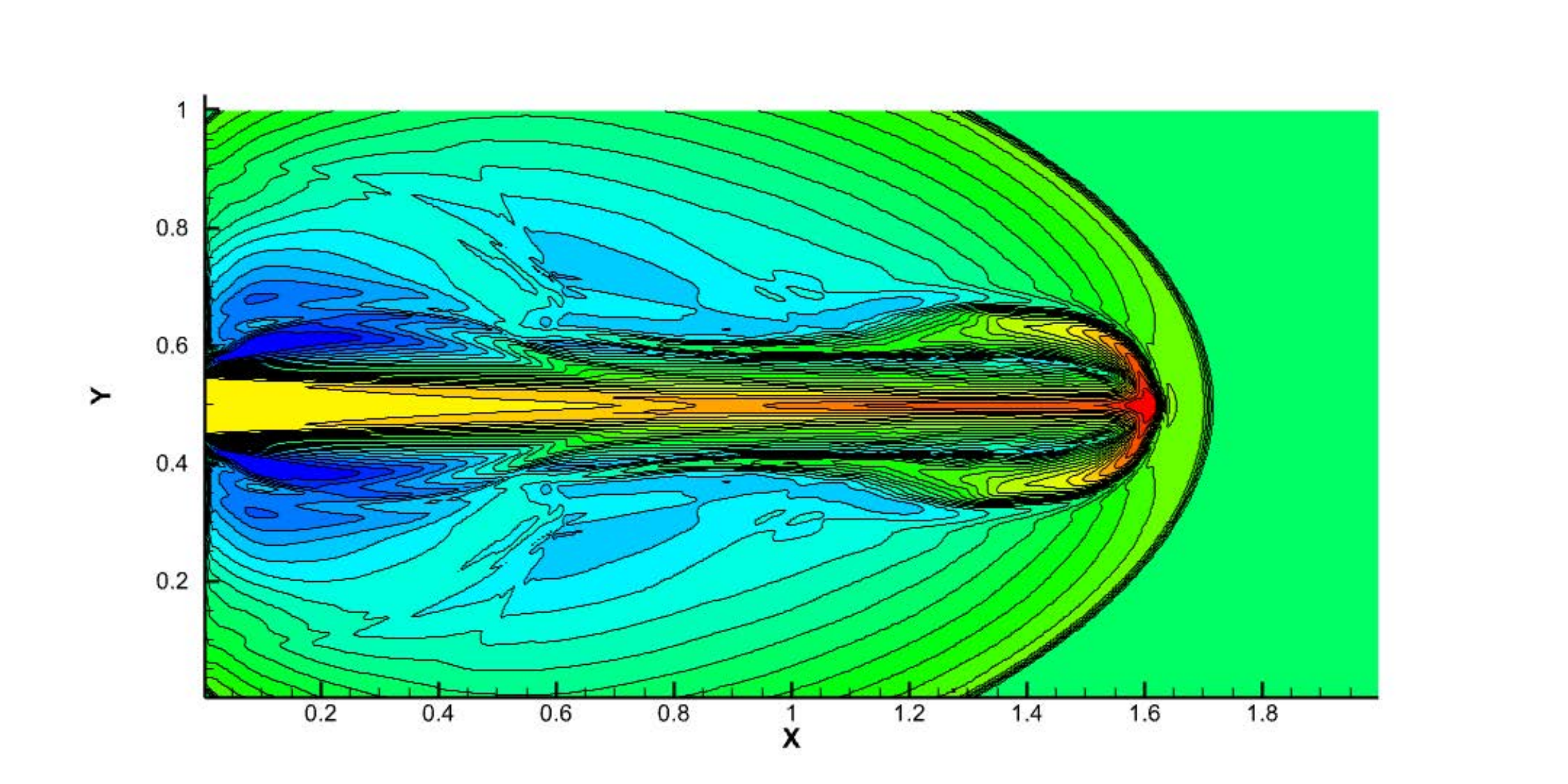}
\includegraphics[width=0.4\textwidth]{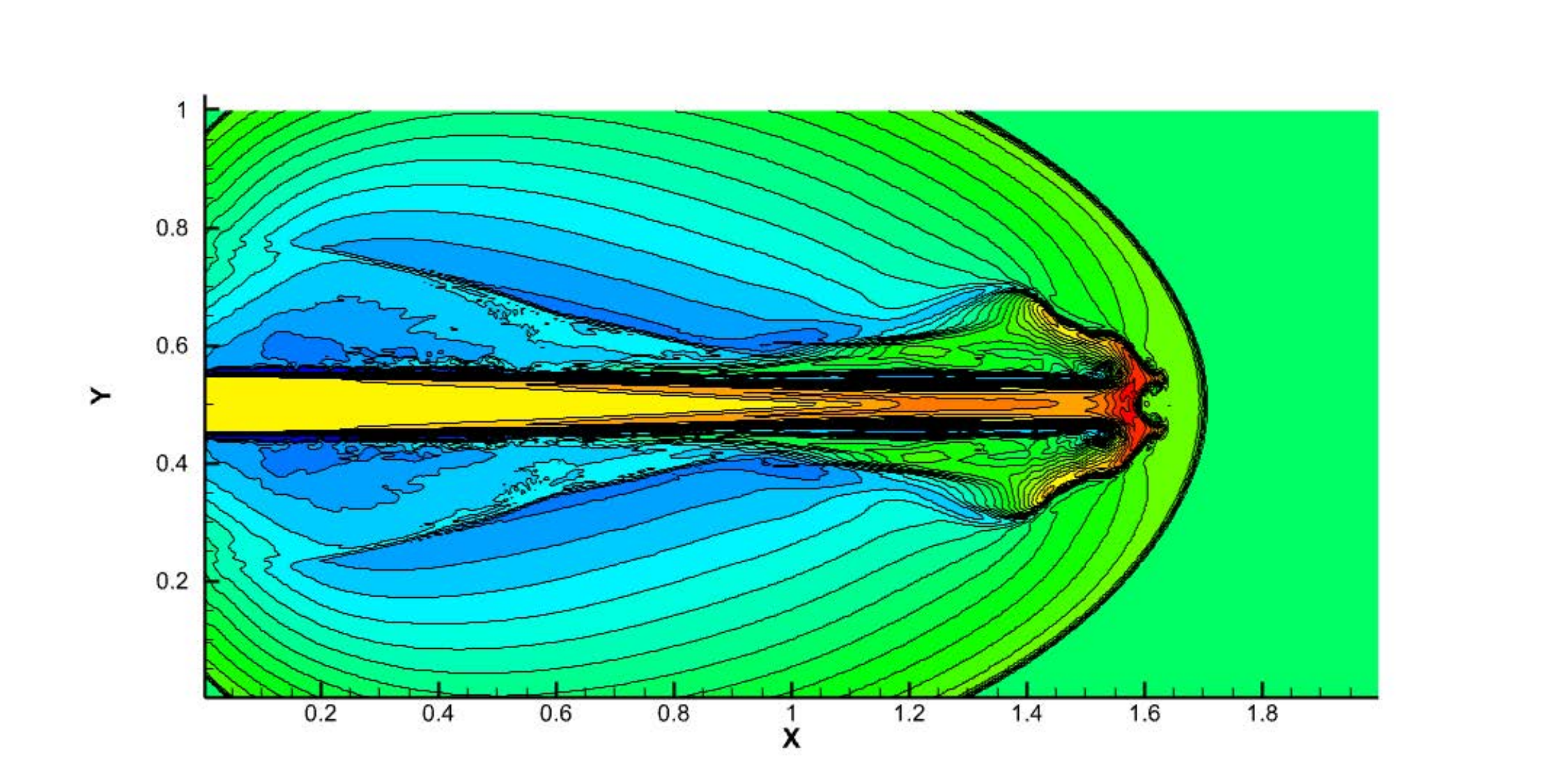}
\includegraphics[width=0.4\textwidth]{jetrhoWENO580}
\includegraphics[width=0.4\textwidth]{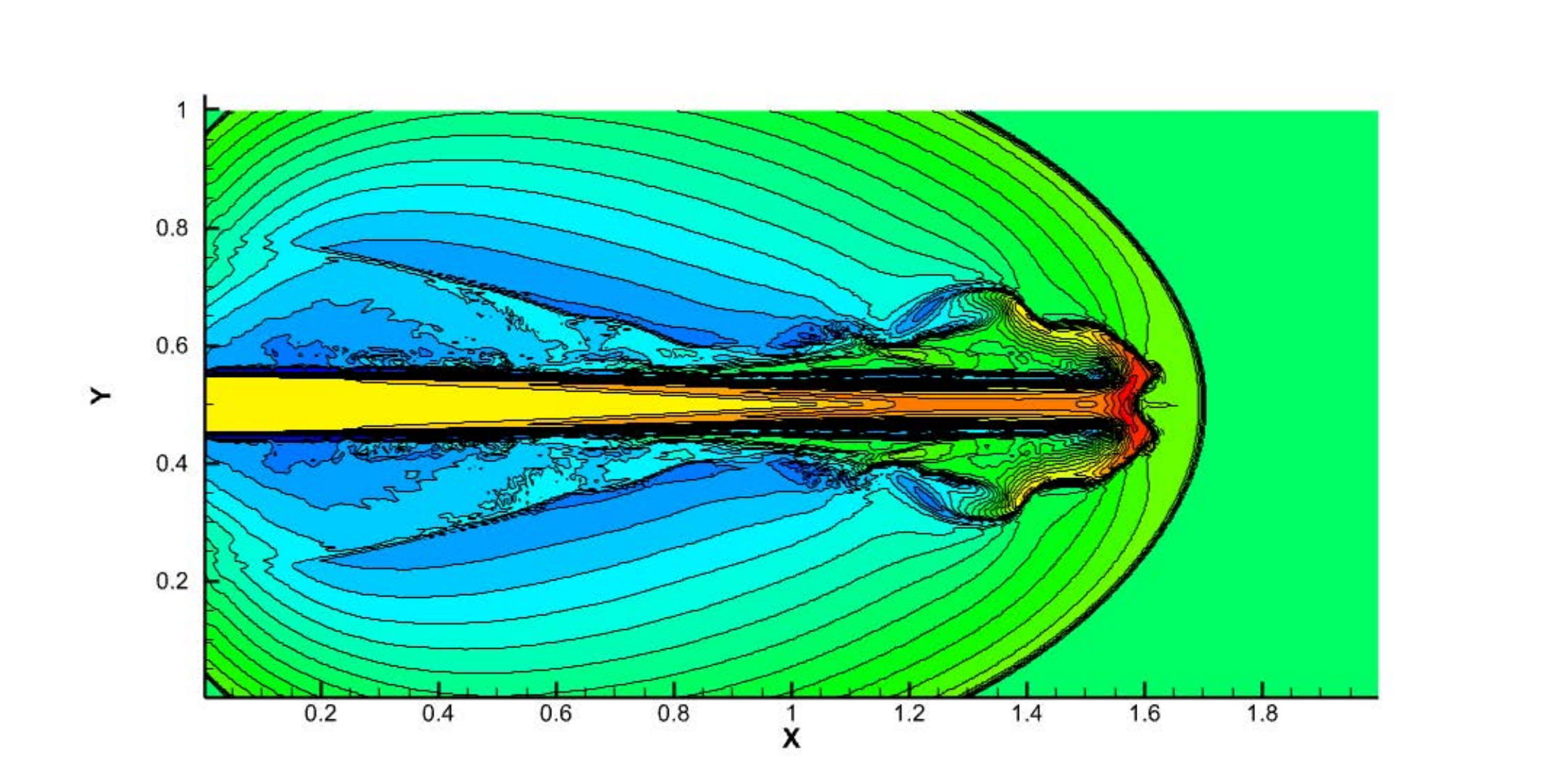}
\includegraphics[width=0.4\textwidth]{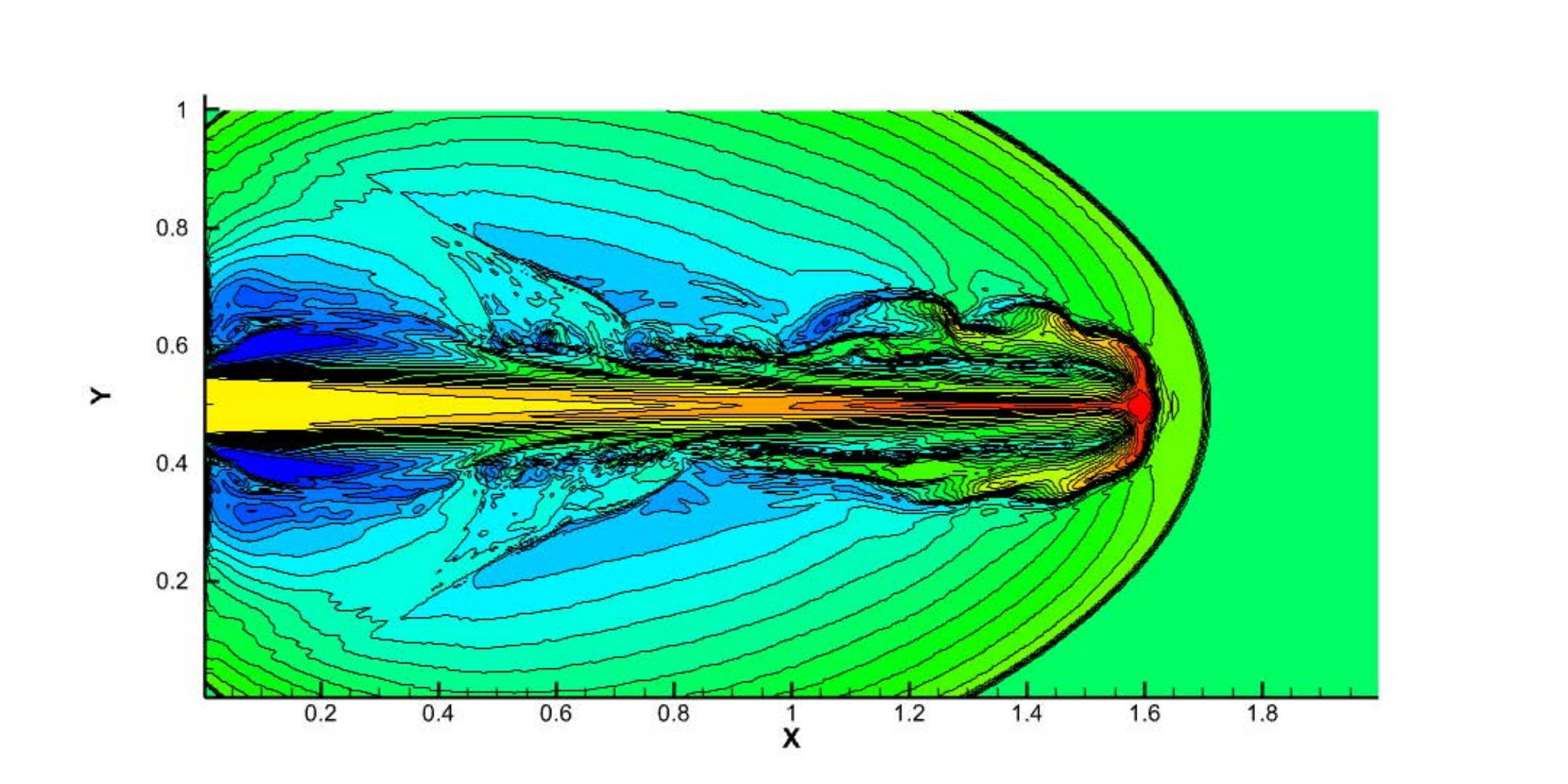}
\includegraphics[width=0.4\textwidth]{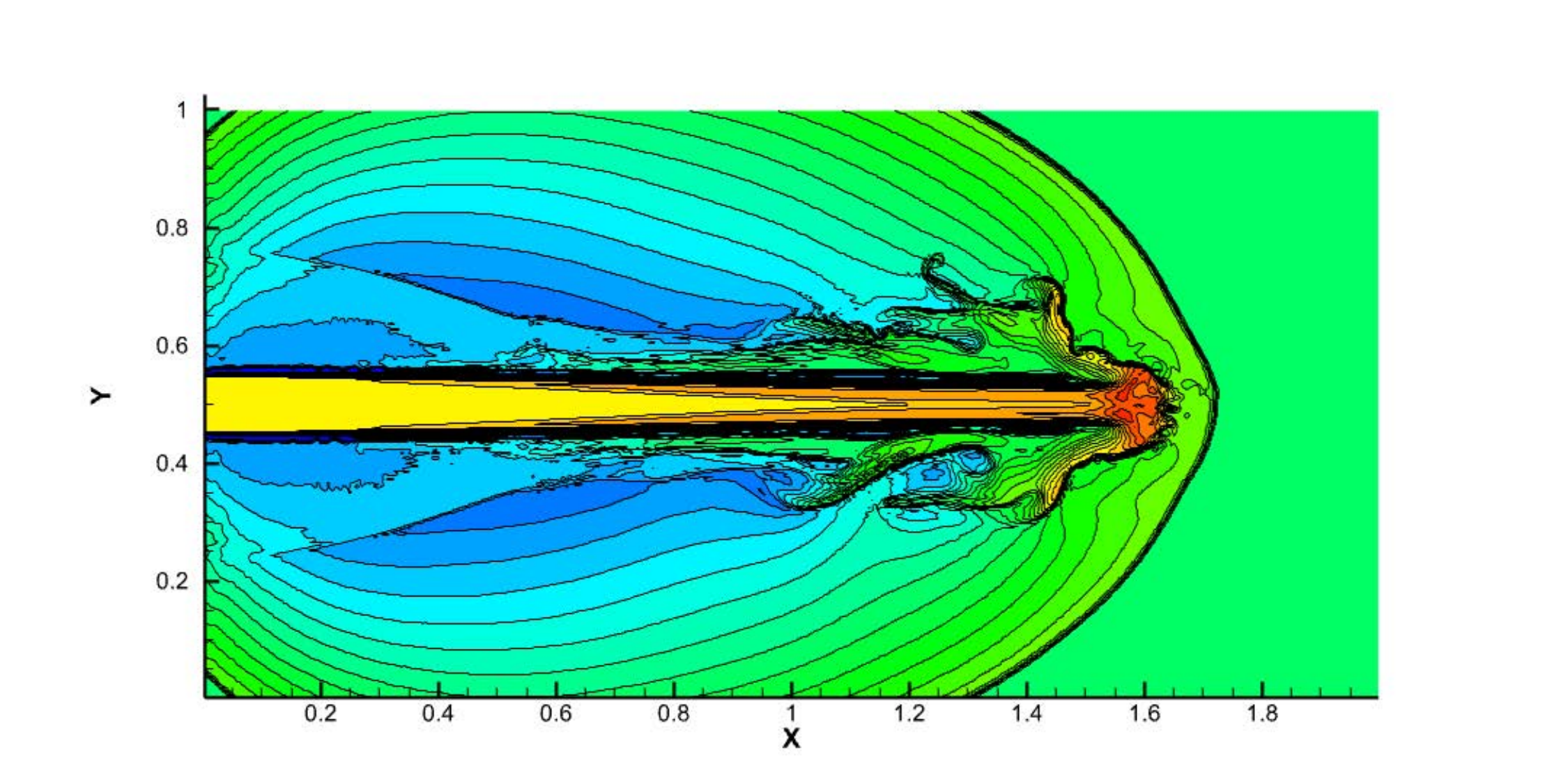}
\vspace{-4mm} \caption{\label{figjet80}
High Mach number astrophysical jet: density contours from non compact scheme with LF(left) and  compact gas-kinetic scheme(right): ASE-DFF(5,3) (top),
ASE-DFF(7,5,3) (middle) and ASE-DFF(9,7,5,3) (bottom) in logarithmic scale. The Mach number is 80 and the
resolution is $448 \times 224$.}
\end{figure}
\begin{figure}[htp]	
\centering
\includegraphics[width=0.4\textwidth]{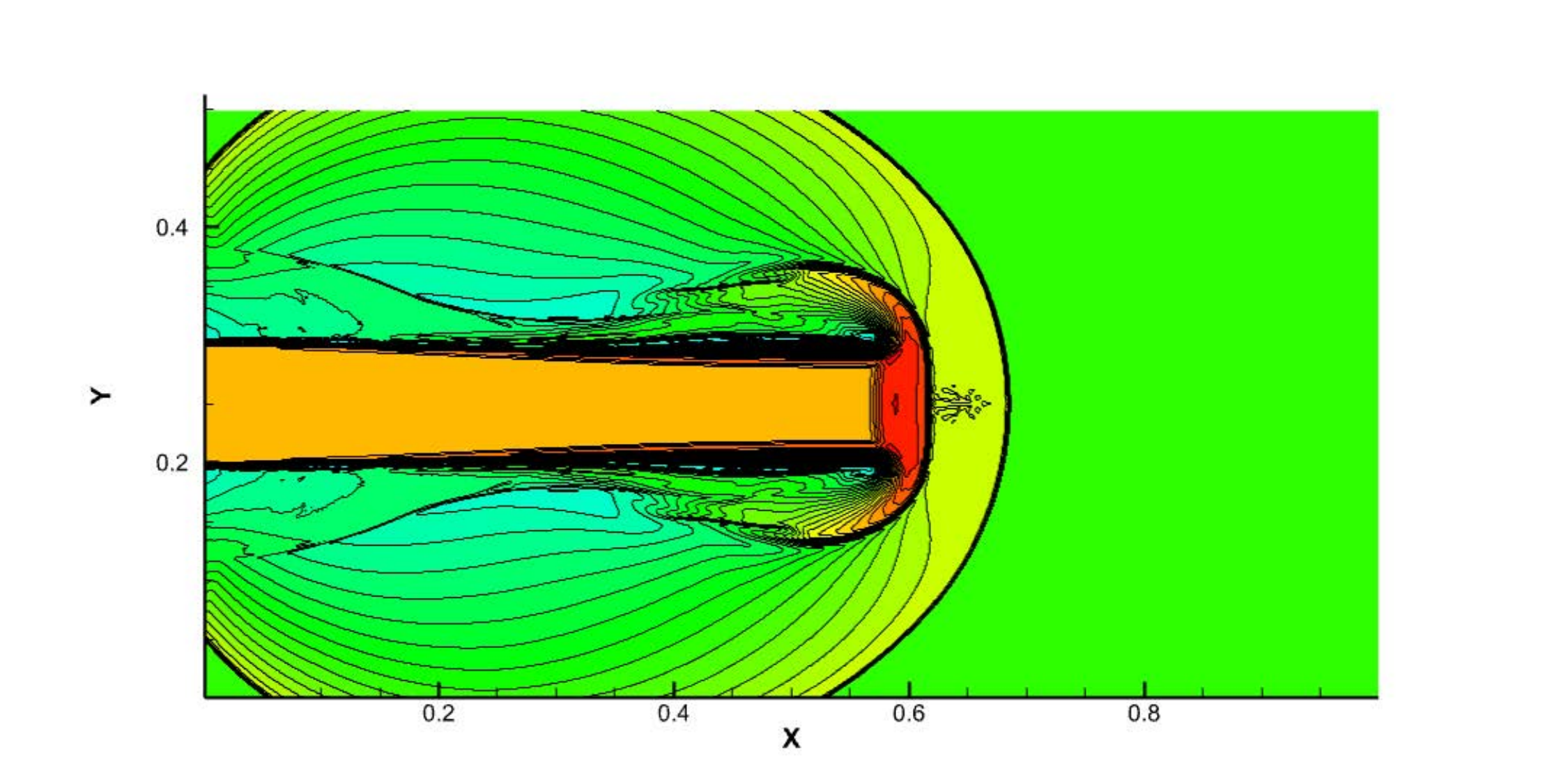}
\includegraphics[width=0.4\textwidth]{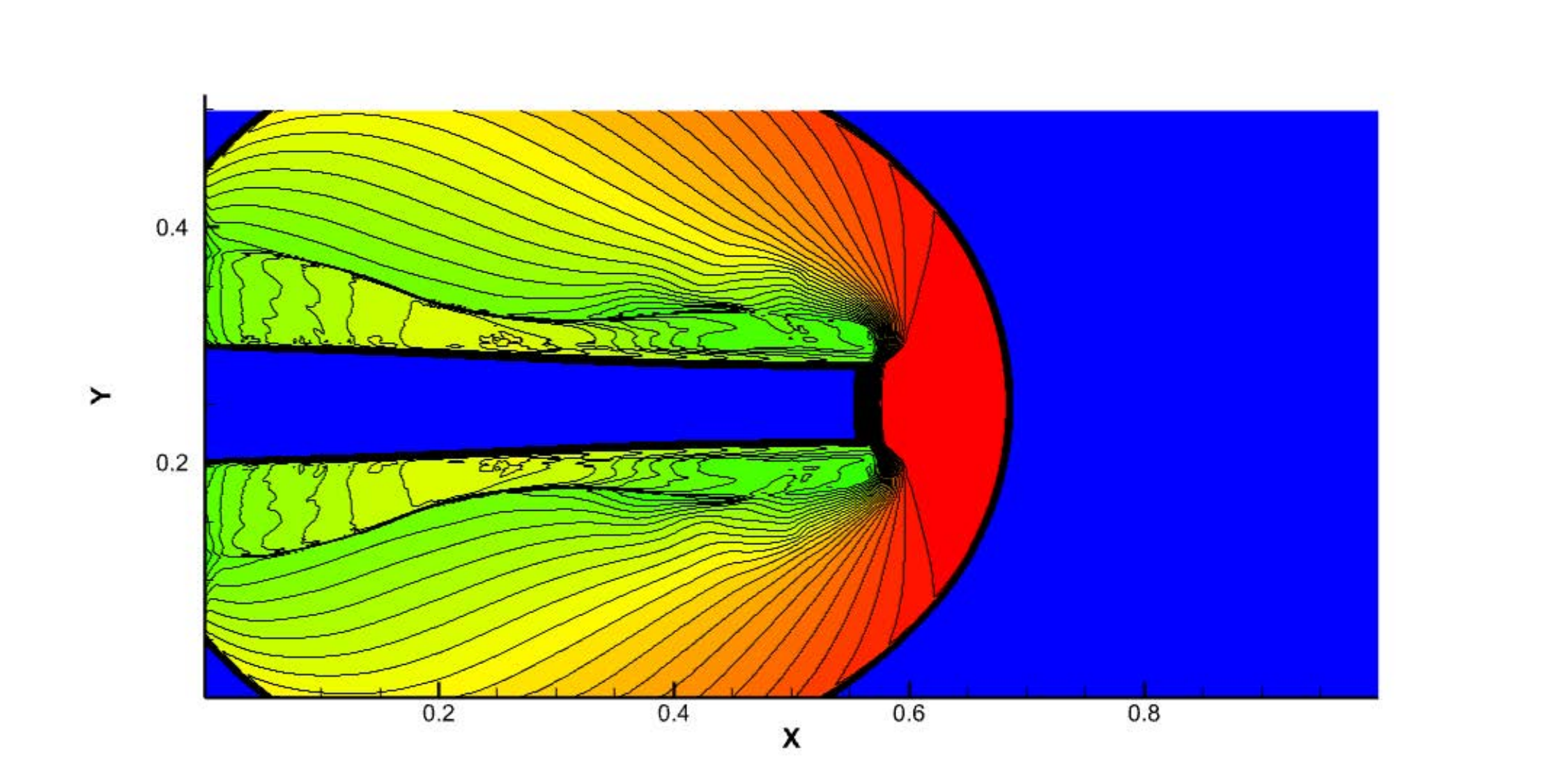}
\includegraphics[width=0.4\textwidth]{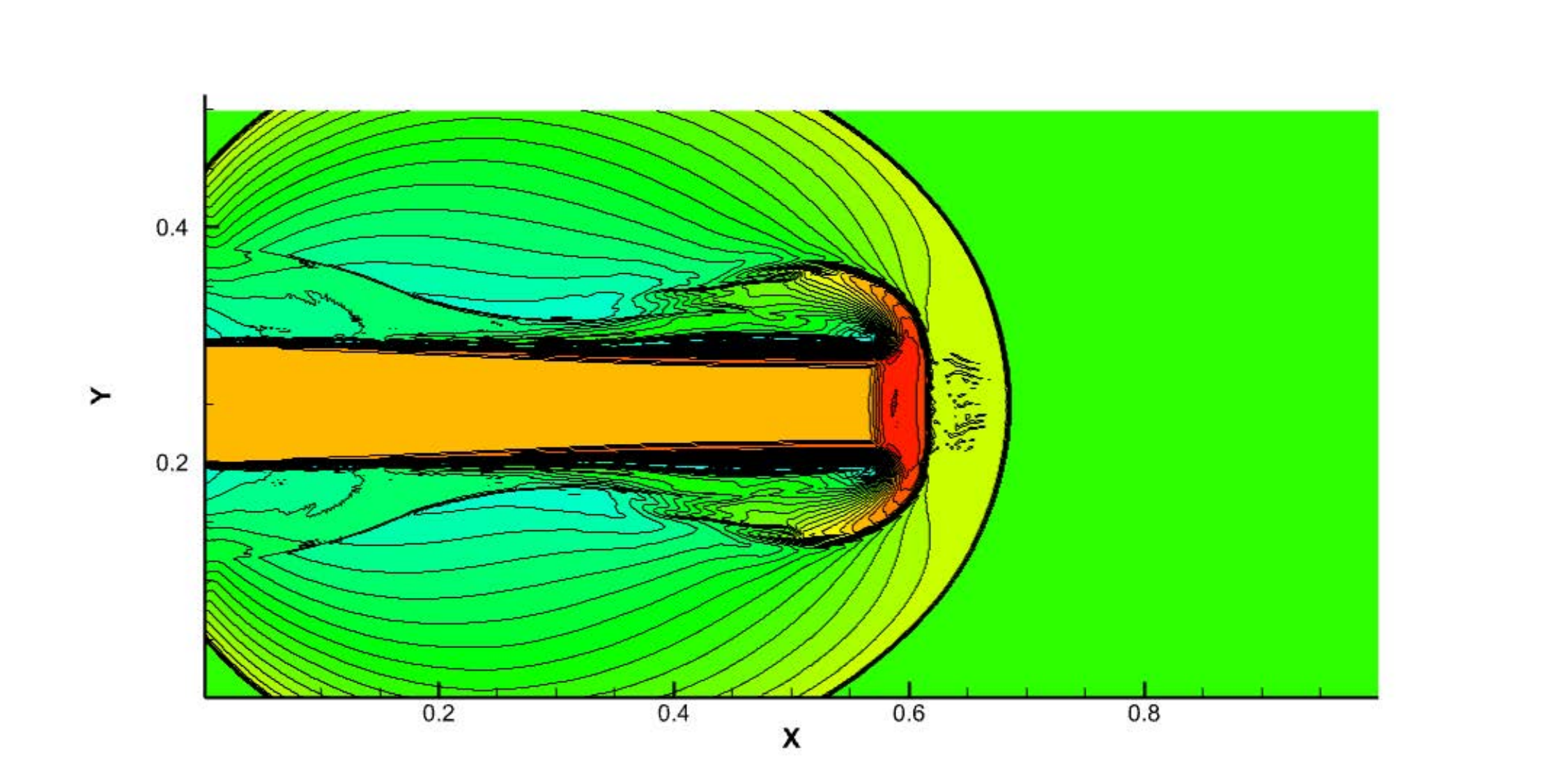}
\includegraphics[width=0.4\textwidth]{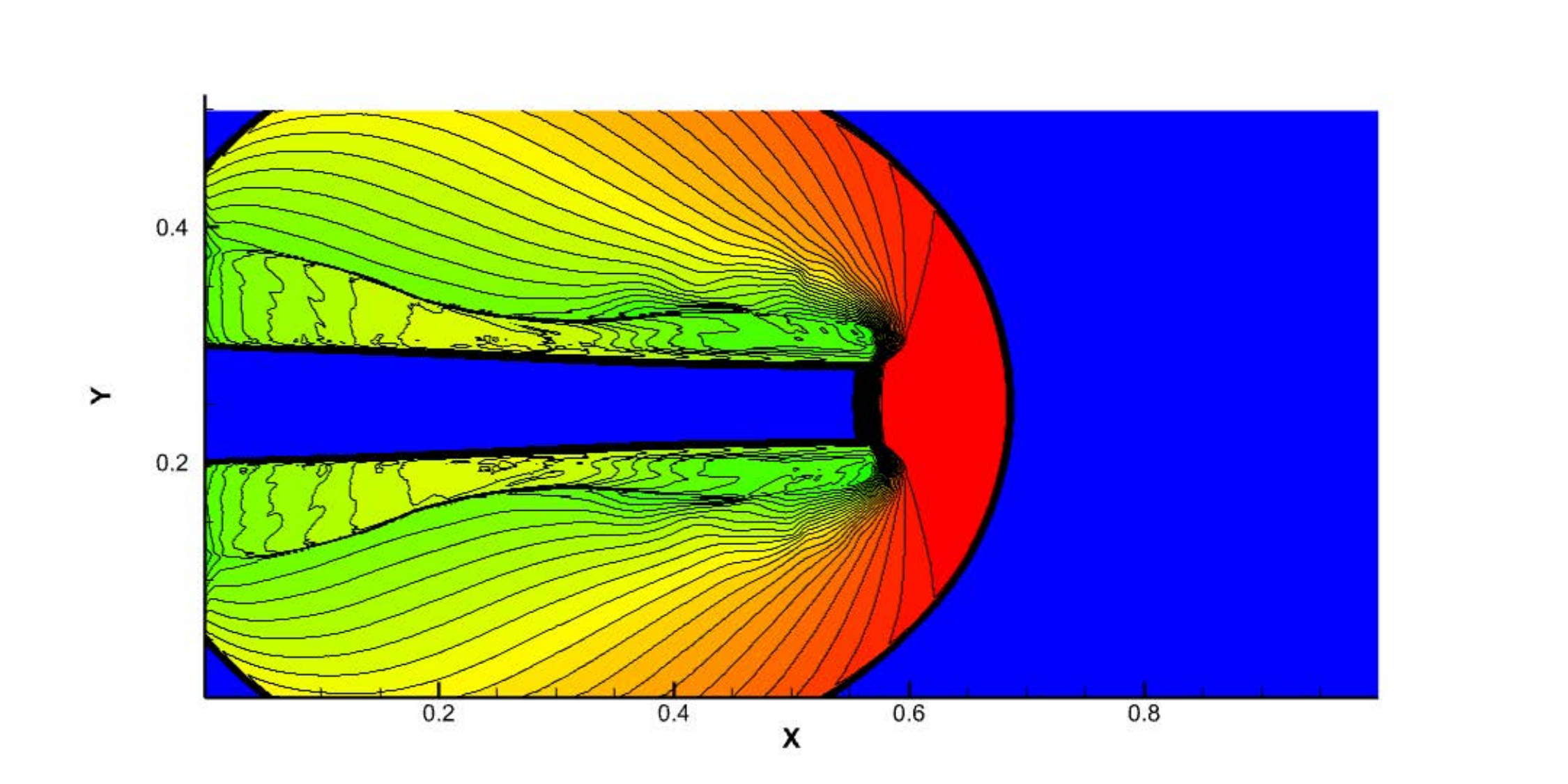}
\includegraphics[width=0.4\textwidth]{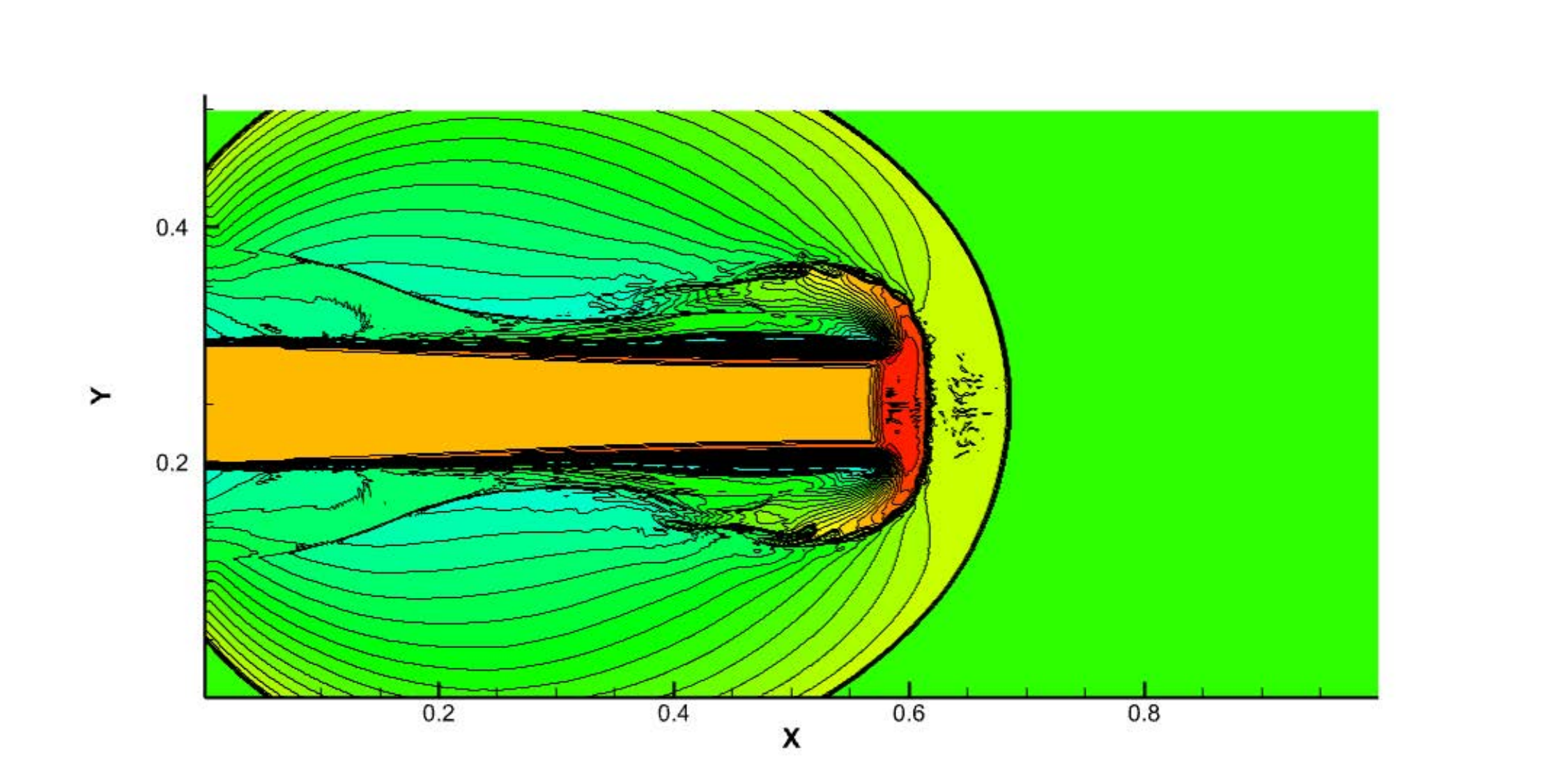}
\includegraphics[width=0.4\textwidth]{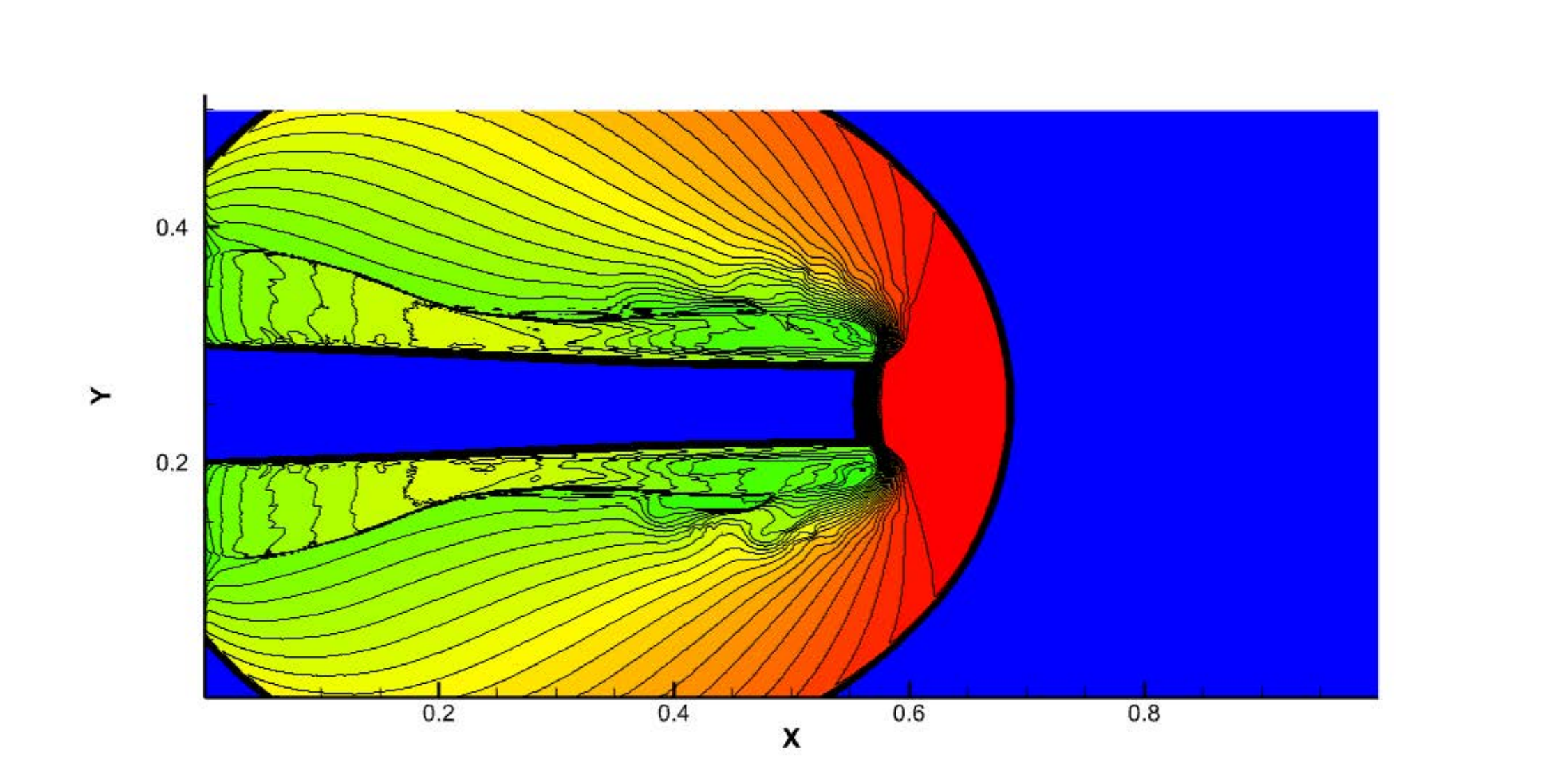}
\vspace{-4mm} \caption{\label{figjet2000}
High Mach number astrophysical jet: density (left) and pressure (right) contours from ASE-DFF(5,3)-CGKS (top),
ASE-DFF(7,5,3)-CGKS (middle) and ASE-DFF(9,7,5,3)-CGKS (bottom) in logarithmic scale. The Mach number is 2000 and
the resolution is $800 \times 400$.}
\end{figure}

As shown in Fig.~\ref{figjet80} and Fig.~\ref{figjet2000}, the ASE-DFF(5,3)-CGKS, ASE-DFF(7,5,3)-CGKS and
ASE-DFF(9,7,5,3)-CGKS have identified similar large-scale characteristic structures. Generally speaking,
this type of problem has very high requirements for the positivity of the flux solver and the robustness of
the reconstruction method. For non-compact GKS, this problem cannot be solved no matter what discontinuity feedback factor is used~\cite{zhang2024adaptivereconstructionmethodarbitrary}.
For compact-GKS, even without using the discontinuity feedback factor, this test cannot be passed. Therefore, the
robustness of the new CGKS method based on the discontinuity feedback factor has been greatly improved.
At the same time, the resolution of the higher-order CGKS has also been relatively improved. Furthermore in Fig.~\ref{figjet80}, compared with ASE-DFF-CGKS,
the non-compact results based on ASE-DFF-LF not only have the lower resolution at the top of the jet,
but also exhibit greater oscillation in the flow field. In Fig.~\ref{figjet2000},the density and pressure contours of high Mach number astrophysical jet
from ASE-DFF-CGKS is shown and remain satisfactory.\par
Based on the two-dimensional Riemann problem(configuration 3), the computational efficiency from different schemes will be evaluated.
The results of CPU time are recorded after running 20 time steps for each scheme with a single processor of Intel Xeon Gold 6133 CPU @ 2.50GHz.
In Table ~\ref{cpu}, we present the CPU time of reconstruction process for the ASE-DFF-CGKS and the HWENO-AO-CGKS respectively.
Compared with HWENO-CGKS, ASE-DFF-CGKS can save up to 50$\%$ of the CPU time in the reconstruction process.
Therefore ASE-DFF-CGKS can enhance the computational efficiency as analyzed in Section ~\ref{sec3.2}.\par
\begin{table}[htp]
\centering
\caption{The CPU time(s) of reconstruction process for the ASE-DFF-CGKS and the HWENO-AO-CGKS.}
\label{cpu}
\small
\begin{tabular}{p{2cm} p{4cm} p{4cm} p{4cm}}
\hline
Mesh size & ASE-DFF(5,3)-CGKS & ASE-DFF(7,5,3)-CGKS & ASE-DFF(9,7,5,3)-CGKS \\
\hline
100$\times$100  & 0.913 & 1.382 & 1.942 \\
200$\times$200 & 5.244 & 6.048 & 7.685 \\
400$\times$400 & 15.764 & 28.722 & 30.228 \\
800$\times$800 & 63.766 & 126.264 & 119.484 \\
\hline
Mesh size & HWENO(5,3)-CGKS & HWENO-AO(7,5,3)-CGKS & HWENO-AO(9,5,3)-CGKS \\
\hline
100$\times$100 & 1.187 & 3.105 & 4.448 \\
200$\times$200 & 6.589 & 15.064 & 17.069\\
400$\times$400 & 24.146 & 63.387 & 67.848 \\
800$\times$800 & 97.128 & 256.688 & 284.287 \\
\hline
\end{tabular}
\end{table}
The discontinuity feedback factor not only has the less computational cost but also enhances the ability of schemes to solve
shock waves and rarefaction waves. As shown in Table ~\ref{robust},, the test results of HWENO-AO-CGKS and ASE-DFF-CGKS are summarized. It can
be seen that in the Le Blanc problem and the high Mach number astrophysical jets with Ma = 80 and 2000, ASE-DFF-CGKS can pass the test.
 But HWENO-AO-CGKS fails to pass these two extreme cases.
\begin{table}[htp]
\centering
\caption{Robustness verification summary of both HWENO-type and DFF-enhanced CGKSs.}
\label{robust}
\begin{tabular}{l|c|c}
\hline
\diagbox{Test case}{Scheme} & HWENO-AO-CGKS & ASE-DFF-CGKS \\
\hline
Shock-tube problem               & $\checkmark$ & $\checkmark$ \\
Shock-density wave interaction   & $\checkmark$ & $\checkmark$ \\
Interacting blast waves          & $\checkmark$ & $\checkmark$ \\
Double rarefaction wave problem  & $\checkmark$ & $\checkmark$ \\
Le Blanc problem                 & $\times$ & $\checkmark$     \\
Two-dimensional Riemann problems & $\checkmark$ & $\checkmark$ \\
Double Mach reflection problem   & $\checkmark$ & $\checkmark$ \\
Viscous shock tubes problem      & $\checkmark$ & $\checkmark$ \\
High Mach number astrophysical jet & $\times$ & $\checkmark$   \\
\hline
\end{tabular}
\end{table}
\section{Conclusion}

This paper presents a novel, very high-order compact gas-kinetic scheme (CGKS) featuring adaptive stencil extension with a discontinuity feedback factor. The proposed scheme retains the two-stage fourth-order time-stepping method of the original CGKS and updates both the conservative variables and their gradients within each cell at every stage. With the incorporation of the discontinuity feedback factor, the CGKS achieves fifth-, seventh-, and ninth-order spatial accuracy. Notably, for compact stencils, the fifth-order compact stencil is equivalent in size to a third-order non-compact stencil, while seventh- and ninth-order compact stencils correspond to those of a fifth-order non-compact scheme.
The elimination of the need to compute smoothness indicators for high-order reconstruction polynomials significantly reduces computational cost. Furthermore, a new two-dimensional reconstruction framework simplifies the construction of non-equilibrium and equilibrium states required for the CGKS flux evolution. This approach is consistent with commonly used high-order gas-kinetic schemes, utilizing a kinetic weighting method to construct all equilibrium states.
One- and two-dimensional accuracy tests demonstrate that the CGKS with the discontinuity feedback factor can achieve up to ninth-order accuracy. Moreover, a series of test cases involving shock and rarefaction waves confirm that this compact, high-order gas-kinetic scheme possesses excellent shock-capturing capabilities and robustness. Its low-dissipation characteristics enable high resolution at discontinuities, and the CFL time step used in numerical tests remains relatively large.
Overall, the newly developed and simplified CGKS based on the discontinuity feedback factor achieves up to ninth-order accuracy, demonstrating significant potential for attaining even higher accuracy on compact stencils in future applications involving unstructured meshes.

\section*{Acknowledgment}
The authors would like to thank Haolin Liu, Qingdian Zhang, Yixiao Wang, Hongyu Liu, Cong shi,  Qihui Gao, YaLing Wu and Zhiwen Zhuang for helpful discussions.
The current research is supported by the National Natural Science Foundation of China (12302378, 12172316, 92371201, and
92371107), the Funding of National Key Laboratory of Computational
Physics, the National Key R\&D Program of China (Grant
No. 2022YFA1004500), the Natural Science Basic Research Plan in Shaanxi Province of China (No.
2025SYS-SYSZD-070), and the Hong Kong Research Grant Council (16301222, and 16208324).
\bibliographystyle{plain}
\bibliography{mujlbib}
	
\end{document}